\documentclass[aps,prl,reprint,superscript address]{revtex4-1}

\usepackage{graphicx}
\usepackage[normalem]{ulem}
\usepackage{amssymb,amsthm,amsfonts,mathrsfs,dsfont,ctable,commath,sansmath,mathtools}
\usepackage{physics}
\usepackage{color}
\usepackage{paralist}
\usepackage{verbatim}

\usepackage[caption=false]{subfig}
\usepackage[unicode=true,pdfusetitle,
 bookmarks=true,bookmarksnumbered=false,bookmarksopen=false,
 breaklinks=false,pdfborder={0 0 0},pdfborderstyle={},backref=false,colorlinks=true]
 {hyperref}

\usepackage{xcolor}

\newtheorem{theorem}{Theorem}

\newtheorem{definition}{Definition}
\newtheorem{proposition}{Proposition}

\newtheorem{lemma}{Lemma}

       \newcommand{\tH}{\mathcal{H}}  

\newcommand{\Z}{\mathbb{Z}}
\newcommand{\R}{\mathbb{R}}

\newcommand{\spc}[1]{\mathcal{#1}}

\def\d{{\rm d}}

\newcommand{\Span}{{\mathsf{Span}}}
\newcommand{\Lin}{\mathsf{Lin}}

\def\>{\rangle}
\def\<{\langle}

\newcommand{\st}[1]{\mathbf{#1}}

\newcommand{\Vac}[1]{\mathrm{Vac}}

\newcommand{\map}[1]{\mathcal{#1}}
\newcommand{\Herm}{\mathsf{Herm}}

\newcommand{\St}{{\mathsf{St}}}
\newcommand{\Eff}{{\mathsf{Eff}}}
\newcommand{\Pur}{{\mathsf{Pur}}}
\newcommand{\Transf}{{\mathsf{Transf}}}

\newtheorem{theo}{Theorem}

\newtheorem{prop}{Proposition}
\newtheorem{cor}{Corollary}
\newtheorem{defi}{Definition}

\newcommand{\Proof}{{\bf Proof. \,}}

\begin{document}

	\title{Bell Nonlocality in Classical Systems Coexisting with other System Types
}

\author{Giulio Chiribella}
\email{giulio.chiribella@cs.ox.ac.uk}
\affiliation{\footnotesize QICI Quantum Information and Computation Initiative, Department of Computer Science, The University of Hong Kong, Pok Fu Lam Road, Hong Kong}
\affiliation{\footnotesize Quantum Group, Department of Computer Science, University of Oxford, Wolfson Building, Parks Road, Oxford, United Kingdom}
\affiliation{\footnotesize Perimeter Institute for Theoretical Physics, 31 Caroline Street North, Waterloo, Ontario, Canada}

\author{Lorenzo Giannelli}
\email{giannell@connect.hku.hk}
\affiliation{\footnotesize QICI Quantum Information and Computation Initiative, Department of Computer Science, The University of Hong Kong, Pok Fu Lam Road, Hong Kong}
\affiliation{\footnotesize HKU-Oxford Joint Laboratory for Quantum Information and Computation}

\author{Carlo Maria Scandolo}
\email{carlomaria.scandolo@ucalgary.ca}
\affiliation{\footnotesize Department of Mathematics \& Statistics, University of Calgary, 2500 University Drive NW, Calgary, Alberta, Canada}
\affiliation{\footnotesize Institute for Quantum Science and Technology, University of Calgary, 2500 University Drive NW, Calgary, Alberta, Canada}

\begin{abstract}
	The realistic interpretation of classical theory assumes that every  classical system has well-defined properties,  which may be unknown to the observer but are nevertheless part of reality and can, in principle, be revealed by measurements. 
	Here we show that this interpretation can, in principle, be falsified if classical systems coexist with other types of physical systems. To make this point, we construct a toy theory that {\em (i)} includes  classical theory as a subtheory and {\em (ii)} allows classical systems to be entangled with another type of systems, called anticlassical. We show that our toy theory allows for the violation of Bell inequalities in  two-party scenarios where one of the  settings   corresponds to a local measurement performed on a classical system alone. Building on this fact, we show  that measurement outcomes  in classical theory cannot, in general, be regarded as predetermined by the state of an underlying reality.  
\end{abstract}

\maketitle

\paragraph{Introduction.} Since the early days of Galileo and Newton, classical theory has been  regarded as the golden standard of a physical  theory that describes reality 
without any fundamental uncertainty.  
In this view, every classical system is assumed to be in a well-defined state, which  may be unknown to the observer, but is nevertheless  part of the physical reality. Statistical mixtures only arise from the observer's ignorance about the true state of the system, and, in principle, this ignorance can always be overcome by performing  measurements. In modern terminology, the  view that classical systems are fundamentally in well-defined (pure) states can be summarized by the statement  that   classical pure states are   {\em ontic}, while classical mixed states are  {\em epistemic}~\cite{hardy2004quantum,spekkens2005contextuality,leifer2014quantum}.    This statement,  combined with the idea that classical measurements reveal some preexisting  properties  of the measured systems, lies at the core of the realistic interpretation of classical theory.  

\begin{figure}[t!]
	\hspace*{-.65cm}
	\includegraphics[width=0.55\textwidth]{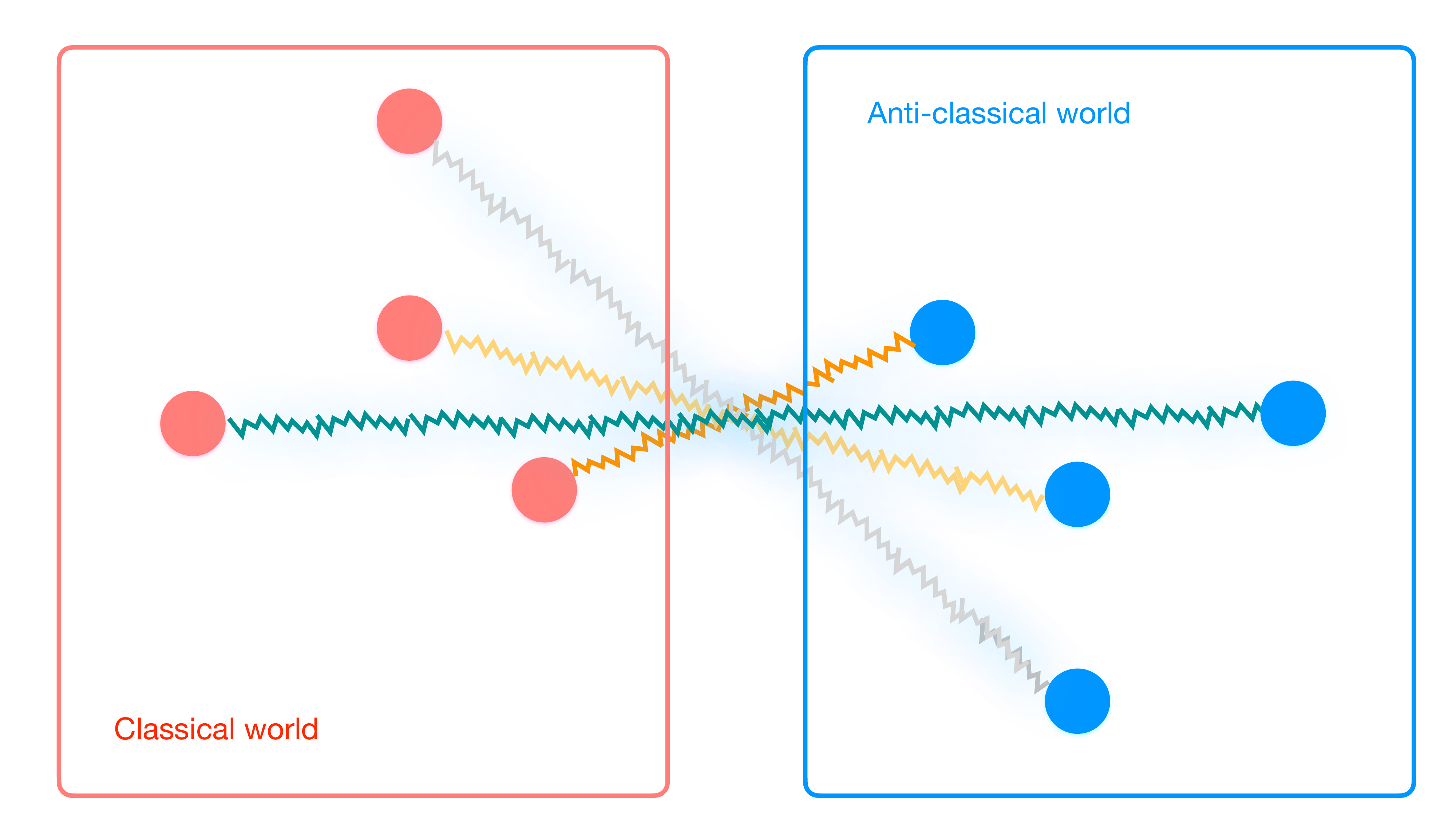} 
	\caption{In a  universe described by our toy theory, an observer who has access only to classical systems  (represented by red disks on the left) would see a  world described by classical theory.    The same situation applies to an observer with access only to anticlassical systems (blue disks on the right).  In contrast, observers with  access to both types of systems  can observe Bell nonlocality and other nonclassical features.}
	\label{fig:classicalanticlassical}
\end{figure}

In this Letter we show that, contrary to widespread belief, 
a realistic interpretation of classical theory is not always logically possible:  while such interpretation is consistent with all experiments involving only classical systems, it can become, in principle, falsifiable if classical systems are considered alongside other types of physical systems. To make this point, we construct a toy theory  that includes classical theory as a subtheory, meaning that it coincides with classical  theory when restricted to  a subset of the possible physical systems.  In addition to all classical systems, the toy theory includes another type of systems, called anticlassical, as illustrated in Fig.~\ref{fig:classicalanticlassical}.  An observer who has access only to classical systems cannot see any difference between classical theory and our toy theory:    all measurements are, in principle, compatible, all pure states are perfectly distinguishable through measurements, and all the states of all composite systems are separable.   In contrast, we show  that  observers with joint access to both types of systems can, in principle, observe nonclassical features such as Bell nonlocality \cite{brunner2014bell}.

Crucially, we show that our toy theory allows for a maximal violation of the Clauser-Horne-Shimony-Holt (CHSH) inequality \cite{CHSH, CHSH-game} in scenarios where     where one of the settings     corresponds to a local  measurement  performed on a classical system. Building on this result, we prove that  measurement outcomes in classical theory cannot, in general, be regarded as predetermined. Finally, we show that, under mild  assumptions, the predictions of our toy theory cannot be reproduced by any deeper theory that describes reality as a list of individual properties of classical  and anticlassical systems. This result indicates that,  no matter whether the properties of classical systems are accessible through measurements or not, their full specification is not sufficient, in general,  to account for the correlations between classical systems and other types of  physical systems.

While our toy theory is not meant to be a description of  the world,  it makes an important conceptual point:    the realistic interpretation of classical theory 
can, in principle, be falsified  if  classical systems exist  alongside other types of physical systems.  Notably,  our toy theory cannot be ruled out from within classical theory: every  classical phenomenon is, in principle, compatible with the existence of some yet-unobserved type of system that prevents the assignment of  definite values to classical variables prior to measurement.

Our results complement recent works by Gisin and Del Santo \cite{gisin2019indeterminism, delsanto2019physics}, who challenged the determinism of classical physics  on the grounds of the impossibility to specify  real-valued variables like position and momentum  with infinite precision. 
In our work,  the impossibility to assign a predefined value to classical variables arises from correlations with some other physical systems, rather than precision limits in the definition of real numbers. As such, our results apply also to classical bits and other discrete classical variables. It is also worth mentioning that physical arguments in favor of classical indeterminism could also be put forward by setting up a dynamical  interaction between classical and quantum systems (see e.g., \cite{blanchard1995events, diosi1995quantum, oppenheim2023postquantum}.)    In the existing frameworks, however, classical and quantum systems cannot be entangled, and therefore there cannot be any CHSH violation when one of the settings corresponds to a measurement on a classical system alone.   In this respect, our toy theory exhibits a stronger form of indeterminism.

\paragraph{Classical and anticlassical systems.}  To formulate  our toy theory we adopt the framework of general probabilistic theories \cite{hardy2001, PhysRevA.75.032304, PhysRevLett.99.240501, hardy2011foliable, hardy_2013, capitolo-hardy}, in the specific version known  as operational probabilistic theories (OPTs)~\cite{chiribella2010probabilistic,inf-derivation-CDP,qt-from-principles-CDP,capitolo-opt,hardy2013reconstructing, ScandoloPhD}.  An OPT describes a set of physical systems, closed under composition, and a set of transformations thereof, closed under parallel and sequential composition. Mathematically, the compositional structure is underpinned by the graphical language of process theories \cite{category1, category2, category3, coecke_kissinger_2017}.

Classical theory can be regarded as a special case of an OPT  \cite{hardy2011foliable,Objectivity}: precisely, it is the largest OPT where   {\em (i)} the  pure states of every given system are perfectly  distinguishable through a single measurement, {\em (ii)} the pure states of every composite system are the products of pure states of the component systems, and  {\em (iii)} all permutations of the set of pure states are valid physical transformations. For simplicity,  we will focus on the classical theory of discrete systems such as bits and their generalizations.

We now construct a toy theory that includes classical theory as a subtheory, meaning that our toy theory coincides with classical theory when restricted to a subset of  physical systems that includes all discrete classical  systems.   
A classical system with $d$ perfectly distinguishable pure states, conventionally denoted by $0,1,
\dots, d-1 $,  will be called a dit (or a bit in the special case $d=2$.)    The mixed states of a dit  are probability distributions of the form $(p_i)_{i=0}^{d-1}$, with $p_i\ge 0  \, ,\forall \, i$ and $\sum_{i=0}^{d-1}  p_i  =1$.  
The reversible processes acting on the dit are permutations of its pure states, while general noisy processes are described by transition probabilities $p(j|i)$.   Similarly, a (generally noisy) measurement  with outcomes in a set $\sf A$  can be represented by transition probabilities $p(a| i)$, yielding the probability of the outcome $a$ when the  dit is in the state $i$.  

An  equivalent way to represent classical states, processes, and measurements, commonly used in the quantum information literature (see e.g., \cite{heinosaari2011mathematical}),  is provided by diagonal matrices.   Specifically, probability distributions  $(p_i)_{i=0}^{d-1}$  can be equivalently represented by  $d\times d$ diagonal matrices of the form $\rho = \sum_{i=0}^{d-1} p_i |i\>\<i|$, where  $\{|i\>\}_{i=0}^{d-1}$ is the canonical orthonormal basis for $\mathbb{C}^d$.  A general process with transition probabilities $p(j|i)$ is  described by a linear map of the form $\map M  (\rho) =  \sum_{i,j}    \,  p(j|i)  \,  |j\>\<j|  \, \<i|\rho|i\>$.   Finally, a measurement with outcomes in the set $\sf A$ is described by  a positive operator-valued measure (POVM)  $(P_a)_{a\in\sf A}$ of the form  $P_a =   \sum_{i=0}^{d-1} \, p(a|i) \, |i\>\<i|$, and the outcome probabilities  can be computed with the Born rule $p (a|i) = \<i|   P_a  |i\> $.

In our toy theory, classical systems coexist with another type of systems, called anticlassical. The anticlassical systems can be viewed as a mirror image of the classical systems: for every classical system type, there exists a corresponding anticlassical system type with exactly the same state space, the same set of physical transformations, and the same set of measurements. To help intuition, one can think of the  distinction between classical and anticlassical systems as analogous to the distinction between  particles and antiparticles,  which have the same state spaces and yet are distinguishable by some external property, such as their charge.    

While classical and anticlassical systems are  described by classical probability theory when considered separately, composite systems including both types of systems exhibit nonclassical features. In the following, we present the simplest version of our toy theory, which describes  arbitrary composite systems made of $m$ bits and $n$ antibits, hereafter  called $(m,n)$ composites. $(m,0)$ and $(0,n)$ composites will be described by classical  theory, while the non-classical behaviours will emerge when both $m$ and $n$ are non-zero.  The generalization to basic systems of arbitrary dimension, as well as the full specification of the allowed states, measurements, and processes, is provided in Supplemental Material \cite{supp}.

\paragraph{Nonclassical composites.} The simplest nonclassical composite is the $(1,1)$ composite,  consisting of a bit and an antibit. In this case,  the pure states  are represented by rank-one projectors onto  unit vectors  $|\Psi\>$ with well-defined parity, that is, unit vectors satisfying either the condition $\Pi_0 |\Psi\>=  |\Psi\> $ or the condition $\Pi_1 |\Psi\>=  |\Psi\> $, where $\Pi_0$  ($\Pi_1$) is the projector on the subspace spanned by the vectors $\{|0\>  |0\>,  |1\> |1\>\}$   ($\{|0\>   |1\>  ,  |1\>  |0\>\}$). The mixed states  of a bit and an antibit are described by density matrices  of the form  $\rho  =  \sum_j  \,   q_j  \,  |\Psi_j\>\<\Psi_j|$, where  $(|\Psi_j\>)_j$ are pure states and $(q_j)_j$ is a probability distribution.  For  an $(m,m)$ composite,   the most general pure state is a unit vector of the form 
$|\Psi\>  = \left[ I_{(B_1\dots B_m)} \otimes U^{(A_1\dots A_m)} \right]     \,  |\Psi'\>$, where $U^{(A_1\dots A_m)}$  is a unitary operator that permutes $m$ bits (antibits) and $|\Psi'\>$ is a unit vector satisfying the condition 
\begin{align}\label{generalpure}
	|\Psi'\>  =   \left( \Pi_{k_1}^{B_1A_1}  \otimes \cdots \otimes \Pi^{B_m A_m}_{k_m}\right)  \,     |\Psi'\>\, ,
\end{align}
for given vector $(k_1,  \dots, k_m)  \in  \{0,1\}^{\times m}$, where we used the notation $\Pi^{B_i A_i}_{k_i}$ for the projector onto the subspace of the composite system of the $i$th bit and $i$th antibit  with fixed parity $k_i\in \{0,1\}$.   

The pure states of arbitrary $(m,n)$ composites are defined in Supplemental Material \cite{supp}. General mixed states are defined as density matrices that are convex combinations of rank-one density matrices associated with the above pure states. Measurements on system $S$ are defined as POVMs $\{P_i\}_{i=1}^k$ whose operators are linear combinations, with positive coefficients, of the allowed states and satisfy the normalization property $\sum_{i=1}^k  P_i  =  I_S$. The outcome probabilities are then given by the Born rule $p_i = \Tr [ P_i\rho ]$.   
With these definitions, states and measurements satisfy a fundamental consistency condition: when a subset of the systems is measured, the conditional state of the remaining systems is still a valid state allowed by our toy theory.    We call this condition {\em consistency of the conditional states} and prove it in Supplemental Material~\cite{supp}, where we also show that  similar consistency properties hold for all processes in our toy theory.  In particular,  all the  multipartite states, processes, and measurements allowed by our toy theory coincide with  the states, processes, and measurements of classical theory once all the anticlassical systems are eliminated.   

It is worth noting that, unlike classical theory and standard quantum theory on the complex field, our toy theory does not satisfy  local tomography \cite{araki, wootters1990local, hardy2001, PhysRevA.75.032304, d2010probabilistic, purificationT-CDP, hardy2011},  the property that  the states of composite systems are completely characterized by the correlations of local measurements. While this property holds separately for all classical systems and for all anticlassical systems, it fails to hold when classical and anticlassical systems are combined together. 

The violation of local tomography is not an accident, but rather a necessary condition for obtaining   nonclassical composites out of  systems with classical state spaces \cite{classical-with-entanglement} (see also the no-go theorem in \cite{aubrun2021entangleability} where local tomography is implicit in the choice of possible tensor products).     Nevertheless, we show that our toy theory satisfies a  weaker locality property, known as bilocal tomography \cite{hardy2012limited}, for all multipartite systems consisting of bits and antibits: any arbitrary state of $m$ bits and $n$ antibits can be fully characterized by the correlations of  measurements performed on  pairs of bits and antibits.   A proof of this fact is provided in Supplemental Material \cite{supp}.  Other examples of physical theories that violate local tomography but satisfy bilocal tomography are quantum theory on real vector spaces \cite{araki, wootters1990local},  fermionic quantum theory \cite{fermionic, fermionic-2}, and  doubled quantum theory ~\cite{Purity,ScandoloPhD}.

\paragraph{Classical mixtures from entanglement.} It is immediate to see that every mixed state of a classical bit can be obtained from a pure state of the composite system by discarding the antibit.   For example, the generic mixed state $\rho  =  p\,  |0\>\<0|  +  (1-p) \,  |1\>\<1|$ can be obtained from the pure entangled state $|\Psi\>  =  \sqrt{  p}\,  |0\>  |0\>  +  \sqrt{1-p}\, |1\>  |1\>$.    In other words, every mixed state of a classical bit admits a {\em purification} \cite{purificationT-CDP, qt-from-principles-CDP}. 

In the rest of the Letter, we discuss the implications of purification for the interpretation of classical physics. Let us first assume, for the sake of argument, that our toy theory   describes nature at the  fundamental ({\em i.e.}, ontic) level. 	In this setting, the  claim that  every classical system  must be in a pure state at the ontic level would imply that the joint states of a bit-antibit pair are  always of the separable form $\Sigma =  q\,  |0\>\<0|  \otimes \rho_0 +  (1-q)\,  |1\>\<1| \otimes \rho_1$, for some probability $q \in  [0,1]$ and some states $\rho_0$ and $\rho_1$ of the antibit (see Supplemental Material~\cite{supp}). However, this condition is manifestly in contradiction with the existence of pure entangled states. 
Operationally, any pure entangled state of a bit-antibit pair  can  be distinguished from all separable states  by performing a measurement allowed by the toy theory.   For example,     the pure entangled state $|\Psi\>  =  \sqrt{  p}\,  |0\>  |0\>  +  \sqrt{1-p}\, |1\>  |1\>$, $p  \in  (0,1)$, can be distinguished with a guaranteed success probability of at least $\min\{p,1-p\}$ from all separable states  (see Supplemental Material~\cite{supp}). 
Hence, we conclude that, in a world where our toy theory is  fundamental, the belief that classical systems must always be in some (possibly unknown) pure states can be experimentally falsified. 

\paragraph{Bell nonlocality.}   
\begin{figure}[t!]
	\includegraphics[width=0.48\textwidth]{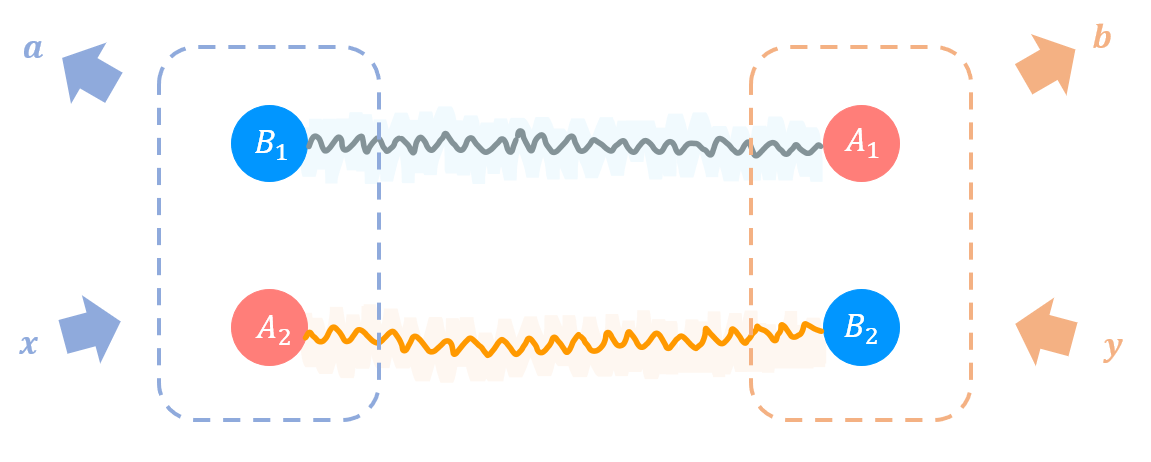} 
	\caption{\textbf{Activation of Bell nonlocality with bit-antibit entangled pairs.}   Alice (left) and Bob (right) perform local measurements on two copies of  an  entangled state of a bit-antibit pair. The first copy (top)  involves bit $B_1$ and antibit $A_1$,  while the second copy (bottom) involves  bit $B_2$ and antibit $A_2$.   Alice's and Bob's laboratories (represented by dotted boxes) contain  systems $B_1A_2$ and $A_2 B_1$, respectively. Their measurements have settings $x$ and $y$, respectively, and produce outcomes $a$ and $b$, respectively. 
	}
	\label{fig:Bell}
\end{figure}

We now use our toy theory to challenge the common belief that the outcomes of classical measurements reveal the values of some preexisting properties of the measured systems.  The starting point of our argument is the observation  that our toy theory exhibits activation of Bell nonlocality \cite{peres96, masanes08, cavalcanti11, navascues11, palazuelos12}. Suppose that a bit $B$ and an antibit $A$ are in the entangled state $|\Phi\>_{BA}  =  ( |0\>_B   |0\>_A  +   |1\>_B |1\>_A)/\sqrt 2$.     This state alone does not give rise to any Bell inequality violation: since the local measurements on a bit and antibit are classical, one can easily construct a local hidden variable model. However, Bell nonlocality arises when we consider the two-copy state  $|\Phi\>_{B_1A_1}\otimes |\Phi\>_{B_2A_2}$, where $B_1B_2$ are bits, and $A_1A_2$ are antibits.   Suppose that two parties, Alice and Bob, play a nonlocal game, such as the CHSH game \cite{CHSH, CHSH-game}, in the scenario where Alice has access to system  $B_1A_2$, while Bob has access to system $B_2A_1$,  as illustrated in  Fig.~\ref{fig:Bell}.

We now show that the  state $|\Phi\>_{B_1A_1}\otimes |\Phi\>_{B_2A_2}$ allows Alice and Bob to reproduce the correlations of  arbitrary  single-qubit  measurements performed locally on a 2-qubit maximally entangled state. More specifically, we show that  a  qubit measurement that projects Alice's qubit on  a given  orthonormal basis   $\left\{|v_0\>,  |v_1\>\right\}$ with $|v_0\>  =  \alpha \,  |0\>  + \beta\,  |1\>$ can be simulated by  a measurement on the bit-antibit pair $B_1A_2$, described by two orthogonal projectors $\{P_0,  P_1\}$ with  
\begin{align} 
	\nonumber P_0  =&  |V^{(0)}_0\>\<V^{(0)}_0  |_{B_1A_2}  +|V^{(1)}_0\>\<V^{(1)}_0|_{B_1A_2} \\
	\nonumber   & \quad |V^{(0)}_0\>_{B_1A_1}    =  \alpha  \,  |0\>_{B_1}|0\>_{A_2}    +  \beta \,   |1\>_{B_1}|1\>_{A_2}\\
	&\quad |V^{(1)}_0\>_{B_1A_2}    =  \alpha  \,  |0\>_{B_1}|1\>_{A_2}    +  \beta \,   |1\>_{B_1}|0\>_{A_2} \,, \label{P0}
\end{align}
and $P_1  =  I_{B_1} \otimes I_{A_2} -  P_0$, acting on the bit-antibit pair $B_1A_2$.     Similarly, a measurement that projects  Bob's qubit on the orthonormal basis $\{|w_0\>  , |w_1\>\}$ with $|w_0\> = \gamma |0\>  +  \delta |1\>$ can be simulated by the projective measurement $\{ Q_0,  Q_1\}$ defined by 
\begin{align} 
	\nonumber Q_0  =  &|W^{(0)}_0\>\<W^{(0)}_0|_{B_2A_1}  +|W^{(1)}_0\>\<W^{(1)}_0|_{B_2A_1}\\
	\nonumber &\quad  |W^{(0)}_0\>  =  \gamma \, |0\>_{B_2}  |0\>_{A_1}  +  \delta \,|1\>_{B_2}  |1\>_{A_1}  \\
	&\quad |W^{(1)}_0\>  =  \gamma \, |1\>_{B_2}  |0\>_{A_1}  +  \delta \,|0\>_{B_2}  |1\>_{A_1} \, ,
\end{align}
and $Q_1  =  I_{B_2}\otimes I_{A_1}  -  Q_0.$ When these measurements are performed on  the  state $\rho  =|\Phi\>\<\Phi|_{B_1A_1}\otimes |\Phi\>\<\Phi|_{B_2A_2}$, Alice and Bob obtain outcomes $a$ and $b$ with probability  
\begin{align}
	p(a,b)   = \Tr  [  (  P_a\otimes Q_b)  \,  \rho]   \equiv  \left|  \<  v_a|  \< w_b|  \, |\Phi\>\right|^2\,,
\end{align}
equal to the outcome probability of the original single-qubit measurements performed on the 2-qubit maximally entangled state $|\Phi\>$ (see  Supplemental Material \cite{supp} for more details).   
In this way, every pair of local  measurements on a maximally entangled 2-qubit quantum state can be simulated by local measurements in our toy theory.  In particular,  Alice and Bob can simulate the optimal strategy in the CHSH game \cite{CHSH,CHSH-game,Cirelson80,Cirelson87}, thereby achieving a maximal violation of the CHSH inequality.

Let us now examine the implications of the above result for the interpretation of classical theory. A first, important consequence is that the value of Alice's classical bit  cannot, in general,  be regarded as predetermined.  This conclusion follows from the fact that the violation of the CHSH inequality can be achieved with setup in which one of Alice's measurements is the canonical measurement on bit $B_1$.  Technically, this follows from the fact that  one of Alice's measurements in the original quantum scenario is a qubit measurement on the computational basis $\{|0\>,  |1\>\}$.  In our simulation, this   measurement corresponds to the projectors  $P_0  =  |0\>\<0|_{B_1} \otimes I_{A_2}$ and $P_1  =  I \otimes I  -  P_0 =   |1\>\<1|_{B_1}  \otimes I_{A_2}$, as one can see from Eq. (\ref{P0}).  Operationally, this measurement is realized by  discarding the antibit $A_2$ and measuring  bit $B_1$ on the basis $\{|0\>,  |1\>\}$.    Since Alice's bit value is a measurement outcome  in a setup that violates the CHSH inequality, we conclude that the bit value cannot be predetermined \cite{barrett2006maximally}: explicitly, in Supplemental Material we show that, if the underlying ontic state determines the  value of Alice's bit up to an error  $\epsilon$, then the CHSH value cannot exceed $2  (1+  2\epsilon)$ and therefore cannot reach the maximum value $2\sqrt 2$ when $\epsilon$ is small.  
In Supplemental Material we also show that  
the above argument applies to all pure entangled states of a dit and an antidit~\cite{supp}. 

Another implication of Bell  nonlocality  is that, even if we replace our toy theory with a more fundamental description of nature, this description cannot, under reasonable assumptions,  assign individual ontic states to  classical systems.  Two  different arguments leading to this  conclusion are  provided in Supplemental Material~\cite{supp}.  In both cases,   the conclusion is that  classical systems in our toy theory cannot be reduced  to independent and uncorrelated degrees of freedom of the underlying reality.

\paragraph{Conclusions.} In this Letter, we have shown  that the realistic interpretation of classical theory can,  in principle, be  falsified   when classical systems coexist with other types of physical systems.  We built  a toy theory in which every classical system can be entangled with a dual, anticlassical system. The entanglement between classical and anticlassical systems gives rise to activation of Bell nonlocality and implies that, in general, the outcomes of measurements on classical systems cannot be interpreted as revealing the values of some preexisting properties of the measured systems.

\medskip{}
\begin{acknowledgments}
	{\bf Acknowledgments.} G. C. and L. G. wish to thank  J. Barrett,  H. Kristj\'ansson, and T. van der Lugt for helpful comments.
	This work was supported by the Hong Kong Research Grant Council through  
	the Senior Research Fellowship Scheme SRFS2021-7S02 and the Research Impact Fund R7035-21F, and by the John Templeton Foundation  through the ID No. 62312 grant, as part of the ‘The Quantum Information Structure of Spacetime’ project (QISS). The opinions expressed in this publication are those of the authors and do not necessarily reflect the views of the John Templeton Foundation. C. M. S. acknowledges the support of the Natural Sciences and Engineering Research Council of Canada (NSERC) through the Discovery Grant “The power of quantum resources” RGPIN-2022-03025
	and the Discovery Launch Supplement DGECR-2022-00119.  Research at the Perimeter Institute is supported by the Government of Canada through the Department of Innovation, Science and Economic Development Canada and by the Province of Ontario through the Ministry of Research, Innovation and Science. 
\end{acknowledgments}


%

\onecolumngrid

\section{SUPPLEMENTAL MATERIAL}

\medskip

\section{Toy theory with basic systems of arbitrary dimension}

We now provide a generalization of our toy theory to the scenario where the basic systems have general dimension $d\ge 2$.

\subsection{Systems}

A classical (anti-classical) system with $d$ perfectly distinguishable pure states, named \textit{dit} (\textit{anti-dit}), is denoted by the letter $D$ ($A$) and is assigned to a  $d$-dimensional Hilbert space $\tH_D \simeq \mathbb{C}^d$ ($\tH_A \simeq \mathbb{C}^d$).   In the following we will  consider composite systems where the dimension $d$ of the elementary systems is fixed.  In other words, we consider  composite systems 
consisting of a given number  $m$ of dits and a given number $n$ of anti-dits of given dimension $d$.  The resulting composite system will be denoted by the pair $(m,n)$ and we will be called a {\em system of type $(m,n)$}.

A composite system $S$ consisting of $m$ dits $D_1\cdots D_m$ and $n$ anti-dits $A_1 \cdots A_n$ will be associated to the Hilbert space 
\begin{align}
	\spc H_S   :  =  \spc H_{D_1} \otimes \cdots \otimes \spc H_{D_m} \otimes \spc H_{A_1} \otimes \cdots \otimes \spc H_{A_n} \simeq  \left( \mathbb{C}^d \right)^{\otimes m}  \otimes \left( \mathbb{C}^d \right)^{\otimes n}  \, . 
\end{align}
In the following, we will denote by $\Lin (\spc H_S)$ (or simply $\Lin (S)$) the vector space of all linear operators on $\spc H_S$. The states of system $S$ will be described by a suitable subset of the operators in $\Lin (\spc H_S)$.

\subsection{Pure states}  

We now specify the pure states of  composite systems of type $(m,n)$, for all possible values  of $m$ and $n$.  In all these cases, the pure states of a system $S$  are mathematically represented by suitable rank-one projectors, of the form $\rho  =  |\psi\>\<\psi|$ for some unit vector $|\psi\>  \in \spc H_S$ satisfying appropriate conditions. The set of pure states of system $S$ will be denoted by $\Pur\St (S)  \subset \Lin  (\spc H_S)$.     With a slight abuse of notation, common in presentations of quantum theory, we will sometime refer to the unit vector $|\psi\>$, rather than the corresponding projector, as a ``pure state.''    

\subsubsection{Composites of type $(m,0)$ and $(n,0)$}

All composites of this type obey the rules of classical  theory, and their states are described by density matrices that are diagonal in a given basis, called the {\em computational basis}.  

The  pure states of a dit $D$  (anti-dit $A$) are $d\times d$ density matrices of the form $\rho  =  |i\>\<i|$, where $\{|i\>\}_{i=0}^{d-1}$ is the computational basis  for $\mathbb{C}^d$.    In the following, we will use the short-hand notation 
\begin{align}
	\Z_d  :  =  \{ 0,\dots,  d-1 \}   \, ,
\end{align}
and will label the computational basis as $\{|i\>\}_{i\in\Z_d}$.

For composite systems consisting only of dits ({\em i.e.} systems of type $(m,0)$) or only of  anti-dits ({\em i.e.} systems of type $(0,m)$), the allowed pure states are density matrices of the form $\rho  =  |{\st i}\>\<{\st i|}$, where  $\st i=  (i_1,\dots,  i_m)$ is a vector with $m$ entries in $\Z_d$, and 
\begin{align}
	|{\bf i}\>  :  =  |i_1\>  \otimes \cdots \otimes |i_m\>  \,. \end{align}
In the following, the set of vectors $(i_1, \dots,  i_m)$ with $m$ entries in $\Z_d$ will be denoted by $\Z_d^m$.

\subsubsection{Composites of type $(m,m)$}
Let us start from the  $(1,1)$ case. The composite system of a dit $D$ and an anti-dit $A$  is   associated to the Hilbert space $\spc H_{DA} : =\spc H_D \otimes \spc H_A$, where  $\spc H_D \simeq    \mathbb{C}^d$ and $\spc H_A \simeq    \mathbb{C}^d$ are  the Hilbert spaces associated to $D$  and $A$, respectively.   To specify the allowed pure states, we introduce a set of orthogonal subspaces, labelled by integers in $\Z_d$.  For a given $q\in  \Z_d$,  we define the subspace
\begin{align}\label{hq}
	\spc H^{q}_{DA}  := \Span\Big\{   |i\>_D |i\oplus q\>_A\,,~  i  \in  \Z_d \Big\} \subset \spc H_{DA} \, ,         \end{align}
where $\oplus$ denotes the sum modulo $d$. 
The integer $q$ will be referred  to as the {\em type} of the subspace.  For $d=2$, the type  $q\in  \{0,1\}$ is simply the parity.   Vectors in the subspace $\spc H_{DA}^{q}$ will be called {\em vectors of type $q$}. 

We are now ready to define the pure states of the $(1,1)$ composites:  
\begin{definition}\label{def:DA}
	The pure states of a dit/anti-dit composite $DA$ are projectors of the form $ |\psi\>\<\psi|$, where $|\psi\>$ is a unit vector of type $q$ for some  $q \in  \Z_d$.    In formula,  
	\begin{align}\label{unionDA}
		\Pur\St  (DA)    :=  \bigcup_{q=\in  \Z_d}   \Pur\St  (DA;  q)  \qquad  {\rm with} \qquad  \Pur\St  (DA;  q)  :  =  \left\{  |\psi\>\<\psi|  ~\Big| ~   |\psi\>  \in \spc H^{q}_{DA}  \, ,  \| \, |\psi\> \,\| = 1\right\}  \, . 
	\end{align}
\end{definition}
We call $\Pur\St  (DA;  q)$  the set of {\em pure states of type  $q$}.  Note that $\Pur\St  (DA;  q)$ coincides with the set of pure states of a quantum system of dimension $d$ (qudit). As we will see later in this Supplemental Material, the relation with quantum theory will play a key role in determining  the properties of our toy theory.

In the special case of a dit and an anti-dit ($d=2$),  the state space of the composite system is the direct sum of two orthogonal sectors, where each sector is isomorphic to a qubit state space. A similar state space also appeared in earlier toy theories, such as Fermionic theory \cite{fermionic, fermionic-2},  doubled quantum theory ~\cite{Purity,ScandoloPhD}, and  extended classical theory~\cite{ScandoloPhD}. These theories,  however, differ  from our toy theory in the definition of multipartite composites with more than two components,  and, most importantly, do not contain the whole classical theory as a subtheory.

Let us now consider the  case of general $m\ge 1$.  A composite  of $m$ dits $D_1 \dots D_m =:  \st D$ and $m$ anti-dits  $A_1\dots A_m  :=   \st A$ is associated to the Hilbert space  $\spc H_{\st D \st A}:   =  \spc H_{\st D} \otimes \spc H_{\st A}$, where 
$\spc H_{\st D} :  =  \spc H_{D_1} \otimes \cdots \otimes \spc H_{D_m}$ and $\spc H_{\st A} :  =  \spc H_{A_1} \otimes \cdots \otimes \spc H_{A_m}$ are the Hilbert spaces associated to the dits  $\st D$ and to the anti-dits $\st A$, respectively.

To define the pure states of the composite system  $\st D \st A$, we pair every dit in $\st D$ with an anti-dit in $\st A$.  The pairing is defined by a permutation $\pi :  \{1,\dots,  m\}  \to \{1,\dots, m\}$, which associates the dit $D_i$ with an anti-dit $A_{\pi (i)}$ for every $i \in  \{  1,\dots,  m\}$.   We then define  
the subspaces
\begin{align}\label{hpiq}
	\spc H_{\st D \st A}^{\pi, \st q}  :  =  \spc H_{D_1A_{\pi(1)}}^{q_1} \otimes \cdots \otimes \spc H_{D_mA_{\pi(m)}}^{q_m}   \, ,  \qquad  {\st  q  }   =  (q_1,\dots,  q_m) \in  \{  0, \dots,  d-1\}^{\times m} \, ,
\end{align}
where, for every $i\in  \{1,\dots, m\}$,  $\spc H_{D_iA_i}^{q_i}$ is defined as in Eq.  (\ref{hq}).    We call $\spc H^{\pi, \st q}_{\st D \st A }$ a {\em subspace of type $(\pi, \st q)$} and refer to the vectors $|\psi\>  \in \spc H^{\pi, \st q}_{\st D \st A} $  as {\em vectors of type $(\pi, \st q)$.}

Note that the order of the Hilbert spaces in the l.h.s. and r.h.s. of  Eq. (\ref{hpiq}). Here and in the following we understand the equality up to an appropriate reordering of the tensor factors according to their labels. This convention, often used in quantum information, yields a more readable notation for tensor products: for example, one can write $O_{S_1S_3}  \otimes O'_{S_2}$ for an operator acting on the Hilbert space $\spc H_{S_1} \otimes \spc H_{S_2} \otimes \spc H_{S_3}$, instead of having to decompose $O_{S_1S_3}$ as $O_{S_1S_3}  =  \sum_i  (F_i)_{S_1}  \otimes (G_i)_{S_3}$, with $(F_i)_{S_1} \in  \Lin  (\spc H_{S_1})$ and $(G_i)_{S_3} \in  \Lin  (\spc H_{S_3})$, and writing  $\sum_{i}   (F_i)_{S_1} \otimes O'_{S_2} \otimes (G_i)_{S_3}$.

With this notation, we are now ready to define the pure states of the $(m,m)$-composites: 

\begin{definition}\label{def:purestatesmm}
	The pure states of an $(m,m)$-composite $\st D \st A$ with $\st D =  D_1 \cdots D_m$ and $\st A  = A_1 \cdots A_m$, are projectors of the form $\rho  =  |\psi\>\<\psi|$, where $|\psi\>$ is a unit vector of type $(\pi, \st q)$,  for some permutation $\pi  :  \in \{1,\dots, m\} \to \{1, \dots  m\}$ and some vector  $q \in  \Z_d^{ m}$.    In formula,  
	\begin{align}\label{unionstDA}
		\Pur\St  (\st D \st A)    :=  \bigcup_{\pi  \in  {\sf S}_m }\bigcup_{ \st q  \in \Z_d^{ m} }   \Pur\St  (\st D \st A;  \pi, \st q)  \qquad  {\rm with} \qquad  \Pur\St  (\st D \st A; \pi,  q)  :  =  \left\{  |\psi\>\<\psi|  ~\Big| ~   |\psi\>  \in \spc H^{\pi ,\st q}_{\st D\st A}  \, ,  \|  \,  |  \psi\>  \,  \|  = 1\right\}  \, , 
	\end{align}
	where ${\sf S}_m$ is the group of permutations of the set $\{1,\dots,  m\}$.  
\end{definition}
We call $\Pur\St  (DA;  \pi , \st q)$  the set of {\em pure states of type  $(\pi, \st q)$}.  Note that $\Pur\St  (\st D\st A;  \pi, \st q)$ coincides with the set of pure states of a quantum system made of $m$ qudits.

An equivalent characterization of the set of pure states is provided by the following proposition:  

\begin{proposition}\label{prop:alternative}
	Let $\st D \st A$  be an  $(m,m)$-composite with $\st D  =  D_1\cdots D_m$ and $\st A  = A_1\cdots A_m$,  let   $|\psi\>_{\st D\st A} \in  \spc H_{\st D \st A}$  be a unit vector.  Then, the following are equivalent:  
	\begin{enumerate}
		\item  $|\psi\>_{\st D \st A}$ represents a pure state of $\st D \st A$
		\item $|\psi\>_{\st D \st A}$ can be written as
		\begin{equation}
			\label{eq:explicitmm}
			\begin{aligned}
				\ket{\psi}_{\st D \st A} =  \sum_{i_1,\dots, i_m}  \lambda_{i_1 \dots i_m}   ~|i_1\>_{D_1} \otimes \cdots \otimes |i_m\>_{D_m} \otimes   |i_1 \oplus q_1\>_{A_{\pi(1)}} \otimes \cdots \otimes |i_m\oplus q_m\>_{A_{\pi(m)}} \, , 
			\end{aligned}
		\end{equation}
		some complex coefficients $(\lambda_{i_1,\dots,  i_{m}})$,  some vector   $\st q  =  (q_1,\dots,  q_m)  \in  \Z_d^{ m}$, and some permutation  $\pi \in  {\sf S}_m$. 
		\item    $|\psi\>_{\st D \st A}$ can be written as
		\begin{equation}
			\label{eq:simplifiedmm}
			\begin{aligned}
				\ket{\psi}_{\st D \st A} =  \sum_{i_1,\dots, i_m}  \lambda_{i_1 \dots i_m}   ~|i_1\>_{D_1} \otimes \cdots \otimes |i_m\>_{D_m} \otimes   |i_1 \oplus q_1\>_{A'_{1}} \otimes \cdots \otimes |i_m\oplus q_m\>_{A'_{m}} \, , 
			\end{aligned}
		\end{equation}
		some complex amplitudes $(\lambda_{i_1,\dots,  i_{m}})$,  some vector   $\st q  =  (q_1,\dots, q_m)  \in  \Z_d^{ m}$, and some relabelling of the anti-dits $A_1'\cdots A_m'$. 
	\end{enumerate}
\end{proposition} 

\Proof  $1 \Rightarrow 2$.    If $|\psi\>_{\st D \st A}$ represents a pure state,  it must belong to the subspace  $\spc H^{\pi,  \st q}_{\st D \st A}$ for some permutation $\pi$ and some vector $\st q$.  The computational basis for this subspace is 
\begin{align}
	\big\{|i_1\>_{D_1}  \otimes \cdots \otimes |i_m\>_{D_m}  \otimes |i_1\oplus q_1\>_{  A_{\pi (1)}}  \otimes \cdots \otimes |i_m \oplus q_m\>_{  A_{\pi (m) }}\big\}_{  (i_1,\dots,  i_m) \in  \Z_d^m} \,.
\end{align} 
Hence,  $|\psi\>_{\st D \st A}$ must be of the form (\ref{eq:explicitmm}), for suitable coefficients $(\lambda_{i_1,\dots,  i_m})_{i_1,  \dots,  i_m}$. 
\smallskip 

$2\Rightarrow 1$.  The vector $|\psi\>_{\st D \st A}$ in Eq. (\ref{eq:explicitmm}) is a  unit vector in   $\spc H_{\st D \st A}^{  \pi,  \st q}$.  By Definition \ref{def:purestatesmm}, all unit vectors in   $\spc H_{\st D \st A}^{\pi,  \st q}$ represent valid pure states.  

\smallskip 

$2\Rightarrow 3$.    Immediate by defining $A_i': =  A_{\pi(i)}$ for every $i\in  \{1,\dots,  m\}$. 

\smallskip 

$3\Rightarrow 2$.   Immediate by observing that every relabeling $A_1'\cdots A_m'$   corresponds to a  permutation $\pi\in  {\sf S}_m$ such that $A_i'  =  A_{\pi  (i)} \, , \forall i\in  \{1,\dots,  m\}$. \qed

\medskip

\subsubsection{Composites of type $(m,n)$ with $m\ne n$}
We  conclude this subsection by defining the pure states of  a general system of $m$ dits $\st D  =  D_1\cdots  D_m$ and $n$ anti-dits $\st A  =  A_1 \cdots A_n $ with $m\ne  n$.

Let us consider first the $m<n$ case. For a permutation $\pi  \in  {\sf S}_n$,  a vector $\st q  \in  \Z_d^m$, and another vector $\st r   \in  \Z_d^{n-m}$, we define the subspace 
\begin{align}\label{hpiqr}
	\spc H_{\st D \st A}^{\pi, \st q, \st r}    &:  =\spc H_{D_1 A_{\pi  (1)}}^{q_1} \otimes \cdots \otimes  \spc H_{D_m A_{\pi  (m)}}^{q_m}  \otimes |r_{1}\>_{A_{\pi (m+1)}} \otimes \cdots  \otimes |r_{n-m}\>_{A_{\pi (n)}} \, ,
\end{align} 
where we used the notation  
\begin{align}
	\spc H  \otimes |w\> :  = \left\{ |v\>\otimes |w\>  ~\Big|~   |v\>  \in  \spc H \right\} \subseteq  \spc H\otimes \spc K \, , 
\end{align}
for  arbitrary Hilbert spaces $\spc H$ and $\spc K$ and for an arbitrary vector $|w\>  \in \spc K$.

We call the triple $(\pi,  \st q, \st r)$ the {\em type} of the subspace and we say that vectors in $\spc H_{\st D \st A}^{\pi,  \st q, \st r}$ are {\em of type $(\pi,  \st q, \st r)$}.  The pure states are defined as follows: 
\begin{definition}\label{def:purestatesm<n}
	The pure state of a composite system $\st D \st A$ with $\st D  = D_1 \dots D_m$ and $\st A =  \st A_1  \cdots \st A_n$, $n>m$ are projectors of the form $|\psi\>\<\psi|$, where $|\psi\>$ is a unit vector of type $(\pi,  \st q, \st r)$, for some permutation $\pi  \in {\sf S}_n$,  and some pair of  vectors $\st q \in  \Z_d^{ m}$ and  $\st r \in  \Z_d^{ (n-m)}$.  In formula, 
	\begin{align}
		\Pur\St  (\st D \st A)   :=  \bigcup_{\pi  \in  {\sf S}_m } \bigcup_{\st q\in  \Z_d^{ m}  }   \bigcup_{\st r\in  \Z_d^{ (n-m)}}  \,\Pur\St(  \st D \st A  ;  \pi, \st q, \st r )  
	\end{align}
	with 
	\begin{align}
		\Pur\St(  \st D \st A  ;  \pi, \st q, \st r )  :=  \left\{ |\psi\>\<\psi| ~\Big|~  |\psi\> \in  \spc H^{  \pi ,  \st q , \st r}_{\st D \st A} \, ,  \|  \, |\psi\> \, \|  = 1\right\} \, .
	\end{align}  
\end{definition}

The pure states in the $m> n$ case are defined in a similar way.  For a permutation $\pi  \in  {\sf S}_m$ and a pair of vectors $\st q =  (q_1,\dots,  q_n)  \in \Z_d^n$ and $\st r=  (r_{n+m},  \dots,  r_n)  \in  \Z_d^{n-m}$ we define the subspace 
\begin{align}
	\nonumber \spc H_{\st D \st A}^{\pi, \st q, \st r}    &:  =\spc H_{ D_{\pi  (1)} A_1}^{q_1} \otimes \cdots \otimes  \spc H_{D_{\pi  (n)}  A_n}^{q_n}  \otimes |r_{n+1}\>_{D_{\pi (n+1)}} \otimes \cdots  \otimes |r_m\>_{D_{\pi (m)}} \, .
\end{align} 
We call the triple $(\pi,  \st q, \st r)$ the {\em type} of the subspace and we say that vectors in $\spc H_{\st D \st A}^{\pi,  \st q, \st r}$ are {\em of type $(\pi,  \st q, \st r)$}.  

Then, the pure states are  defined as follows: 
\begin{definition}\label{def:purestatesm>n}
	The pure state of a composite system $\st D \st A$ with $\st D  = D_1 \dots D_m$ and $\st A =  \st A_1  \cdots \st A_n$, $n<m$ are projectors of the form $  |\psi\>\<\psi|$, where $|\psi\>$ is a unit vector of type $(\pi,  \st q, \st r)$, for some permutation $\pi  \in {\sf S}_m$,  and some pair of  vectors $\st q \in  \Z_d^{ n}$ and  $\st r \in  \Z_d^{ m-n}$.  In formula, 
	\begin{align}
		\Pur\St  (\st D \st A)   :=  \bigcup_{\pi  \in  {\sf S}_m } \bigcup_{\st q\in  \Z_d^{ n}  }   \bigcup_{\st r\in  \Z_d^{ m-n}}  \,\Pur\St(  \st D \st A  ;  \pi, \st q, \st r )  
	\end{align}
	with 
	\begin{align}
		\Pur\St(  \st D \st A  ;  \pi, \st q, \st r )  :=  \left\{ |\psi\>\<\psi| ~\Big|~  |\psi\> \in  \spc H^{  \pi ,  \st q , \st r}_{\st D \st A} \, ,  \|  \, |\psi\> \, \|  = 1\right\}  \, .
	\end{align}
\end{definition}

In general, a system $S$ of type $(m,n)$ with $m\not = n$   can be decomposed as $S   =  S'  S_{\rm rest}$ where subsystem $S'$ is of type $(m',m')$, with  $m' : =  \min\{m,n\}$ and subsystem $S_{\rm rest}$ is  either of type $(m-n, 0)$, if $m>n$, or of type $(0,  n-m)$, if $n<m$.    This decomposition is not unique, because one has the freedom to choose which dits and anti-dits of $S$ go into $S'$ and which ones go into $S_{\rm rest}$. 
\begin{prop}\label{prop:productmnotn}
	Let $S$ be an $(m,n)$-composite with $m\not  =  n$. For a unit vector $|\psi\>_S \in \spc H_S$, the following are equivalent:
	\begin{enumerate}
		\item $|\psi\>_S$ represents a pure state of system $S$
		\item $|\psi\>_S$ can be decomposed as $|\psi\>_S  = |\psi'\>_{S'} \otimes | \st r  \>_{S_{\rm rest}}$, where $|\psi'\>$ is a unit vector representing a pure state of an $(m',m')$-composite $S'$, 
		$m'  =  \min\{m,n\}$, and $|\st r\>_{S_{\rm rest}}$ is a computational basis vector in $\spc H_{S_{\rm rest}}$,  with system $S_{\rm rest}$   consisting  only of  $|m-n|$ dits or only of $|m-n|$ anti-dits. 
	\end{enumerate}
\end{prop}
\Proof  $1\Rightarrow 2$ Let us represent $S$ as  $S= \st D \st A$ for a suitable set of dits $\st D  =   D_1 \dots  D_m$ and a suitable set of anti-dits $\st A  =   A_1 \dots A_n$. For $m<n$,   Definition \ref{def:purestatesm<n} stipulates that $|\psi\>_S$ belongs to a subspace $\spc H_{\st D \st A}^{  \pi,  \st q, \st r  }$, for suitable $\pi  \in {\sf S}_n$, $\st r  \in \Z_d^m$ and $\st r \in  \Z_d^{n-m}$.  
By relabelling the anti-dits as $A_i'  :=  A_{\pi  (i)}$, for $i\in  \{1,\dots  m\}$, the subspace   $\spc H_{\st D \st A}^{  \pi,  \st q, \st r  }$ can be decomposed as 
\begin{align}
	\spc H_{\st D \st A}^{  \pi,  \st q, \st r  }  =   \spc H_{\st D  \st A'_{\rm first}}^{{\rm id}, \st q}  \otimes  |\st r\>_{\st A'_{\rm last}}  \, ,
\end{align}
where ${\rm id}$ is the identity permutation ${\rm id}  (x)=  x$ for every $x\in  \{1,\dots, n\}$,  $\st A'_{\rm first}  :  =  A_1' \cdots  A_m'$, $\st A'_{\rm last}  :=  A'_{m+1} \cdots  A'_n$, and $|\st r\>_{\st A'_{\rm last}}   :  =  |r_{1}\>_{A'_{m+1}} \otimes \cdots \otimes |r_{n-m}\>_{A'_n}$.    Hence, $|\psi\>_{\st D \st A}$ can be decomposed as $|\psi\>_{\st D\st A}  =   |\psi'\>_{\st D \st A_{\rm first}}  \otimes |\st r\>_{\rm A'_{\rm last}}$.  Hence, the desired statement follows by setting $S'   :  =  \st D \st A'_{\rm first}$ and $S_{\rm rest}  :=  \st A'_{\rm last}$.   

The proof for $m>n$ is analogous.  

$2\Rightarrow 1$ Suppose that $|\psi\>_S$ can be decomposed as $|\psi\>_S  = |\psi'\>_{S'} \otimes |\st r\>_{S_{\rm rest}}$, where $|\psi'\>_{S'}$ represents  a state of system $S'$  and $|\st r\>_{S_{\rm rest}}$ is a computational basis vector.  Suppose first that $m<n$.    In this case, there exists a labelling   $\st A'   = A_1' \cdots  A_n'$  such that  $S'  =  \st D \st A'_{\rm first}$, with $\st A_{\rm first}' : = A_1' \cdots A_m'$, and $S_{\rm rest}  =  A_{m+1}' \cdots  A_n'  =:  \st A_{\rm last}' $.  Since $|\psi'\>_{S'}$ represents a  pure state of $S'$, it belongs to a subspace $\spc H_{\st D \st A'_{\rm first}}^{\pi',  \st q}$ for suitable  $\pi'  \in  {\sf S}_m$ and $q\in  \Z_d^m$.   
Hence, we have 
\begin{align}
	|\psi\>_S  =  |\psi'\>_{S'} \otimes |\st r\>_{S_{\rm rest}}  \in  \spc H_{\st D \st A'_{\rm first}}^{(\pi', \st q)}  \otimes |\st r\>_{\st A_{\rm last}}   \, .
\end{align}
Defining a permutation $\pi  \in  {\sf S}_n$ such that $A'_{\pi' (i)}   =  A_{\pi  (i)}$ for every $i\in  \{1,\dots ,m\}$ and $A'_i  =  A_{\pi  (i)}$ for every $i  \in  \{m+1, \dots, n\}$, we finally obtain the equality  
\begin{align}
	\spc H_{\st D \st A'_{\rm first}}^{\pi', \st q}  \otimes |\st r\>_{\st A_{\rm rest}}    =  \spc H_{\st D  \st A}^{ \pi,  \st q,  \st r} \, .
\end{align}
Since $|\psi\>_S$ is an element of $ \spc H_{\st D  \st A}^{  \pi,  \st q,  \st r}$, it represents a pure state of $S$.

The proof for $m>n$ is analogous.  \qed 

\subsubsection{Closure under tensor product}  

We now show that the set of  pure states of our toy theory is closed under tensor product. In other words, our toy theory satisfies the property of Pure Product States  \cite{chiribella2021process}.    Precisely, we show the following
\begin{proposition}\label{prop:pureproduct}
	For every pair of systems $S$ and $T$, and for every pair of pure states $ |\psi\>\<\psi|_{S}$ and $ |\phi\>\<\phi|_{T}$, the operator $ |\psi\>\<\psi|_{S} \otimes  |\phi\>\<\phi|_{T}$  is a pure  state of the composite system $S T$.
\end{proposition}

\Proof  Let us  first consider the case where the systems $S$ and $T$ are of types $(m,m)$ and $(k, k)$ for two integers $m$ and $k$, respectively.  In this case, Eq. (\ref{eq:explicitmm}) in Proposition \ref{prop:alternative} implies that the vectors $|\psi\>_{S}$ and $|\phi\>_{T}$ can be written as 
\begin{equation}
	\begin{aligned}
		\ket{\psi}_S =  \sum_{i_1,\dots, i_m}  \lambda_{i_1 \dots i_m}   ~|i_1\>_{D_1} \otimes \cdots \otimes |i_m\>_{D_m} \otimes   |i_1 \oplus q_1\>_{A_{\pi(1)}} \otimes \cdots \otimes |i_m\oplus q_m\>_{A_{\pi(m)}} \, , 
	\end{aligned} 
\end{equation}
and 
\begin{equation}
	\begin{aligned}
		\ket{\phi}_T =  \sum_{j_1,\dots, j_k}  \mu_{j_1 \dots j_k}   ~|j_1\>_{D_1} \otimes \cdots \otimes |j_k\>_{D_k} \otimes   |j_1 \oplus g_1\>_{A_{\tau(1)}} \otimes \cdots \otimes |j_k\oplus g_k\>_{A_{\tau(k)}} \, , 
	\end{aligned} 
\end{equation}
respectively, where $(\lambda_{i_1\dots i_m})$  and $(\mu_{j_1\dots j_k})$ are the amplitudes    of the two states, $\st q= (q_1,\dots,  q_m)  \in  \Z_d^m$ and $\st  h =  (h_1,
\dots,  h_k)  \in \Z_d^k$,      $\pi \in {\sf S}_m$, and $\tau \in {\sf S}_k$.

Now, let us define  complex amplitudes $\gamma_{i_1\dots  i_{m+n}}  :=  \lambda_{i_1\dots  i_m}  \,  \mu_{i_{m+1}  \dots  i_{m+n}}$, a vector $\st g\in  \{0,
\dots,  d-1\}^{\times (m+n)}$  with $g_i  =  q_i$ for $i\le m$ and $g_i  = h_{i-m} $ for $i>m$, and a permutation  $\theta:  \{1,\dots, m+n\}  \to \{1,\dots, m+n\}$  satisfying the relation  
\begin{align}
	\theta  (x)   =
	\left\{  
	\begin{array}{lll}
		\pi (x)    &  & \forall x\le m\\
		\tau (x-m)  +  m  \qquad &  \qquad &\forall x>m \, .
	\end{array}
	\right.
\end{align}
With these definitions,   the product vector $|\psi\>_{S} \otimes |\phi\>_T$ can be written as 
\begin{equation}
	\begin{aligned}
		|\psi\>_S \otimes \ket{\phi}_T =  \sum_{i_1,\dots, i_{m+n}}  \gamma_{i_1 \dots i_{m+n}}   ~|i_1\>_{D_1} \otimes \cdots \otimes |i_{m+k}\>_{D_{m+k}} \otimes   |i_1 \oplus g_1\>_{A_{\theta(1)}} \otimes \cdots \otimes |i_{m+n}\oplus g_{m+n}\>_{A_{\theta(m+n)}} \, . 
	\end{aligned} 
\end{equation}
This vector is of the form prescribed  in Eq. (\ref{eq:explicitmm}) in Proposition \ref{prop:alternative}, and therefore represents a valid pure state. This concludes the proof in the special case where systems $S$ and $T$ are of types $(m,m)$ and $(k,k)$ for positive integers $m$ and $k$, respectively.

Finally, let us extend the proof to the general case where system $S$ is of type $(m,n)$  and system $T$ is of type $(k,l)$. By Proposition \ref{prop:productmnotn}, the states $|\psi\>_S$ and $|\phi\>_T$ can be decomposed as 
\begin{align}
	|\psi\>_S  =  |\psi'\>_{S'} \otimes |\st r\>_{S_{\rm rest}}  \qquad {\rm and} \qquad   |\phi\>_T  =  |\phi'\>_{T'} \otimes |\st s\>_{T_{\rm rest}}\, ,
\end{align}
where $|\psi'\>_{S'}$ ($|\phi'\>_{T'}$)  is a pure state of a system $S'$  ($T'$) of type $(m',m')$ ($(k',k')$)  with  $m'=  \min \{m,n\}$  ($k' =  \min\{k,l\}$), and $|\st r\>_{S_{\rm rest}}$  ($|\st s\>_{T_{\rm rest}}$) is a computational basis state of a system $S_{\rm rest}$ ($T_{\rm rest}$) consisting only of dits, or only of anti-dits.  Using the above equation, we can decompose the product vector as 
\begin{align}
	|\psi\>_S \otimes |\phi\>_T  =  |\psi'\>_{S'} \otimes |\phi'\>_{T'} \otimes |\st r\>_{S_{\rm rest}}  \otimes |\st s\>_{T_{\rm rest}} \, .  
\end{align}
By the first part of the proof, we have that the product vector $|\psi'\>_{S'} \otimes |\phi'\>_{T'}$ is a valid pure state of system $S'T'$. Moreover, $|\st r\>_{S_{\rm rest}}  \otimes |\st s\>_{T_{\rm rest}}$ is a computational basis vector for system $S_{\rm rest}  T_{\rm rest}$.    Hence,  Proposition \ref{prop:productmnotn} implies that the product $\big( |\psi'\>_{S'} \otimes |\phi'\>_{T'}\big)  \otimes  \big( |\st r\>_{S_{\rm rest}}  \otimes |\st s\>_{T_{\rm rest}} \big)$ is a pure state of system $(S'T')  (S_{\rm rest}  T_{\rm rest})  \equiv  ST$.  \qed

\subsection{Mixed states}  

For every  composite system in our toy theory, every random mixture of pure states represents a valid mixed state. Hence,  the space of all normalized  states of the system  is the convex hull of the set of corresponding pure states, defined in the previous section. 

\begin{definition}[Normalized states]
	The normalized  states of a $(m,n)$-composite are finite convex combinations of the pure states of the same composite. Specifically, a normalized mixed state is a density matrix of the form $ \rho = \sum_j p_j \ket{ \Psi_j }\bra{ \Psi_j } $ where  $(q_j)_j$ is a probability distribution, and each vector $\ket{ \Psi_j }$ is a valid  pure state of the $(m,n)$-composite. 
\end{definition}
In the following, the set of normalized  states of system $S$ will be denoted as  $\St_1 (S)$.  The set of all (generally subnormalized) states will be 
\begin{align}\St (S)  :=  \{   p \,  \rho ~|~  \rho  \in  \St_1  (S)  \, ,  p  \in  [0,1] \} \, .
\end{align}
Operationally, a subnormalized state  can be interpreted as a probabilistic preparation of the corresponding normalized state.  For composite systems of type $(m,0)$ and  $(0,n)$, it is straightforward to see that the state space assigned by our toy theory is exactly the classical state space.

We now show two  properties of the state spaces in our theory.  The first is a basic consistency property:  
\begin{proposition}\label{prop:product}
	For every pair of systems $S$ and $T$, and for every pair of states $\rho_{S}  \in \St (S)$ and $\sigma_{T}  \in \St (T)$, the operator $\rho_{S} \otimes \sigma_{T}$ is a valid state of the composite system $ST$, that is, $\rho_{S} \otimes \sigma_{T}  \in  \St  (ST)$. 
\end{proposition}

\Proof  The proof follows from Proposition \ref{prop:pureproduct},  combined with  the bilinearity of the tensor product and  the convexity of the state space. \qed
\medskip 

The above proposition guarantees that the product of two valid states is a valid state, as required in the framework of operational-probabilistic theories.

Another property is that the  state spaces in our toy theory are  closed under  partial trace. Denoting by $\Tr_{X}$  the partial trace  over the Hilbert space of an arbitrary system $X$, and by  $S  \setminus  R$ the composite system consisting of all the dits and anti-dits  that are in $S$ but not in $R$,   we have the following: 
\begin{theorem}\label{theo:marginal}  
	Let $S=  D_1  \cdots D_m  A_1 \cdots A_n$ be a composite of $m$ dits and $n$ anti-dits, and let $R$ be a subsystem of $S$.  For every every state $\rho_S  \in\St (S)$,
	the operator 
	\begin{align}
		\sigma_{S\setminus R}   :  = \Tr_{R}  [  \rho_S]  
	\end{align}
	belongs to the state space $\St (S\setminus R)$.  Moreover, if $\rho_S$ belongs to the set of normalized states $\St_1 (S)$, then $\sigma_{S\setminus R}$ belongs to the set of normalized states $\St_1 (S\setminus R)$.   
\end{theorem}

To prove the theorem, we use a  technical lemma: 
\begin{lemma}\label{lemma:removeone}
	Let $S  =  D_1\cdots  D_m A_1\cdots A_n$ be a composite of $m$ dits and $n$ anti-dits. For every pure state $|\psi\>_S  \in \spc H_S$ and every integer $i\in  \Z_d$,   one has that each  vector 
	\begin{align}  
		\Big( I_{ S\setminus D_t} \otimes  \< i|_{D_t}   \Big)\,  |\psi\>_S \, ,\qquad t\in  \{1,\dots,  m\}
	\end{align} 
	is proportional to a valid pure state of system $S\setminus D_t$. 
	Similarly, each vector 
	\begin{align}  
		\Big( I_{ S\setminus A_t} \otimes  \< i|_{A_t}   \Big)\,  |\psi\>_S  \qquad t\in  \{1,\dots,  n\}
	\end{align} 
	is proportional to a valid pure state of system $S\setminus A_t$.  
\end{lemma}

\Proof  Suppose  first  that  $S$ is a system of type $(m,m)$.   In this case, we can use Eq. (\ref{eq:explicitmm}), which yields the relation 
\begin{align}
	\nonumber  \Big( I_{ S\setminus D_t} \otimes  \< i|_{D_t}   \Big)\,   \ket{\psi}_S & =  \Big( I_{ S\setminus D_t} \otimes  \< i|_{D_t}   \Big)\,  \left( \sum_{i_1,\dots, i_m}  \lambda_{i_1 \dots i_m}   ~|i_1\>_{D_1} \otimes \cdots \otimes |i_m\>_{D_m} \otimes   |i_1 \oplus q_1\>_{A_{\pi(1)}} \otimes \cdots \otimes |i_m\oplus q_m\>_{A_{\pi(m)}}  \right) \\
	\label{line1} & = \Bigg( \sum_{i_1,\dots, i_{t-1},  i_{t+1} \dots  i_m}  \lambda_{i_1 \dots i_{t-1} \, i \, i_{t+1} \dots  i_m}   ~|i_1\>_{D_1} \otimes \cdots \otimes  |i_{t-1}\>_{D_{t-1}} \otimes |i_{t+1}\>_{D_{t+1}}\otimes \cdots \otimes  |i_m\>_{D_m} \\
	\label{line2} &  \quad  \otimes   |i_1 \oplus q_1\>_{A_{\pi(1)}} \otimes \cdots \otimes |i_{t-1} \oplus q_{t-1}\>_{A_{\pi(t-1)}} \otimes |i_{t+1} \oplus q_{t+1}\>_{A_{\pi(t+1)}} \otimes   \cdots \otimes  |i_m\oplus q_m\>_{A_{\pi(m)}}  \Bigg) \\
	\label{line3} &  \quad \otimes  |i \oplus q_{t}\>_{A_{\pi(t)}}\, .
\end{align} 
Now the vector in the r.h.s. of Eqs. (\ref{line1}) and (\ref{line2}) is proportional to a valid pure state of a system of type $(m-1,m-1)$, while the computational basis state in the r.h.s. of Eq. (\ref{line3}) is, of course, a valid state of the anti-dit $A_{\pi(t)}$.  Using Proposition \ref{prop:pureproduct}, we conclude that the product of these two vectors is proportional to a valid pure state of system $S\setminus D_t$.  

The proof that $\Big( I_{ S\setminus A_t} \otimes  \< i|_{A_t}   \Big)\,   \ket{\psi}_S$ is a proportional to a valid pure state of $S\setminus A_t$ is analogous to the above.  This observation concludes the proof in the case where $S$ is of type $(m,m)$. 

Now, suppose that $S$ is of type $(m,n)$ with $m\not  = n$.   In this case, Proposition \ref{prop:productmnotn} ensures that the pure state $|\psi\>_s$ is of the form $|\psi\>_s  =  |\psi'\>_{S'} \otimes |\st r\>_{S_{\rm rest}}$, where $|\psi'\>_{S'}$ is a pure state of a system $S'$ of type $(m',m')$, with $m': =  \min\{m,n\}$, and  $|\st r\>_{S_{\rm rest}}$ is a computational basis state of a system $S_{\rm rest}$, consisting of $m-n$ dits if $m>n$, or of $n-m$  anti-dits if $m<n$.   If $D_t$ is one of the  dits in $S'$, then the first part of the proof guarantees that  $\Big( I_{ S'\setminus D_t} \otimes  \< i|_{D_t}   \Big)\,   \ket{\psi'}_{S'}$ is proportional to a valid pure state of system $S'\setminus D_t$.  Tensoring  it with  the state $|\st r\>_{S_{\rm rest}}$  then gives a vector proportional to a valid pure state   of system $S\setminus D_t$.  If, alternatively, $D_t$ is a dit in $S_{\rm rest}$,  it is immediate that $(  I_{S_{\rm rest} \setminus D_t}  \otimes \<i|_{D_t})  |\st r\>_{S_{\rm rest}}$ is proportional to a pure state of $S_{\rm rest} \setminus D_t$, and tensoring with the pure state $|\psi'\>_{S'}$ yields a vector proportional to a valid pure state of system $S\setminus D_t$.  In either cases, the resulting vector $\Big( I_{ S\setminus D_t} \otimes  \< i|_{D_t}   \Big)\,   \ket{\psi}_{S}$  is proportional to a valid pure state of   $S\setminus D_t$.  

The proof that $\Big( I_{ S\setminus A_t} \otimes  \< i|_{A_t}   \Big)\,   \ket{\psi}_S$ is a proportional to a valid pure state of $S\setminus A_t$ is analogous to the above. \qed 

\medskip 

We are now ready to prove Theorem \ref{theo:marginal}.

{\bf Proof of Theorem \ref{theo:marginal}.}    Note that it is sufficient to prove the theorem in the pure state case $\rho_S  = |\psi\>\<\psi|_S$, since the statement for mixed states follows by linearity of the partial trace and by convexity of the state space. 

Consider first the case in which the subsystem $R$ consists of a single dit, or of a single anti-dit. In this case, we have 
\begin{align}
	\nonumber  \Tr_R  [  |\psi\>\<\psi|_S]  &  =   \sum_{i}   \,  \Big( I_{ S\setminus R} \otimes  \< i|_{R}   \Big)\,   \ket{\psi}\<\psi|_{S}\,  \Big( I_{ S\setminus R} \otimes  | i\>_{R}   \Big) \, .
\end{align}
By Lemma \ref{lemma:removeone}, each of the terms in the r.h.s. is proportional to a valid pure state of system $S\setminus R$, with a nonnegative proportionality constant. Then,   the convexity of $\St_1  (S\setminus R)$ implies that $\Tr_R [|\psi\>\<\psi|_S]$ is proportional to a valid state. Since $\Tr\big(\Tr_R [|\psi\>\<\psi|_S]\big)=\Tr[ |\psi\>\<\psi|_S]$, the proportionality constant is 1, and therefore  we have  $\Tr_R [|\psi\>\<\psi|_S] \in \St (S\setminus R)$.  Im particular, if $\Tr[|\psi\>\<\psi|_S]=1$, then $\Tr_R [|\psi\>\<\psi|_S$ is a normalized state.  By linearity, we then obtain that $\Tr_R  [  \rho_S]$  is a normalized state whenever $\rho_S$ is a normalized state.  This concludes the proof in the case where the subsystem $R$ consists of a single dit or a single anti-dit.  

The generalization to arbitrary subsystems $R$ follows by decomposing the partial trace over $R$ into a sequence of partial traces over the individual dits and anti-dits in $R$. \qed

\medskip  

\begin{definition} Let $ST$ be a composite system, and   $\rho_{ST}$ be a state of $ST$.  The state $\rho_{S}   :  = \Tr_T [\rho_{ST}]$ is called the {\em marginal of state $\rho_{ST}$ on system $S$.}
\end{definition}

\subsection{Purification}

We now prove that every classical (anti-classical)  state in our toy theory  can be purified, that is, it can be obtained as the marginal of a pure state of  an appropriate composite system, called the {\em purifying system} \cite{chiribella2010probabilistic}.   

The purifying system is constructed by combining the  classical (anti-classical) system  $S$ with its {\em anti-system}, defined as follows: 
\begin{definition}
	For a system $S$ of type $(m,n)$, the anti-system $\overline S$ is a system of type $(n,m)$.  
\end{definition}
In the particular case $n=0$, the anti-system of an $m$-dit composite is a composite of $m$ anti-dits. Similarly, the anti-system of a composite of $n$ anti-dits is a composite of $n$ dits. 

\begin{proposition}[Existence of a purification]
	\label{prop:purification}
	For every classical (anti-classical) system $S$ and for every normalized state $\rho_S \in  \St_1 (S)$, there exists a unit vector  $|\Psi\>_{S\overline S}  \in  \spc H_S  \otimes \spc H_{\overline S}$  such that   $\rho_S     =\Tr_{ \overline S}  [  |\Psi\>\<\Psi|_{S\overline S}]$, where $\Tr_{\overline S}$ denotes the partial trace over the Hilbert space $\spc H_{\overline S}$. 
\end{proposition}

\Proof
We show the existence of purifications for the  $(m,0)$-composite, the case of $(0,n)$-composites being completely analogous.  
Consider an arbitrary mixed state of system $S = D_1\cdots D_m$, that is, an arbitrary  density matrix of the form 
\begin{align}
	\label{eq:classical-state}
	\rho_{ S} = \sum_{\st i  \in  \Z_d^{ m}}  p_{\st i}  \ket{\st i} \bra{\st i}_S,
\end{align}
where $(p_{\st i})$ is a probability distribution.  Then, the unit vector 
\begin{align}
	\label{eq:purification-state}
	\ket{\Psi}_{S\overline S}:=   \sum_{\st i}  \, \sqrt{p_{\st i}}  \,  |\st i\>_S  \otimes |\st i \>_{\overline S}  =  \sum_{i_1,
		\dots,  i_m  }  \sqrt{p_{i_1 \dots  i_m}}  \,  |i_1\>_{D_1} \otimes \cdots \otimes |i_m\>_{D_m} \otimes |i_1\>_{A_1} \otimes \cdots \otimes |i_m\>_{A_m}   \, ,  
\end{align}
is a vector of type $({\rm id}, \st q_0)$ where ${\rm id}$ is the identity permutation, and $\st q_0  =  (0,\dots,  0)$. By Definition \ref{def:purestatesmm}, it is a valid pure state of system $S\overline S$, with $\overline S  = A_1\cdots A_m$.  Moreover, it is immediate to see that $|\psi\>_{S\overline S}$  is a  purification of $\rho_S$. \qed
\medskip

\subsection{Measurements}

We now specify the measurements allowed by our toy theory, starting from the case in which the measured system is discarded after the measurement.

\begin{definition}[Measurements]\label{def:measurements}
	For a system $S$,   a possible measurement  with outcomes in a finite set $\sf X$ is described by a (finite) positive operator-valued measure (POVM), that is, a tuple $(P_i)_{i  \in {\sf X}}$ of positive operators on the Hilbert space $\spc H_S$,  satisfying the conditions 
	\begin{enumerate}
		\item for every $i\in  \sf X$, $P_i  = \lambda_i \,  \rho_i$, for some nonnegative real number $\lambda_i\ge 0$ and some state  $\rho_i  \in  \St (S)$,  
		\item $\sum_{i\in {\sf X}}  P_i=  I_S$.  
	\end{enumerate}
\end{definition}

When the measured system is in the state $\rho\in \St_1  (S)$,  the probability distribution of the outcomes is determined by   the Born rule, which assigns probability  $p_i = \Tr [ P_i\rho]$ to the outcome $i$,  as in quantum theory.    Definition \ref{def:measurements} guarantees that the outcome probabilities are non-negative and sum up to 1 for every state $\rho \in \St_1  (\rho)$.

The individual operators $P_i$ in a given POVM are  called {\em effects}, and the set of all possible effects is denoted by $\Eff (S)$. 
Definition \ref{def:measurements} implies that our toy theory enjoys the property of {\em self-duality} \cite{barnum2008teleportation,janotta2011limits}:  every effect $P\in \Eff (S)$ is a non-negative multiple of some state $\rho  \in \St (S)$, and {\em vice-versa}.

In our toy theory, self-duality guarantees an important consistency property, namely that distinct states give rise to distinct probability distributions for at least one measurement:  
\begin{prop}\label{prop:separatingeffects}
	For every system $S$ and for every pair of states  $\rho  \in \St_1  (S)$ and $\sigma\in \St_1 (S)$, the condition $\rho  \not  =  \sigma$ implies that there exists at least one effect $P  \in  \Eff (S)$ such that $\Tr[  P  \rho]  \not  =  \Tr [P \sigma]$.   
\end{prop}    
\Proof  The proof is by contrapositive: we show that the condition $\Tr[P \rho]  = \Tr [P \sigma]  \, \forall P  \in \Eff (S)$ implies $\rho  = \sigma$.   

This implication  is immediate due to self-duality: since there exist a non-zero effect proportional to $\rho$ and a non-zero effect proportional to $\sigma$,    the condition  $\Tr [P  (\rho  - \sigma)]  = 0\, , \forall P \in \Eff (S)$ implies the conditions  $\Tr[ \rho  (\rho- \sigma)]  =  0$ and   $\Tr[ \sigma  (\rho- \sigma)]  =  0$, which in turn imply 
$\Tr [  (\rho  -\sigma)^\dag  (\rho-\sigma)]=0$, and therefore $\rho-\sigma=0$. \qed  
\medskip  

Similarly, two distinct effects $P$ and $P'$ must assign distinct probabilities to at least one state: 
\begin{prop}\label{prop:separatingstates}
	For every system $S$, and every pair of effects  ${}  \in \Eff  (S)$ and $P'\in \Eff (S)$, the condition $P  \not  =  P'$ implies that there exists at least one state $\rho  \in  \St_1 (S)$ such that $\Tr[  P  \rho]  \not  =  \Tr [P' \rho]$.   \end{prop}    
The proof is analogous to the proof of the previous proposition. 

\subsection{Conditional states}

In general, a measurement can be performed locally on a part of a composite system, while another part is not measured. In a well-defined theory, these local measurements should be compatible with an assignment of valid states to the unmeasured part of the system. We refer to these states as the {\em conditional states}. In the following, we show that our toy theory assigns well-defined conditional states in the state space of the unmeasured system.  

Consider a composite system consisting of two subsystems $S$ and $T$, initially in the state $
\rho_{ST}   \in \St_1  (ST)$.  Than, suppose that system $S$ undergoes a measurement described by the POVM $(P_i)_{i  \in {\sf X}}$, and that the measurement gives outcome $i$.  In this case, the probability of the outcome is given by  
\begin{align}
	p_i  =  \Tr [  (P_i  \otimes I_T) \,  \rho_{ST}]   \qquad \forall i\in  \sf X \, ,
\end{align}
and, for $p_i \not  =0$, the state of system $T$ is described by the operator
\begin{align}\label{conditionalstate}
	\rho_{T, i}   :=  \frac{  \Tr_S  [ (P_i \otimes I_T)  \, \rho_{ST}]}{p_i} \, .
\end{align}
We refer to the operator $\rho_{T,i}$ as the {\em conditional state} associated to the state $\rho_{ST}$ and to the effect $P_i$.  

For the above definition to be consistent, the operator $\rho_{T,i}$ should be an element of $\St_1  (T)$. We call this property {\em consistency of conditional states} and show that it holds in our toy theory.  Consistency of conditional states is equivalent to the following theorem, formulated in terms of unnormalized states:
\begin{theorem}\label{theo:consistency}
	For every pair of systems $S$ and $T$, every state $\rho_{ST} \in  \St (ST)$ and every effect $P_S\in \Eff (S)$, one has 
	\begin{align}
		\Tr_S [  (P_S\otimes  I_T )\, \rho_{ST}]  \in  \St  ( T)  \, .
	\end{align}
\end{theorem}
The proof is rather technical  and is postponed to Section \ref{sec:consistency} at the end of this Supplemental Material.   

In particular, the  consistency of conditional states requires that if the unmeasured system $T$ is a classical system, then the conditional states must be valid classical states. This condition puts a constraint on the type of entangled states allowed by our theory: entanglement should not allow an experimenter to steer a classical system to non-classical states. This constraint is similar in spirit to a constraint put forward  in a different context  by Layton and Oppenheim \cite{layton2023classical}.  There, the authors considered two interacting quantum systems and defined conditions for one of them to have a classical limit. In this context, the constraint that classical systems cannot be steered to non-classical states  is equivalent to the requirement that the two systems become unentangled in the limit.   In our toy theory, instead, the consistency of conditional states is ensured by a suitable construction of the tensor product between classical and anti-classical systems.

\subsection{Physical transformations}

The consistency of conditional states provides a simple recipe for constructing physical transformations.  Given three arbitrary systems $S_{\rm in}$, $S_{\rm out}$, and $S_{\rm aux}$ in our toy theory,  an arbitrary state $\sigma_{\rm aux, out}$ of system  $  S_{\rm aux}  S_{\rm out}$, and a POVM operator  $P_{\rm in, aux}$ on system $S_{\rm in}  S_{\rm aux}$,  one can define a {\em conditional transformation} $\map T_{\sigma,  P}$ via the relation
\begin{align}\label{tsigmaP}
	\map T_{\sigma, P}  (\rho)   :  =  \Tr_{S_{\rm in}S_{\rm aux}} [    (\rho_{\rm in} \otimes  \sigma_{\rm aux, out})  \,  (P_{\rm  in , aux } \otimes I_{\rm out}  )]   
\end{align}
for every possible input state $\rho$. 

\begin{lemma}\label{lem:tsigmaP}  Let $\map T_{\sigma, P}$ be a conditional transformation with input system $S_{\rm in}$ and output system $S_{\rm out}$, as defined in Eq. (\ref{tsigmaP}), and let $R$ be an arbitrary system.  Then, the map $\map T_{\sigma, P} \otimes \map I_R$ transforms states of system $S_{\rm in}  R$ into (generally subnormalized) states of system $S_{\rm out}R$.  
\end{lemma}

\Proof
Immediate from Theorem \ref{theo:consistency}.\qed    
\medskip

More generally, the set of all possible physical transformations for a given pair of input/output  systems is defined as follows:  
\begin{definition}
	A linear map from $\Lin  (S_{\rm in})$   to $\Lin  (S_{\rm out})$ is a physical transformation with input  $S_{\rm in}$ and output $S_{\rm out}$ if it is  trace non-increasing and proportional to a  conditional transformation of the form (\ref{tsigmaP}). Explicitly, the set of all physical transformations is defined as 
	\begin{equation} \label{transfinout}
		\begin{aligned}
			\Transf  (S_{\rm in} \to S_{\rm out}) :  = 
			\{   \lambda\,  \map T_{\sigma,  P}  ~ | ~ \lambda\,  \map T_{\sigma,  P}~\text{is trace non-increasing} \, ,   \lambda \ge 0 & \,  , \\ \sigma_{\rm aux, out} \in  \St  ( S_{\rm aux}  S_{\rm out}) \, , 
			P_{\rm in, aux} \in  \Eff  (  S_{\rm in}  S_{\rm aux})  ~{\rm for~some~auxiliary~system~} &S_{\rm aux} \}\,.
		\end{aligned}
	\end{equation}
	
\end{definition}

Note  that every classical process can be generated by the above construction. For example, a process from $m$ dits to $n$ dits is described by a conditional probability distribution $p(\st y|  \st x)$, specifying the probability that the output is in the state described by the $n$-dit string $\st y$ when the input is in the state described by the $m$-dit string $\st x$.  This process can be obtained from the conditional transformation  (\ref{tsigmaP}), setting
\begin{align}
	\sigma_{\rm aux, out} : = \frac 1{d^m}\sum_{\st x,\st y}    p(\st y |\st x) |\st x\>\<\st x|  \otimes |\st y\>\<\st y|  \, ,   \qquad    P_{\rm  in, aux}   =  \sum_{\st x}   |\st x\>\<\st x| \otimes |\st x\>\<\st x| \,, \qquad {\rm and}  \qquad \lambda  =  d^m\,.
\end{align}  

The following observation will become useful later:
\begin{lemma}\label{lem:inclusion}
	The set  $\Transf  (S_{\rm in} \to S_{\rm out})$ defined in  Eq. (\ref{transfinout}) is contained in the set of completely positive trace non-increasing transformations that map states allowed by our toy theory into (generally subnormalized) states allowed by our toy theory, even when acting locally on part of a composite system. \end{lemma}

\Proof
Complete positivity is immediate from the fact that all conditional transformations $\map T_{\sigma, P}$ are  a subset of the transformations allowed in quantum theory (which are all completely positive), and that  the scaling factor $\lambda$ in Eq. (\ref{transfinout}) is non-negative.   The trace non-increasing property is demanded explicitly in Eq. (\ref{transfinout}). Finally, the condition that valid states of our toy theory are mapped into valid (generally subnormalized) states of our toy theory is guaranteed by the fact that each conditional transformation $\map T_{\sigma, P}$  produces a valid subnormalized state (Lemma \ref{lem:tsigmaP}), and that the  condition of trace non-increase  guarantees that the state remains subnormalized even after multiplication by the scaling factor $\lambda$. \qed

\medskip

We now prove that our toy theory admits an operational version of  the Choi isomorphism \cite{choi1975completely}.   For every  system $S=  D_1  \cdots  D_m A_1 \cdots A_n $, we consider the anti-system  $ \overline S_{\rm in} \coloneqq   \overline D_1 \cdots \overline D_n  \overline A_1  \cdots \overline A_m $ and  the  maximally entangled state 
\begin{align}
	|\Phi \>_{S\overline S}  :=  |\Phi^+\>_{D_1\overline A_1}   \otimes \cdots  |\Phi^+\>_{D_m\overline A_m} \otimes   |\Phi^+\>_{\overline D_1 A_1}   \otimes \cdots \otimes |\Phi^+\>_{\overline D_n A_n}  \, ,
\end{align}
where $|\Phi^+\>_{DA}  =   \sum_{i=0}^{d-1} |i\>_D|i\>_A / \sqrt d$ is the canonical Bell state of a general dit/anti-dit composite $DA$.  

The Choi operator of a generic transformation $\map T $,  with input system $S_{\rm in}  = D_1\cdots D_m  A_1  \cdots A_n$ and output system $S_{\rm out}=  D_1\cdots D_{m'}  A_1  \cdots  A_{n'}$,  is the  operator ${\sf Choi }  (\map T) \in \Lin  (S_{\rm out} \overline S_{\rm in})$  defined by 
\begin{align}\label{choidef}
	{\sf Choi }  (\map T)   :   =     (\map T  \otimes \map I_{\overline S_{\rm in}})   (|\Phi\>\<\Phi|)_{S_{\rm in} \overline S_{\rm in}} \, .
\end{align}
Since the states and transformations allowed by our toy theory are a subset of the sets of quantum states and quantum transformations, respectively, the correspondence $\map T  \mapsto  {\sf Choi}  (\map T)$ is injective: different transformations are mapped into different Choi states.  

As in quantum theory, the inverse of the Choi correspondence can be interpreted as conclusive teleportation: the transformation $\map T$ can be probabilistically extracted from the  the Choi state ${\sf Choi}  (\map T)$ by a teleportation-like scheme where the input system $S_{\rm in}$ is measured jointly with its copy $\overline S_{\rm in}$ via a projective measurement on the maximally entangled state. Specifically, one can define the inverse of the Choi correspondence as 
\begin{align}\label{inversechoi}
	{\sf Choi}^{-1}  (\sigma)   =   (d^{m+n})^2  \,   \map T_{\sigma,  P_{\rm ent}} \, ,
\end{align}
where   $ P_{\rm ent}  :=  |\Phi\>\<\Phi|_{\overline {\rm in},  {\rm in}}$ is the projector on the maximally entangled state, and $\map T_{\sigma, P_{\rm ent}}$ is defined as in Eq.  (\ref{tsigmaP}).

The above equation shows that  the correspondence  $\map T  \mapsto  \Phi_{\map T}$, viewed as a map between the unnormalized  transformations and the unnormalized states of our toy theory, is also surjective:  for every  unnormalized state $\sigma $ of system $S_{\rm out}  \overline S_{\rm in}$, there is an  unnormalized conditional transformation $\map T  :  =  {\sf Choi}^{-1}  (\sigma)$ such that ${\sf Choi} (\map T)  =  \sigma$.

We conclude this section by providing an alternative characterization of the physical transformations allowed by our toy theory:  

\begin{theo}\label{theo:transformations}
	The set of physical transformations $\Transf (S_{\rm in}\to S_{\rm out})$, defined in Eq. (\ref{transfinout}), coincides with the set of all completely positive trace non-increasing transformations mapping states allowed by our toy theory into (generally subnormalized) states allowed by our toy theory, even when acting locally on part of a composite system.
\end{theo}

\Proof
Lemma \ref{lem:inclusion} already proved that the set  $\Transf (S_{\rm in}\to   S_{\rm out})$ is included in the set of completely positive, trace non-increasing maps that transform valid states into valid (generally subnormalized) states. The converse inclusion follows from the Choi isomorphism. Let $\map M$ be a completely positive, trace non-increasing map that transforms valid states into valid (subnormalized) states, even when acting locally on part of a composite system.  For such a map, the Choi operator ${\sf Choi} (\map M) \eqqcolon \sigma$ must be a valid subnormalized state. Then, the Choi isomorphism guarantees that $\map M$ is proportional to a  transformation in $\Transf (S_{\rm in} \to S_{\rm out})$. Finally, since $\map M$ was trace non-increasing, it is actually an element of $\Transf (S_{\rm in}\to S_{\rm out})$. \qed

\medskip

\subsection{Channels and instruments}

A physical transformation that happens with unit probability on  every normalized state is called  a {\em channel}.   

In our toy theory, the channels are described by trace-preserving maps, similarly as in quantum theory:
\begin{definition}
	A physical transformation $\map C  \in  \Transf  ( S_{\rm in}  \to S_{\rm out})$ is a channel if it is trace-preserving.   
\end{definition}

Other physical transformations arise in measurement processes that induce an evolution of the system depending on the measurement outcome.  These measurement processes are described by {\em instruments}, namely  collections of physical transformations that sum up to a channel. 

In our toy theory, the instruments are defined as follows: 
\begin{definition}
	An instrument 
	with input system $S_{\rm in}$, output system $S_{\rm out}$, and outcomes in the set $\sf X$, is a tuple  $( \map M_x)_{x\in\sf X}$, where $\map M_x$ is an element of $\Transf (  S_{\rm in} \to S_{\rm out}) $ for every $x\in\sf X$,  and $\sum_{x\in\sf X} \, \map M_x$ is trace-preserving.  
\end{definition}

\subsection{Generalization to systems of arbitrary dimension}

Here we formulate a version of our toy theory where the dimension of the composite systems is not necessarily the power of a fixed integer $d$.   To this purpose, we consider composite systems including  dits and anti/dits with different values of $d$. The basic type of dit (anti-dit) of dimension $d_i$ will be denoted by $D^{(i)}$  ($A^{(i)}$).  In this version of the toy theory, we take the  dimensions of the basic system types to be prime numbers, and we generate  systems of non-prime dimension by composing the basic system types. 

The state spaces  of the composite systems are defined in a similar way as in the previous subsections.  
For example, the pure states of a composite system $S  =  \st D^{(1)} \st A^{(1)}  \cdots \st D^{(k)}  \st A^{(k)}$, consisting of $m_1$ dits/anti-dits of dimension $d_1$,    $m_2$ dits/anti-dits of dimension $d_2$, and so on,  are projectors of the form $|\psi\>\<\psi|$, where $|\psi\>$ is a unit vector in the subspace 
\begin{align}
	\spc H_{ \st D^{(1)} \st A^{(1)}  \cdots \st D^{(k)}  \st A^{(k)}}^{(\pi_1, \st q_1), \dots,  (\pi_k,  \st q_k)}  :  =  \spc H_{\st D^{(1)} \st A^{(1)}}^{\pi_1, \st q_1} \otimes \cdots \otimes \spc H_{\st D^{(k)} \st A^{(k)}}^{\pi_k, \st q_k} \, ,  
\end{align}
where, for every $i\in  \{1,\dots, k\}$,   $\pi_k$ is a permutation in ${\sf S}_{m_i}$,  and $\st q_i  $ is a vector in $\Z_d^{m_i}$.

When the number of dits of some type is different from the number of anti-dits of the same type,  the pure states are obtained by putting the excess systems in computational basis states, as we did earlier in Definitions \ref{def:purestatesm<n} and \ref{def:purestatesm>n}.  Then, the  mixed states are  obtained as  convex combinations of the pure states.  Measurements, effects, and transformations are also defined as in the previous subsections.

This version of our toy theory satisfies all the properties discussed earlier in this section:  the sets of pure states are closed under tensor product, the sets of mixed states are closed under partial trace, and all mixed  states of purely classical (or purely anti-classical) systems can be purified.  The theory still satisfies the property of consistency of the conditional states, and admits a version of the Choi isomorphism, using which it is possible to show that all trace non-increasing maps that transform valid states in to valid (possibly subnormalized) states correspond to valid transformations allowed by our toy theory.  The proofs are more cumbersome than the ones presented earlier, due to the presence of multiple types of basic systems, but all the arguments are essentially the same.  

\section*{Bilocal Tomography}

Here we show that our toy theory satisfies Bilocal Tomography \cite{hardy2012limited} for all multipartite systems in which all subsystems  consist either of dits or anti-dits.  Mathematically, Bilocal Tomography can  be formalized as follows:  
\begin{definition}
	Let $S=  S_1\cdots  S_k$ be a $k$-partite system consisting of $k\ge 2$  subsystems.  
	For even  $k$, we say that  system $S$ satisfies Bilocal Tomography if, for every pair of distinct states $\rho$ and $\sigma$ in $\St(S)$,   there exists at least one permutation  $\pi  \in  {\sf S}_k$  and  $k/2$  effects   $P_{12} \in  \Eff (S_{\pi  (1)}  S_{\pi  (2)})\, ,  P_{34} \in \Eff (S_{\pi  (3)}  S_{\pi (4)}), \dots,   P_{k-1,k} \in \Eff (S_{\pi  (k-1)}  S_{\pi (k)})$ such that 
	\begin{align}
		\nonumber &\Tr \left[\,  \Big((P_{1 2})_{S_{\pi(1)}  S_{\pi(2)}} \otimes  (P_{3 4})_{S_{\pi(3)} S_{\pi(4)}}  \otimes \cdots \otimes  (P_{k-1,  k})_{S_{\pi(k-1)}  S_{\pi(k)}} \Big) \, \rho \right]  \\
		& \qquad \qquad \not  = \Tr \left[  \Big( (P_{1 2})_{S_{\pi(1)}  S_{\pi(2)}} \otimes  (P_{3 4})_{S_{\pi(3)}  S_{\pi(4)}}  \otimes   \cdots \otimes (P_{k-1, k})_{S_{\pi(k-1)} S_{\pi(k)}} \Big) \, \sigma\right] \, .
	\end{align}
	For odd $k$,  we say that  system $S$ satisfies Bilocal Tomography if, for every pair of distinct states $\rho  \in \St(S)$ and $\sigma \in \St (S)$, there exists at least one permutation  $\pi \in {\sf S}_k$,  $(k-1)/2$  effects  $P_{12} \in  \Eff (S_{\pi  (1)}  S_{\pi  (2)})\, ,  P_{34} \in \Eff (S_{\pi  (3)}  S_{\pi (4)}), \dots,   P_{k-2,k-1} \in \Eff (S_{\pi  (k-1)}  S_{\pi (k)})$, and one effect $P_k  \in \Eff(  S_{t(k)})$ such that 
	\begin{align}
		\nonumber &\Tr \left[\,  \Big((P_{1 2})_{S_{\pi(1)}  S_{\pi(2)}} \otimes  (P_{3 4})_{S_{\pi(3)}  S_{\pi(4)}}  \otimes \cdots \otimes  (P_{k-2, k-1})_{S_{\pi(k-2)}  S_{\pi(k-1)}} \otimes (P_k)_{S_{\pi(k)}} \Big) \, \rho \right]  \\
		& \qquad \qquad \not  = \Tr \left[  \Big( (P_{1 2})_{S_{\pi(1)} S_{\pi(2)}} \otimes  (P_{3 4})_{S_{\pi(3)}  S_{\pi(4)}}  \otimes  \cdots  (P_{k-2, k-1})_{S_{\pi(k-2)}  S_{\pi(k-1)}}  \otimes (P_k)_{S_{\pi(k)}} \Big) \, \sigma\right] \, .  \label{differentonbilocal}
	\end{align}
\end{definition}
The main result of this section is the following theorem:
\begin{theo}\label{theo:bilocal}
	Every   $k$-partite system $S= S_1\cdots S_k$ in which subsystem $S_i$ is either a dit or an anti-dit for every $i\in  \{1,\dots,  k\}$ satisfies Bilocal Tomography.    
\end{theo}

The intuition at the basis of the proof is  that  every dit/anti-dit pair is associated to a set of $d$-dimensional quantum systems. Exploiting this fact, the property of Bilocal Tomography in our toy theory can be reduced to the property of Local Tomography in ordinary quantum theory.   

The proof of Theorem \ref{theo:bilocal} is based on a few technical lemmas, and on the following notations.   For an arbitrary system $S$ and  an arbitrary set of linear operators $  {\sf  O}  \subseteq \Lin  (\spc  H_S)$, we define the vector space 
\begin{align}
	\Span_\R   \big(  {\sf  O} \big) : =  \left\{ \sum_i  \,c_i\,  O_i  ~\Big| ~ c_i  \in \R  \, ,  O_i \in {\sf O}  \right\}\, ,
\end{align}
consisting of finite linear combinations of elements in $\sf O$.   For $m$ sets of linear operators on $\spc  H_S$ ${\sf O}_1,  \dots , {\sf O}_m$, we denote by  ${\sf O_1} \otimes \cdots \otimes {\sf O_m}$ the set of all elements of the form $O_1\otimes \cdots \otimes O_m$, with $O_i\in  {\sf O_i}$ for every $i\in  \{1,\dots,  m\}$.  For a set $\sf O$ and a fixed operator $F$, we denote by ${\sf O} \otimes F$ the set of all operators of the form $O\otimes F$, with $O\in \sf O$.

\begin{lemma}\label{lem:globallocal}
	Let $S= \st D \st A$ be a system of type $(m,m)$, with $\st D  =  D_1\cdots  D_m$ and $\st A  =   A_1  \cdots A_m$. Then, for every permutation $\pi \in {\sf S}_m$ and for  every vector $\st q  \in  \Z_d^{ m}$, the space of pure states of type $(\pi,\st q)$  satisfies the condition
	\begin{align}\label{samespan-sector}
		\Span_\R  \left\{  \Pur\St   \big(S; \pi , \st q\big)  \right\} =   \Span_\R  \left\{  \Pur\St   \big(D_1  A_{\pi(1)}; q_1\big) \otimes \cdots \otimes \Pur\St   \big(D_m  A_{\pi(m)}; q_m\big)     \right\}   \, .
	\end{align}
\end{lemma}
\Proof By Definition \ref{def:purestatesmm},  $\Pur\St   \big(S; \pi, \st q \big)$ is the set of all projectors on  vectors   in  $\spc H^{\pi,\st q}_{\st D \st A}$.  Hence, its linear span    is the set of all Hermitian operators on  $\spc H^{\pi,\st q}_{\st D \st A}$; in formula, 
\begin{align}
	\nonumber \Span_\R    \, \left\{   \Pur\St   \big(S; \pi, \st q \big) \right\}  &=  \Herm \left(  \spc H^{\pi,\st q}_{\st D \st A} \right) \\
	\nonumber & = \Herm \left( \spc H^{q_1}_{D_1 A_{\pi(1)}}\otimes \cdots \otimes \spc H^{q_m}_{D_m A_{\pi(m)}}\right)\\
	& = \Herm \left( \spc H^{q_1}_{D_1 A_{\pi(1)}} \right)  \otimes \cdots \otimes \Herm\left(\spc H^{q_m}_{D_m A_{\pi(m)}}\right)\,,  \label{spanglobal}
\end{align}
where the subspaces $\spc H^{q_1}_{D_1  A_{\pi  (1)}},  \dots,  \spc H^{q_m}_{D_m  A_{\pi  (m)}}$ are defined as in Eq. (\ref{hq}), and   the second equality  follows from the definition of $\spc H_{\st D \st A}^{\pi, \st q}$ in Eq. (\ref{hpiq}).  

On the other hand,  for every $i$, $\Pur\St   \big(D_iA_{\pi(i)}; q_i)$ is the set of all  projectors on vectors in $\spc H^{q_i}_{D_i A_{\pi(i)} }$ (Definition \ref{def:DA}). Hence,  its linear span is  the set of all Hermitian operators on  $\spc H^{q_i}_{D_i A_{\pi(i)} }$; in formula,  
\begin{align}
	\Herm \left( \spc H^{q_i}_{D_i A_{\pi(i)}} \right)   = \Span_\R  \left(  \Pur\St   \big(D_i  A_{\pi(i)}; q_i\big) \right)  \qquad \forall i\in  \{1,\dots,  m\} \, .
\end{align}
Hence,  we have 
\begin{align}
	\nonumber  \Herm \left( \spc H^{q_1}_{D_1 A_{\pi(1)}} \right)  \otimes \cdots \otimes \Herm\left(\spc H^{q_m}_{D_m A_{\pi(m)}}\right)    &   =  \Span_\R  \left(  \Pur\St   \big(D_1  A_{\pi(1)}; q_1\big)    \right)\otimes \cdots \otimes \Span_\R\left(  \Pur\St   \big(D_m  A_{\pi (m)}; q_m\big) \right)\\
	&   =  \Span_\R  \left(  \Pur\St   \big(D_1  A_{\pi(1)}; q_1\big)    \otimes \cdots \otimes  \Pur\St   \big(D_m  A_{\pi (m)}; q_m\big) \right)
	\label{spanlocal}
\end{align}
Combining Eqs.    (\ref{spanglobal}) and (\ref{spanlocal}) we then obtain the desired result. \qed  
\medskip

We now generalize the above lemma to the $m\not  =  n$ case. 

\begin{lemma}\label{lem:globallocalmn}
	Let $S= \st D \st A$ be a system of type $(m,n)$, with $\st D  =  D_1\cdots  D_m$ and $\st A  =   A_1  \cdots A_n$.  For every $m<n$,  every permutation $\pi \in {\sf S}_n$,  and every pair of  vectors $\st q  \in  \Z_d^{ m}$ and $\st r\in  \Z_d^{n-m}$, the space of pure states of type $(\pi,\st q, \st r)$  satisfies the condition
	\begin{align}
		\nonumber &\Span_\R  \left\{  \Pur\St   \big(S; \pi , \st q, \st r \big)  \right\}\\
		& \qquad  =   \Span_\R  \left\{  \Pur\St   \big(D_1  A_{\pi(1)}; q_1\big) \otimes \cdots \otimes \Pur\St   \big(D_m  A_{\pi(m)}; q_m\big)  \otimes |r_{1}\>\<r_{1}|_{A_{\pi(m+1)}} \otimes \cdots \otimes |r_{n-m}\>\<r_{n-m} |_{A_{\pi(n)}}    \right\}   \, .  \label{samespan-sectorm<n}
	\end{align}
	For every $m>n$,  every permutation $\pi \in {\sf S}_m$,  and every pair of  vectors $\st q  \in  \Z_d^{ m}$ and $\st r\in  \Z_d^{m-n}$, the space of pure states of type $(\pi,\st q, \st r)$  satisfies the condition
	\begin{align}
		\nonumber &\Span_\R  \left\{  \Pur\St   \big(S; \pi , \st q, \st r \big)  \right\}\\
		& \qquad  =   \Span_\R  \left\{  \Pur\St   \big(D_{\pi(1)}  A_1; q_1\big) \otimes \cdots \otimes \Pur\St   \big(D_{\pi(n)}  A_n; q_n\big)  \otimes |r_{1}\>\<r_{1}|_{D_{\pi(n+1)}} \otimes \cdots \otimes |r_{m-n}\>\<r_{m-n} |_{D_{\pi(m)}}    \right\}   \, .  \label{samespan-sectorm>n}
	\end{align}
\end{lemma}
\Proof By Definition \ref{def:purestatesm<n},  $\Pur\St   \big(S; \pi, \st q ,\st r \big)$ is the set of all projectors on  vectors   in  $\spc H^{\pi,\st q , \st r}_{\st D \st A}$.  Hence, its linear span    is the set of all Hermitian operators on  $\spc H^{\pi,\st q , \st r}_{\st D \st A}$. The definition of $\spc H^{\pi,\st q , \st r}_{\st D \st A}$ in Eq. (\ref{hpiqr}) implies the equality
\begin{align}
	\Herm  \left(   \spc H_{D_1A_{\pi(1)}}^{q_1} \otimes \cdots \otimes \spc H_{D_mA_{\pi(m)}}^{q_m} \right)  \otimes |r_1\>\<r_1|_{A_{\pi (m+1)}}  \otimes \cdots \otimes |r_{n-m}\>\<r_{n-m}|_{A_{\pi (n)}}  
	\, .
\end{align}
Hence, we have 
\begin{align}
	\nonumber \Span_\R    \, \left\{   \Pur\St   \big(S; \pi, \st q,\st r \big) \right\}  &=  \Herm \left(  \spc H^{\pi,\st q,\st r}_{\st D \st A} \right)  \otimes   |r_1\>\<r_1|_{A_{\pi (m+1)}}  \otimes \cdots \otimes |r_{n-m}\>\<r_{n-m}|_{A_{\pi (n)}}\\
	\nonumber & = \Herm \left( \spc H^{q_1}_{D_1 A_{\pi(1)}}\otimes \cdots \otimes \spc H^{q_m}_{D_m A_{\pi(m)}}\right) \otimes |r_1\>\<r_1|_{A_{\pi (m+1)}}  \otimes \cdots \otimes |r_{n-m}\>\<r_{n-m}|_{A_{\pi (n)}}\\
	& = \Herm \left( \spc H^{q_1}_{D_1 A_{\pi(1)}} \right)  \otimes \cdots \otimes \Herm\left(\spc H^{q_m}_{D_m A_{\pi(m)}}\right) \otimes |r_1\>\<r_1|_{A_{\pi (m+1)}}  \otimes \cdots \otimes |r_{n-m}\>\<r_{n-m}|_{A_{\pi (n)}}\,,  \label{spanglobalmn}
\end{align}
where the subspaces $\spc H^{q_1}_{D_1  A_{\pi  (1)}},  \dots,  \spc H^{q_m}_{D_m  A_{\pi  (m)}}$ are defined as in Eq. (\ref{hq}), and   the second equality  follows from the definition of $\spc H_{\st D \st A}^{\pi, \st q}$ in Eq. (\ref{hpiq}).  

In the proof of the previous lemma, we have  shown that 
\begin{align}
	\Herm \left( \spc H^{q_1}_{D_1 A_{\pi(1)}} \right)  \otimes \cdots \otimes \Herm\left(\spc H^{q_m}_{D_m A_{\pi(m)}}\right)   
	=  \Span_\R  \left(  \Pur\St   \big(D_1  A_{\pi(1)}; q_1\big)    \otimes \cdots \otimes  \Pur\St   \big(D_m  A_{\pi (m)}; q_m\big) \right)
	\label{spanlocalmn}
\end{align}
Combining Eqs.    (\ref{spanglobalmn}) and (\ref{spanlocalmn}) we then obtain the desired result. 

The proof for $m>n$ is analogous. \qed

\medskip

\begin{lemma}\label{lem:samespan}
	Let $S  =  \st D \st A$ be a system of type $(m,n)$, with $\st D=  D_1\dots,  D_m$ and $\st A  = A_1\dots A_n$.  
	One has  the equality
	\begin{align}\label{lgstates}
		\Span_\R    \left(     \St(S) \right)  =
		\left\{
		\begin{array}{ll}  
			\Span_\R\left(   \bigcup_{\pi \in {\sf S}_m}   \St   (D_1  A_{\pi(1)}) \otimes  \cdots \otimes \St  (D_m  A_{\pi(m)}) \right)  &  {\rm for~} m=n  \\
			&\\
			\Span_\R\left(   \bigcup_{\pi \in {\sf S}_n}  \St   (D_1  A_{\pi(1)}) \otimes  \cdots \otimes \St   (D_m  A_{\pi(m)}) \otimes \St  (A_{\pi (m+1)}) \otimes \cdots \otimes \St   (A_{\pi (n)})    \right)  \qquad  & {\rm for~} m<n\\
			&\\
			\Span_\R\left(  \bigcup_{\pi \in {\sf S}_m} \St   (D_1  A_{\pi(1)}) \otimes  \cdots \otimes \St  (D_n  A_{\pi(n)}) \otimes \St  (D_{\pi (n+1)}) \otimes \cdots \otimes \St   (D_{\pi (m)})   \right)  & {\rm for~} m>n \, .
		\end{array}
		\right.
	\end{align}
	Similarly,  
	\begin{align}\label{lgeffects}
		\Span_\R    \left(     \Eff(S) \right)  =
		\left\{
		\begin{array}{ll}  
			\Span_\R\left(   \bigcup_{\pi \in {\sf S}_m}   \Eff   (D_1  A_{\pi(1)}) \otimes  \cdots \otimes \Eff  (D_m  A_{\pi(m)}) \right)  &  {\rm for~} m=n  \\
			&\\
			\Span_\R\left(   \bigcup_{\pi \in {\sf S}_n}  \Eff   (D_1  A_{\pi(1)}) \otimes  \cdots \otimes \Eff   (D_m  A_{\pi(m)}) \otimes \Eff  (A_{\pi (m+1)}) \otimes \cdots \otimes \Eff   (A_{\pi (n)})    \right)  \qquad  & {\rm for~} m<n\\
			&\\
			\Span_\R\left(  \bigcup_{\pi \in {\sf S}_m} \Eff   (D_1  A_{\pi(1)}) \otimes  \cdots \otimes \Eff  (D_n  A_{\pi(n)}) \otimes \Eff  (D_{\pi (n+1)}) \otimes \cdots \otimes \Eff   (D_{\pi (m)})   \right)  & {\rm for~} m>n \, .
		\end{array}
		\right.
	\end{align}
\end{lemma}

\Proof Let us start by proving Eq. (\ref{lgstates})  for $m=n$.
Using Lemma \ref{lem:samespan}, we obtain  
\begin{align}
	\nonumber  \Span_\R  \left(  \St (S) \right)  &=  \Span_\R  \left(  \Pur\St (S)\right)  \\  
	\nonumber &=  \Span_\R  \left(  \bigcup_{\pi  \in  {\sf S}_m}  \bigcup_{\st q \in  \Z_d^m}  \Pur\St (S;  \pi, \st q)  \right) \\  
	\nonumber &   =  \Span_\R  \left(  \bigcup_{\pi  \in  {\sf S}_m}  \bigcup_{\st q \in  \Z_d^m}  \Pur\St   \big(D_1  A_{\pi(1)}; q_1\big)    \otimes \cdots \otimes  \Pur\St   \big(D_m  A_{\pi (m)}; q_m\big) \right)  
	\\
	\nonumber &   =  \Span_\R  \left( \bigcup_{\pi  \in  {\sf S}_m}    \left(  \bigcup_{q_1  \in  \Z_d}  \Pur\St   \big(D_1  A_{\pi(1)}; q_1\big)~\right)     \otimes \cdots \otimes   \left(   \bigcup_{q_m  \in  \Z_d}  \Pur\St   \big(D_m  A_{\pi (m)}; q_m\right)  \right)  
	\\  
	\nonumber &   =  \Span_\R  
	\left( \bigcup_{\pi  \in  {\sf S}_m} 
	\Pur\St  (D_1  A_{\pi(1)}) 
	\otimes \cdots \otimes  
	\Pur\St   (D_m  A_{\pi (m)})    \right)\\
	&   =  \Span_\R  
	\left( 
	\bigcup_{\pi  \in  {\sf S}_m}     \St  (D_1  A_{\pi(1)})  
	\otimes \cdots \otimes  
	\St   (D_m  A_{\pi (m)})   \right) \,.
\end{align} 
This concludes the proof of Eq. (\ref{lgstates}) for $m=n$.  The proofs for $m>n$ and $m<n$ are analogous, the only difference being that they use the  additional relation $\Span_\R  \left( \cup_{r \in  \Z_d}  |r\>\<r|_{X}\right) =  \Span_\R  (\St (X))$, where $X$ is either a dit or an anti-dit.  

Finally, Eq. (\ref{lgeffects}) follows from Eq. (\ref{lgstates}) and from the self-duality of our toy theory. 
\qed 

\medskip  

\begin{lemma}\label{lem:bilocalstatespace}  
	Let system $S  =  S_1 \cdots  S_k$ be a $k$-partite system in which, for every $i\in  \{1,\dots,  k\}$, the  subsystem $S_i$ is either a dit or an anti-dit.   Then,  one has the equalities  
	\begin{align}
		\Span_\R \left ( \St  (S_1 \cdots S_k) \right)   =
		\begin{cases}
			\Span_\R\left( \bigcup_{\pi  \in  {\sf S}_k}  \St   (S_{\pi(1)} S_{\pi(2)} ) \otimes  \cdots \otimes \St  (S_{\pi(k-1)} S_{\pi(k)} )\right)   \qquad    &  {\rm for~even~}k  \\  & \\
			\Span_\R\left( \bigcup_{\pi  \in  {\sf S}_k}    \St  (S_{\pi(1)} S_{\pi(2)} ) \otimes  \cdots \otimes \St_\R  (S_{\pi(k-2)} S_{\pi(k-1)} ) \otimes \St (S_{\pi(k)})  \right)  \qquad  &{\rm for~odd}~k  \, .
		\end{cases}
	\end{align}
	and 
	\begin{align}
		\Span_\R \left ( \Eff  (S_1 \cdots S_k) \right)   =
		\begin{cases}
			\Span_\R\left( \bigcup_{\pi  \in  {\sf S}_k}  \Eff   (S_{\pi(1)} S_{\pi(2)} ) \otimes  \cdots \otimes \Eff  (S_{\pi(k-1)} S_{\pi(k)} )\right)   \qquad    &  {\rm for~even~}k  \\  & \\
			\Span_\R\left( \bigcup_{\pi  \in  {\sf S}_k}    \Eff  (S_{\pi(1)} S_{\pi(2)} ) \otimes  \cdots \otimes \Eff_\R  (S_{\pi(k-2)} S_{\pi(k-1)} ) \otimes \Eff (S_{\pi(k)})  \right)  \qquad  &{\rm for~odd}~k  \, .
		\end{cases}
	\end{align}
\end{lemma}  
\Proof  Immediate from Lemma \ref{lem:samespan} and from the fact that the set  ${\sf S}_k$ includes all the permutations acting only on the dits and all the permutations acting only on the anti-dits.  \qed  

\medskip

We are finally ready to prove Theorem \ref{theo:bilocal}. 

\smallskip 

{\bf Proof of Theorem \ref{theo:bilocal}.}   We provide the proof in the even $k$ case, because the odd $k$ case is analogous.    Proposition \ref{prop:separatingeffects}  guarantees that two distinct states $\rho$ and $\sigma$ give rise to different probabilities $\Tr[P \rho]$ and $\Tr  [P  \sigma]$ for at least one effect $P  \in  \Eff (S)$.  On the other hand, Lemma \ref{lem:bilocalstatespace} implies the inclusion 
\begin{align}
	\Eff  (S)  \subset   \Span_\R  \left(   \bigcup_{\pi  \in  {\sf S}_k} \Eff   (S_{\pi(1)}  S_{\pi(2)})  \otimes  \cdots \otimes  \Eff   (S_{\pi(k-1)}  S_{\pi(k)})  \right)  \,.
\end{align}
Hence, there must exist a permutation $\pi\in {\sf S}_k$  and  a  set of   effects $P_{1, 2}  \in \Eff  (S_{\pi(1)  \, \pi(2)}) \, ,\dots \,  , P_{k-1, k}  \in \Eff  (S_{\pi(k-1)  \, \pi(k)}) $  such that Eq. (\ref{differentonbilocal}) holds. \qed

\section*{Impossibility to assign individual pure states to classical systems}

In this section we report the complete proof already sketched in the main article that, under the assumption that our theory describes nature at the fundamental level, it is incorrect to assume that every classical system is in a pure state at the ontological level. To do this, we first show that a) the latter claim would imply that bipartite state can only be separable, and then that b) entangled states exist in our theory, hence arriving at a contradiction.

a) The first point is straightforward. If only pure states can represent classical systems at the ontological level, then pure entangled bipartite states are inadmissible. By contradiction, let us consider such a state to represent the state of an arbitrary $(1,m)$-composite system. Since it is entangled and pure, its marginal on the classical system is necessarily a mixed state. Furthermore, since our theory describes nature at the fundamental level, such mixture cannot be interpreted as epistemic, because it derives from a pure state. In conclusion, we are left with a mixed state that describes the classical system at the ontological level. Finally, the most general separable state of the $(1,m)$-composite is the state
\begin{align*}
	\Sigma = \sum_{i=0}^1 \gamma_{i} \ket{i} \bra{i} \otimes \rho_i
\end{align*}
where $\rho_i$ are arbitrary states of the $(0,m)$-composite (we could have considered the $(n,m)$-composite as well, and then tracing out all the $(n-1,m)$-partite systems, however, to keep notation simple, we chose $n=1$. For the same reason, in the following we will only consider $m=1$).

b) We now start proving that not only entangled states exist, but also that they are not a mere mathematical representation of our framework: they can be in principle distinguished from separable states by repeatedly performing measurements allowed in the theory on identical copies of the entangled state.

We start by writing the most general separable state of the $(1,1)$-composite:
\begin{align*}
	\rho_{BA} = \sum_{i,j=0}^1 \gamma_{ij} \ket{i} \bra{i} \otimes \ket{j} \bra{j}
\end{align*}
while entangled pure states $\ket{\Psi}_{BA}$ can have the form
\begin{align*}
	\ket{\Psi} = \sqrt{p} \ket{00} + \sqrt{1-p} \ket{11} \text{ or } \ket{\Psi} = \sqrt{p} \ket{01} + \sqrt{1-p} \ket{10},
\end{align*}
where $p\in(0,1/2]$, $\ket{ij}$ is short notation for $\ket{i} \otimes \ket{j}$ for $i,j=0,1$, and we ignored any relative phase. We just consider entangled states of the former kind in the following, the other case being analogous.

In order to distinguish $\ket{\Psi}\bra{\Psi}$ from any possible $\rho_{BA}$,  we can suppose to perform the following POVM: $\{P_{\rm yes}, P_{\rm no}\}$ where $P_{\rm yes} =  |\Psi\>\<\Psi|$ and $P_{\rm no}  =  I\otimes I- P_{\rm yes}$. Such measurement is admissible in the theory, indeed, $P_{\rm no}$ can be written as a linear combination, with positive coefficients, of allowed states, namely $\ket{\Psi^\perp} \bra{\Psi^\perp} + \ket{01}\bra{01} + \ket{10}\bra{10}$, where $\ket{\Psi^\perp} = \sqrt{1-p} \ket{00} - \sqrt{p} \ket{11}$.

Clearly, $p(\text{yes}|\Psi) = 1 $ and $p(\text{no}|\Psi) = 0 $, where $p(\text{yes/no} |\Psi)$ is the probability of getting outcome yes/no on the state $\ket{\Psi}\bra{\Psi}$ given by the Born rule $p(\text{yes/no}|\Psi) = \Tr[P_{\textrm{yes/no}} \ket{\Psi}\bra{\Psi}]$. On the other hand, $p(\text{yes}|\rho) = \Tr[P_{\mathrm{yes}} \rho] = \gamma_{00}p+(1-p)\gamma_{11}$ and $p(\text{no}|\rho) = 1-  p(\text{yes}|\rho)= 1 - \gamma_{00}p - (1-p)\gamma_{11}$. Worst case scenario happens when $p(\text{yes}|\rho)$ is as close as possible to 1, namely when $\gamma_{01} = \gamma_{10} = 0$, hence $1-\gamma_{11} = \gamma_{00} \eqqcolon q$. In this case,
\begin{equation*}
	p(\text{no}|\rho) =
	\begin{cases}
		p + q (1-2p) \ge p \text{ if }p\in(0,1/2]\\
		(1-p) + (1-q) (2p-1) \ge (1-p) \text{ if }p\in[1/2,1).\\
	\end{cases}
\end{equation*}
In conclusion, when the POVM is performed on an arbitrary separable state, it gives outcome ``no'' with probability at least $\min \{p, 1-p\}$. Therefore, repeatedly performing such measurement on an entangled state would allow use to rule out that it is in a separable form if the ``no'' outcome is never observed.

\section*{Simulation of local measurements on  a two-qubit maximally entangled state} 

Here we consider the situation where two parties, Alice and Bob, perform local measurement on a composite system consisting of two bits $B_1B_2$ and two anti-bits $A_1A_2$,  jointly in the state  $|\Phi\>_{B_1A_1}   |\Phi\>_{B_2A_2} $, where
\begin{align}
	|\Phi\>  =  \frac{ |0\>   |0\>  +   |1\> |1\>}{\sqrt 2} \,.
\end{align}
is the maximally entangled state of the bit/anti-bit composite.  In this Bell activation scenario, Alice has access to bit $B_1$ and anti-bit $A_2$, while Bob has access to bit $B_2$ and anti-bit $A_1$.  Our goal is to show that Alice and Bob can simulate the correlations of arbitrary local measurements performed on a two-qubit maximally entangled state.

Suppose that Alice and Bob want to simulate qubit measurements that project on basis vectors  $\left\{|v_a\>\right\}$, where $|v_a\>  =  v_{a,0}\,|  0\>  +  v_{a,1}\,|1\>,  $  for Alice, and $\left\{|w_b\>\right\}$, where $|w_b\>  =  w_{b,0}\, |0\>  +  w_{b,1}\, |1\>$ for Bob.  When Alice's and Bob's measurements are performed on the two-qubit maximally entangled state $|\Phi^+\> =   ( |0\>   |0\>  +   |1\>  |1\>)/\sqrt 2$, the probability of the outcomes $a$ and $b$ is   
\begin{align}
	p_{\rm quantum}(a,b)   = \frac{|  v_{a,0}  w_{b,0}  +  v_{a,1}  w_{b,1}|^2}2 \,.
\end{align}

To reproduce this probability distribution, Alice and Bob measure their bits and anti-bits with two-outcome measurements described by the operators  $P_a  := \sum_{l\in\{0,1\}} |V^{(l)}_a\>\<V^{(l)}_a|$  and  $Q_b  := \sum_{l\in\{0,1\}} |W^{(l)}_{b}\>\<W^{(l)}_{b}|$, with 
\begin{align}
	|V^{(l)}_a\>    =  v_{a,0}  \,  |\phi_0^{(l)}\>  +  v_{a,1} \,     \,  |\phi_1^{(l)}\> \, 
\end{align}
and 
\begin{align}|W^{(l)}_{b}\>    =  w_{b,0}  \,  |\phi_{l}^{(l)}\>  +  w_{b,1} \,     \,  |\phi_{l\oplus 1}^{(l)}\> ,\,
\end{align}
having used the notation  
\begin{align}|\phi_i^{(j)} \>_{BA}  :=  |i\>_B |i \oplus j\>_A \,,
\end{align} 
for $i,j\in\{0,1\}$. 
It is easy to check that these operators form two valid quantum measurements in our toy theory:  they sum up to the identity matrix, and each operator is a linear combination, with positive coefficients, of projectors on pure states.

Using the notation $\Phi =  |\Phi\>\<\Phi|$, the  probability distribution of the outcomes $a$ and $b$ can be written as
\begin{align}
	\nonumber p_{\rm toy}  (a,b)    =   \Tr \left[    (  P_a  )_{B_1A_2}   \otimes (Q_b)_{B_2 A_1}   ~  (\Phi_{B_1A_1} \otimes  \Phi_{B_2A_2}) \right] & \\
	\nonumber   =  \sum_{l,l'}    \left|   \<\Phi|_{B_1A_1}  \<\Phi|_{B_2A_2}  \,   |V_a^{(l)}  \>_{B_1A_2} | W^{(l')}_{b} \>_{B_2A_1}  \,   \right|^2 & \\
	\label{zzz}   =   \sum_{l,l'}  \Big|  \sum_{k,k'}    v_{a,  k}  \, w_{b ,     k'  }   \<\Phi|_{B_1A_1}  \<\Phi|_{B_2A_2}  \,  |\phi_k^{(l)} \>_{B_1A_2}   |\phi^{(l')}_{k'\oplus l'}\> \Big|^2 &\,.
\end{align} 

Now, we use the fact that the joint state  of the total system can be equivalently rewritten as 
\begin{align}\label{bipartite}
	|\Phi\>_{B_1A_1}   |\Phi\>_{B_2A_2}   =\frac 12  \,  \sum_{k,l  \in  \{0,1\}}    |\phi_k^{(l)}\>_{B_1A_2}    |\phi_{k \oplus l}^{(l)}\>_{B_2 A_1}  \, ,   
\end{align}
where $\oplus$ denotes the addition modulo 2. 
Using the above expression, we obtain the relation  
\begin{align}
	\<\Phi|_{B_1A_1}  \<\Phi|_{B_2A_2}  \,  |\phi_k^{(l)} \>_{B_1A_2}   |\phi^{(l')}_{k'\oplus l'}\>_{B_2A_1}  =   \frac 1{2}\, \delta_{l,l'}  \, \delta_{k,  k'} \, ,
\end{align}
which, inserted into Eq. (\ref{zzz}), yields the conclusion  
\begin{align}
	\nonumber  p_{\rm toy}  (a,b)  &  =  \frac 14 \, \sum_l \, \left| \sum_{k}   \,    v_{a,  k}  \, w_{b,     k  }     \right|^2  \\   &\equiv  p_{\rm quantum }  (a,b) \, ,
\end{align}
In summary, every pair of local  measurements on a maximally entangled two-qubit quantum state can be simulated by local measurements in our toy theory on two copies of the state $|\Phi\>$. 

\section*{No violation of two-party  Bell inequalities when all but one  settings of one party have  predetermined outcomes}
	
	Here we review the known fact that the violation of the CHSH inequality implies that none of the outcomes involved in the experiment can be predetermined. In other words, if just one outcome (associated to one of the  two settings of one of the two parties) is predetermined, then the CHSH inequality cannot be violated. This fact holds in general for every two-party  Bell inequality when all but one settings of one party have predetermined outcomes. These results can be derived from a general argument based the monogamy of nonlocal correlations \cite{barrett2006maximally}. For convenience of the reader, however,  here we provide an elementary step-by-step proof.  
	
	Consider a general two-party scenario, with $x$ and $y$ ($a$ and $b$) denoting Alice's and Bob's settings (outcomes), respectively.  We denote by $\sf X$ and $\sf Y$  ($\sf A$ and $\sf B$)  the sets of all possible settings (all possible outcomes) of Alice and Bob, respectively.   In the CHSH case, all the sets are binary, namely $|\sf X| = |\sf Y|  =  |\sf A|  =  |\sf B |  = 2$.     In general, the cardinality of the sets $\sf X, \sf Y, \sf A ,\sf B$ can be any positive integer.

	We now provide the main definitions used in the rest of this section.  
	
	\begin{defi} 
		A conditional probability distribution  $p_{AB}( a,b|   x,y )$ is {\em no-signalling}  if it satisfies the constraints
		\begin{align}
			\label{nosignallingAB}
			\sum_{a\in\sf A}  p_{AB}( a,b|   x,y )   &=    \sum_{a\in\sf A}  p_{AB}( a,b|   x',y )  \qquad \forall  x,x'\in  \sf X\\ 
			\label{nosignallingBA}  \sum_{b\in\sf B}  p_{AB}( a,b|   x,y )   &=    \sum_{b\in\sf B}  p_{AB}( a,b|   x,y' )  \qquad \forall  y,y'\in  \sf Y \, .
		\end{align}
	\end{defi}
	
	For a no-signalling distribution $p_{AB}( a,b|   x,y ) $, we denote  its marginals   by 
	\begin{align}
		\label{pA}
		p_A  (a|x)   &:= \sum_{b\in\sf B}  p_{AB}( a,b|   x,y )     \qquad \forall y\in  \sf Y\\
		\label{pB}
		p_B  (b|y)   &:= \sum_{a\in\sf A}  p_{AB}( a,b|   x,y )     \qquad \forall x\in  \sf X \, .
	\end{align}

	\begin{defi}
		A {\em no-signalling ontic model} for a conditional probability distribution  $p_{AB}( a,b|   x,y )$  is a triple $(\Lambda,  q(\d\lambda),   q_{AB}(a,b|  x,y,\lambda) )$ consisting of a random variable  $\lambda$ with sample space $\Lambda$,  a probability distribution $q(\d \lambda)$,  and  a family of no-signalling  probability distributions $q_{AB}( a,b|   x,y,\lambda )$ indexed by $\lambda\in \Lambda$ such that   
		\begin{align}
			p_{AB}(a,b|x,y)= \int \, q(\d \lambda) ~ q_{AB}(  a,b|  x,y,\lambda) \, .   
		\end{align}    
	\end{defi}
	
	Note that the  probability distributions $q_{AB}  (a,b|  x,y,\lambda)$ in the above definition are required to satisfy  the no-signalling conditions  for every possible ontic state $\lambda $, namely  
	\begin{align}
		\label{nosignallingqAB}
		\sum_{a\in\sf A}  q_{AB}( a,b|   x,y,\lambda )   &=    \sum_{a\in\sf A}  q_{AB}( a,b|   x',y,\lambda )  \qquad \forall  x,x'\in  \sf X, \, \forall \lambda\in \Lambda\\ 
		\label{nosignallingqBA}  \sum_{b\in\sf B}  q_{AB}( a,b|   x,y ,\lambda)   &=    \sum_{b\in\sf B}  q_{AB}( a,b|   x,y' ,\lambda)  \qquad \forall  y,y'\in  \sf Y ,  \,   \forall \lambda \in \Lambda \, .
	\end{align}
	The marginals will be denoted by 
	For a no-signalling distribution $p_{AB}( a,b|   x,y ) $, we denote  its marginals   by 
	\begin{align}
		\label{qA}
		q_A  (a|x,\lambda)   &:= \sum_{b\in\sf B}  q_{AB}( a,b|   x,y,\lambda )     \qquad \forall y\in  \sf Y, \, \forall \lambda\in \Lambda\\
		\label{qB}
		q_B  (b|y,\lambda)   &:= \sum_{a\in\sf A}  q_{AB}( a,b|   x,y,\lambda )     \qquad \forall x\in  \sf X, \, \forall \lambda\in \Lambda \, .
	\end{align}

	\begin{defi}
		The probability distribution  $p_{AB}( a,b|   x,y )$ admits  {\em predetermined outcomes} for settings  $\sf X_{\rm pre} \subseteq \sf X$ if $p_{AB}( a,b|   x,y )$ has  a no-signalling ontic model  $(\Lambda,  q(\d\lambda),   q_{AB}  (a,b|  x,y,\lambda) )$    such that, for every $\lambda \in \Lambda$ and for every $x\in\sf X_{\rm pre}$, there exists an outcome  $\alpha(x,\lambda)\in\sf A$ satisfying the condition 
		\begin{align}
			q_A ( \alpha(x,\lambda)   |  x ,\lambda   )   = 1    \, .
		\end{align}  
	\end{defi}

	\begin{defi}
		A {\em local realistic  model} for a conditional probability distribution  $p_{AB}( a,b|   x,y )$  is a no-signalling ontic model  $(\Lambda',  q'(\d \lambda'),  q'_{AB}  (a,b|x,y,\lambda'))$ where, for every $\lambda' \in\Lambda'$,  the probability distributions $q'_{AB}  (a,b|x,y,\lambda')$ have the product form $q'_{AB}  (a,b|x,y,\lambda')  = q_{A}'  (a|x,\lambda')  \, q_B'  (b|y, \lambda') $  for some probability distributions $q_{A}'  (a|x,\lambda')$ and $q_B'  (b|y, \lambda')$. 
	\end{defi}
	
	We now show that if all except one of Alice's settings have predetermined outcomes, then the probability distribution $p(a,b|x,y)$ admits a local realistic model. 
	\begin{prop}\label{prop:allbutone}
		Let  $p_{AB}( a,b|   x,y )$ be a no-signalling probability distribution  and let $x_* \in  \sf X$ be one of Alice's settings.  If  $p_{AB}( a,b|   x,y )$ admits predetermined outcomes for all of Alice's settings  except $x_*$, then   $p_{AB}( a,b|   x,y )$ has a local realistic model and therefore does not violate any Bell inequality. 
	\end{prop}
	\Proof  
	The predetermination condition implies the existence of a no-signalling ontic model  $(\Lambda,  q(\d \lambda),  q_{AB}  (a,b|x,y,\lambda))$ such that $q_A (\alpha(x,\lambda)|x,\lambda) =1 \, \forall \lambda\in \Lambda , \, \forall x\in {\sf X}  \setminus \{x_*\}$. This condition  implies  $q_A (a|x, \lambda) =0$   for every $a\not =  \alpha(x,\lambda)$. Hence, the  definition of $q_A (a|x,\lambda)$  [Eq. (\ref{qA})] implies  
	\begin{align}\label{pAB11}
		q_{AB}  (a,b|  x,y,\lambda)   =0 \qquad \forall a \not =    \alpha (x,\lambda) ,   \, \forall b\in  \sf B ,  \,\forall x\in  \sf X\setminus \{x_*\}   ,\,  \forall y\in \sf Y, \, \forall \lambda\in \Lambda \,.
	\end{align} 
	Hence, one has
	\begin{align}\label{pAB12}
		q_{AB}   (\alpha(x,\lambda)  ,  b|  x,  y, \lambda)     = \sum_{a\in\sf A}  q_{AB}  (a,b|  x,y,\lambda)   =  q_B  (b|x,\lambda)   \qquad \forall b\in  \sf B,   \,  \forall x\in  \sf X\setminus \{x_*\}   , \,   \forall  y\in\sf Y, \, \forall \lambda\in \Lambda  \, .
	\end{align}  
	Combining Eqs.  (\ref{pAB11}) and (\ref{pAB12}) then yields 
	\begin{align}\label{productprob} 
		q_{AB}  (a,b| x,y,\lambda )  =  \delta_{a, \alpha(x,\lambda)}  \,  q_B(b|y,\lambda)   \qquad \forall a \in  \sf A,  \forall b  \in  \sf B, \, \forall x\in  \sf X\setminus \{x_*\} \, ,\forall y\in\sf Y , \, \forall \lambda \in \Lambda\, .
	\end{align} 
	Now, define a new  random variable $\lambda':=   (\lambda,  c)$ with sample space $\Lambda' :  =\Lambda \times  \sf A$  and  probability distribution 
	\begin{align}
		q'(\d \lambda'):=   q(\d \lambda)  ~  q_A  (  c  |x_*,  \lambda) 
	\end{align} 
	normalized as $\int q(\d \lambda)  \sum_{c\in\sf A}  q_A  (  c  |x_*,  \lambda)  =1$.  
	A local realistic  model for $p_{AB}  (a,b| x,y )$ is then constructed by setting 
	\begin{align}
		q_A'  (a|x, \lambda')  :  =  \left\{  
		\begin{array}{ll}
			\delta_{a,c}  \qquad&  {\rm for}~x=x_* \\
			&\\
			\delta_{a , \alpha(x,\lambda)}  &  {\rm for}~{\rm for}~x\not = x_*  \, .  
		\end{array} 
		\right.
	\end{align}
	and, for every $c$ such that $q_A(c|x_*,  \lambda) \not  = 0$,   
	\begin{align}
		q_B'  (b|y, \lambda')  :  =\frac{ q_{AB}   (c,  b  |  x_*  ,  y,\lambda )}{q_A(c|x_*, \lambda)} \, .
	\end{align}
	Notice that the probability distribution $q_B'  (b|y, \lambda') $ is normalized: $\sum_{b\in\sf B}  q_B'  (b|y, \lambda')   =  \sum_{b\in\sf B}   q_{AB}   (c,  b  |  x_*  ,  y,\lambda ) / q_A(c|x_*, \lambda)  = q_A(  c|  x_*,\lambda)/q_A(c|x_*,\lambda)   = 1$.   
	
	One can easily verify that the above probability distributions yield a local realistic model for  $p_{AB}   (\alpha (x,\lambda)  ,  b|  x,  y)$: indeed, for the setting $x_*$ one has 
	\begin{align}
		\nonumber  \int  q'(\d \lambda') ~   q_{A}'  (a|x_*,\lambda') ~q_B'  (b|y, \lambda')   &  = \int  q(\d \lambda)   \sum_{c\in\sf A}     \,   q_A(c|x_*,  \lambda)       \,    \delta_{a,c}  \,   \frac{ q_{AB}   (c,  b  |  x_*  ,  y,\lambda )}{q_A(c|x_*, \lambda)}  \\
		\nonumber   &   = 
		\int \, q(\d \lambda)  ~  q_{AB}   (a,  b  |  x_*  ,  y,\lambda ) \\
		&  = p_{AB}  (a,b|x_*,  y) \, ,
	\end{align}
	and for all the other settings $x\not  =  x_*$ one has 
	\begin{align}
		\nonumber  \int  q'(\d \lambda') ~   q_{A}'  (a|x,\lambda') ~q_B'  (b|y, \lambda')   &    = \int  q(\d \lambda)   \sum_{c\in\sf A}     ~  q_A(c|x_* ,  \lambda)       \,    \delta_{a,\alpha(x,\lambda)}  \,   \frac{ q_{AB}   (c,  b  |  x_*  ,  y,\lambda )}{q_A(c|x_*, \lambda)}  \\
		\nonumber &  =  \delta_{a,\alpha (x,\lambda)}~      \int  q(\d \lambda)    \, \sum_{c\in\sf A}   \, q_{AB}   (c,  b  |  x_*  ,  y,\lambda )  \\
		\nonumber &  =  \delta_{a,\alpha (x,\lambda)}~      \int  q(\d \lambda)    \,    \, q_{B}   ( b  |    y,\lambda )  \\
		\nonumber &  =        \int  q(\d \lambda)    \,       q_{AB}  (a,b|x,y,\lambda) \\
		&  =        p_{AB}  (a,b|x,y)  \, ,
	\end{align}
	the second to last equality following from Eq. (\ref{productprob}). \qed
	
	\medskip 
	
	When Alice has only two possible settings ($|\sf X|  =  2$), Proposition \ref{prop:allbutone} implies  the following corollary:  
	\begin{cor}
		Let  $p_{AB}( a,b|   x,y )$ be a probability distribution satisfying the no-signalling constraints (\ref{nosignallingAB}) and (\ref{nosignallingBA}). If Alice has two settings and one of them has a predetermined outcome, then the probability distribution   $p_{AB}( a,b|   x,y )$ admits a local realistic model and therefore does not violate any Bell inequality. 
	\end{cor}

	It is worth noting that
	\begin{enumerate}
		\item  Proposition \ref{prop:allbutone}  can be straightforwardly generalized to a Bell scenario with { more than two parties} where all settings except one have predetermined outcomes for all parties except one, 
		\item Proposition \ref{prop:allbutone} can be extended  to the case  where the condition of predetermination is only satisfied approximately. This extension is explicitly provided  in the following. 
	\end{enumerate}
	
	\begin{defi}
		The probability distribution  $p_{AB}( a,b|   x,y )$ admits  {\em predetermined outcomes  up to error $\epsilon$} for settings  $\sf X_{\rm pre} \subseteq \sf X$ if $p_{AB}( a,b|   x,y )$ has  a no-signalling ontic model  $(\Lambda,  q(\d\lambda),   q_{AB}  (a,b|  x,y,\lambda) )$    such that, for every $\lambda \in \Lambda$ and every $x\in\sf X_{\rm pre}$, there exists an outcome  $\alpha(x,\lambda)\in\sf A$ satisfying the condition 
		\begin{align}
			q_A ( \alpha(x,\lambda)   |  x ,\lambda   )   >   1-\epsilon \, .
		\end{align}  
	\end{defi}
	
	We now show that the violation of two-party Bell inequalities must be small whenever  all but one of Alice's settings   have approximately predetermined outcomes: 
	
	\begin{prop}\label{prop:approx}  
		Let    $\omega  (p_{AB}) : =  \sum_{a,b,x,y}  \,   \omega_{a,b,x,y}  \,   p_{AB}  (a,b|x,y) $ be an arbitrary correlation with $\omega_{a,b,x,y}  \in \R \, \forall a,b,x,y$,   let $\omega_*$ be the maximum  of $\omega  (p_{AB})$ over all probability distributions  $p_{AB}  (a,b|x,y)$ admitting a local realistic model, let  $x_* \in \sf X$ be one of Alice's settings, and let $p^*_{AB}  (a,b|x,y)$ be a no-signalling probability distribution admitting predetermined outcomes up to error $\epsilon$ for   all of Alice's settings except $x_*$.  Then, the correlation achieved by  $p_{AB}^* $ is upper bounded by  
		\begin{align}\label{boundapproxpredet}
			\omega(p_{AB}^*)  \le   \omega_*    +  \epsilon    \sum_{x\not =  x_*}  \sum_{y\in\sf Y}   \left(\max_b   |\omega_{\alpha (x,\lambda),b  ,x,y}  |  +  \max_{a\not = \alpha (x,\lambda)} \max_b    |\omega_{a ,b  ,x,y}| \right) 
		\end{align} 
	\end{prop}
	
	\Proof  The predetermination condition up to error $\epsilon$ implies the existence of a no-signalling ontic model $(\Lambda, q(\d\lambda),  q_{AB}  (a,b|x,y,\lambda))$ such that $q_A (\alpha(x,\lambda)|x,\lambda) >1-\epsilon,    ~ \forall \lambda\in \Lambda , \, \forall x\in {\sf X}  \setminus \{x_*\}$.  This condition implies  
	\begin{align}\label{x}
		\sum_{a\not=  \alpha (x,\lambda)}  \, \sum_{b \in \sf B} \,  q_{AB}(  a,b|  x,  y,\lambda)    =  \sum_{a \not  =  \alpha (x,\lambda)}  q_A (a|  x,\lambda)  <\epsilon  \qquad     \forall  x\in\sf X\setminus\{  x_*\} , \, \forall y\in\sf Y, \, \forall \lambda \in \Lambda \, .
	\end{align}
	and 
	\begin{align}
		\nonumber  \sum_{b\in\sf B}  \,  \Big|  q_B (b|y,\lambda)    -    q_{AB}   (\alpha (x,\lambda),  b|  x, y,\lambda) \Big|      &=  \sum_{b \in \sf B} \sum_{a\not =  \alpha (x,\lambda)}   q_{AB}  (a,b| x,y,\lambda)   \\
		\nonumber & =    \sum_{a\not =  \alpha  (x,\lambda)}   q_{A}  (a| x,\lambda) \\
		& <  \epsilon     \qquad   \forall x\in  \sf X \setminus \{x_*\} , \,  \forall y\in \sf Y , \, \forall \lambda\in \Lambda \, .  \label{xstar}
	\end{align}    
	
	Now, for every $\lambda \in  \Lambda$, define the  new probability distribution  $q'_{AB}   (a,  b|  x, y,\lambda)$ as 
	\begin{align}\label{qprimedef}
		q_{AB}'   (\alpha (x,\lambda),  b|  x, y,\lambda) :  =  
		\left\{ 
		\begin{array}{ll}
			q_{AB}  (a,b|  x_*,  y, \lambda)  \qquad &{\rm for}~x=x_*\\
			&  \\
			\delta_{a, \alpha  (x,\lambda)}  ~q_B (b|y,\lambda) &{\rm for}~x\not=  x_* \, .
		\end{array}
		\right.
	\end{align}
	Notice that the probability distribution $q'_{AB}   (\alpha  (x,\lambda),  b|  x, y,\lambda)$  satisfies the no-signalling conditions  (\ref{nosignallingqAB}) and (\ref{nosignallingqBA}). 
	
	Finally, define the probability distribution  
	\begin{align}
		p_{AB}'  (a,b|x,y) :  =  \int  q(\d \lambda) \, q_{AB}'  (a,b|x,y,\lambda)  \, .
	\end{align} 
	By construction,  $p_{AB}'  (a,b|x,y)$ admits predetermined outcomes for all of Alice's settings except $x_*$.  Hence, Proposition \ref{prop:allbutone} implies that $p_{AB}' (a,b|x,y)$ has a local realistic model, and therefore satisfies the bound $ \omega  (p'_{AB})   \le \omega_*$.  
	
	Overall, the correlation achieved by $p_{AB}^*  (a,b|x,y)$ can be bounded as 
	\begin{align} 
		\nonumber \omega(p_{AB}^*)    &  =    \omega(p_{AB}')    +   \omega  (  p_{AB}^*  -  p_{AB}')  \\
		\nonumber &  \le  \omega_*  +  \sum_{a,b,x,y}  \,   \int q(\d \lambda)  ~  \omega_{a,b,x,y}      \,     \Big(  q_{AB}   (a,b|x,y,\lambda)  -  q_{AB}' (a,b|x,y,\lambda)   \Big)  \\
		\nonumber &  \le  \omega_*  +  \sum_{a,b,x,y}  \,  \Big |\omega_{a,b,x,y}\Big|  \,  \int q(\d\lambda)\, \Big|  q_{AB}   (a,b|x,y,\lambda)  -  q_{AB}' (a,b|x,y,\lambda)  \Big| \\
		\nonumber &\le   \omega_*   \\
		\nonumber &  \quad +  \sum_{a,b,y}  \,  \Big|\omega_{a,b,x_*,y}\Big|  \, \int q(\d \lambda)  \, \Big|  q_{AB}   (a,b|x_*,y,\lambda)  -  q_{AB}' (a,b|x_*,y,\lambda)  \Big|    \\  
		\nonumber  & \quad     +  \sum_{x\not  =  x_*}  \sum_{y}  \sum_b  \,  \Big|\omega_{\alpha (x,\lambda),b,x,y}\Big|  \,  \int q(\d \lambda) \,  \Big|q_{AB}   (\alpha (x,\lambda),b|x,y,\lambda)  -  q_{AB}' (\alpha (x,\lambda),b|x,y,\lambda) \Big|  \\
		\nonumber  &  \quad  +  \sum_{x\not  =  x_*}  \sum_{y}  \sum_b  \,   \sum_{a\not  =  \alpha (x,\lambda)}\, \Big|\omega_{a,b,x,y}\Big|  \,  \int q(\d \lambda) \,  \Big|  q_{AB}   (a,b|x,y,\lambda)  -  q_{AB}' (a,b|x,y,\lambda) 
		\Big|  \\
		\nonumber &\le  \omega_*    +     \sum_{x\not  =  x_*}  \sum_{y}    \epsilon \,  \max_{b} \Big|  \omega_{\alpha (x,\lambda),b,x,y} \Big|      +    \sum_{x\not  =  x_*}  \sum_{y}   \epsilon  \,  \max_{a\not  = \alpha (x,\lambda),  b}    \,  \Big|\omega_{a,b,x,y} \Big| \, ,  
	\end{align}
	where the last inequality follows from Eqs. (\ref{qprimedef}),  (\ref{x}), and (\ref{xstar}). 
	\qed  
	
	\medskip 
	In the special case of the CHSH inequality, all the outcomes and settings are binary ($|A|= |B|=  |X|=  |Y|  = 2$) and one has 
	\begin{align}
		\omega_{\rm CHSH}    (p_{AB})   :  =  \sum_{a,b,x,y}   \,  (-1)^{a+b+xy}  \,  p_{AB}  (a,b|x,y) \, . 
	\end{align}
	Hence, the bound (\ref{boundapproxpredet}) becomes 
	\begin{align}
		\omega_{\rm CHSH}    (p_{AB})  \le  2   (1+   2 \epsilon ) \, ,  
	\end{align}
	whenever $p_{AB }  (a,b|x,y)$ is a no-signalling probability distribution such that one of Alice's settings is predetermined up to error $\epsilon$.

\section*{Activation of Bell nonlocality for arbitrary pure states}

Here we show that all pure entangled states of all dit/anti-dit composites give rise to activation of Bell non-locality.  

Let $d\ge 2$ be an  integer, and let $DA$ be the composite system consisting of a dit  $D$ and an anti-dit $A$.  
An arbitrary pure state of system $DA$ is of the form
\begin{align}\label{pureentangled}
	\ket{\psi}_{DA} = \sum_{i=0}^{d-1} \alpha_i \ket{i}_{D}\ket{i \oplus q}_A, 
\end{align}
where $(\alpha_i)_i$ is a normalized set of   coefficients, which we take to be positive without loss of generality,  $q \in \Z_d$ is the type  of the state, and $\oplus$ denotes addition modulo $d$.  The state $|\psi\>_{DA}$ is entangled if and only if at least two of the coefficients $(\alpha_i)_i$ are nonzero. From now on, we will assume that the state is entangled and we will denote the nonzero coefficients by $\alpha_{i_1}$ and $\alpha_{i_2}$, respectively.   

The state (\ref{pureentangled}) alone does not give rise to any Bell inequality violation, because the local measurements on the dit $D$ and on the anti-dit $A$ are purely classical.  We now show that two identical copies of the state  (\ref{pureentangled}) give rise to Bell nonlocality whenever the state is entangled.  

To achieve activation, we consider the composite system $D_1D_2A_1A_2$, consisting of two dits and two anti-dits. The system is initially in the state  $\ket{\psi}_{D_1  A_1} \otimes \ket{\psi}_{D_2  A_2}$, corresponding to two identical copies of the state (\ref{pureentangled}).  Two parties, Alice and Bob, have access to systems $D_1A_2$ and $D_2A_1$, respectively.  The initial state $\ket{\psi}_{D_1  A_1} \otimes \ket{\psi}_{D_2  A_2}$ can be conveniently rewritten as 
\begin{align}
	\nonumber     \ket{\psi}_{D_1  A_1} \otimes \ket{\psi}_{D_2  A_2}   &= \sum_{i\in \Z_d}  \sum_{j\in  \Z_d}  \alpha_i \, \alpha_{ j} \, \ket{i}_{D_1} \ket{j}_{D_2}\ket{i \oplus q}_{A_1}\ket{ j \oplus q}_{A_2}\\
	\nonumber   &  =  \sum_{i\in\Z_d }  \sum_{j\in  \Z_d}  \alpha_i \, \alpha_{ i\oplus l  \ominus k} \, \ket{i}_{D_1} \ket{i\oplus l \ominus q }_{D_2}\ket{i \oplus k}_{A_1}\ket{ i \oplus l}_{A_2} \\
	&  =  \sum_{i\in \Z_d} \sum_{j \in \Z_d} \alpha_i \, \alpha_{ i\oplus l\ominus q} \, \left|\phi^{(l)}_i\right\>_{D_1A_2}  \otimes   \left|\phi^{(2q\ominus l)}_{i  \oplus l  \ominus q}\right\>_{D_2A_1}  \, ,  \label{eq:2-copies}
\end{align}
having used the notation 
\begin{align}
	|\phi^{(l)}_i\>_{DA} :  =  |i\>_D |i \oplus l\>_A \, ,  \qquad \forall i\in \Z_d \, , \forall  l\in  \Z_d \, .
\end{align}

Since the states $\left\{|\phi^{(l)}_i\>\right\}_{i,l}$ are mutually orthogonal for different values of $i$ and $l$, Eq. (\ref{eq:2-copies}) provides a Schmidt decomposition with respect to the bipartition $(D_1A_2)(D_2A_1)$. Note that the state $\ket{\psi}_{D_1  A_1} \otimes \ket{\psi}_{D_2  A_2}$ has Schmidt rank at least 2, since at least the terms with $(i=i_1,l=k)$ and $(i=i_2, l=k)$  in the r.h.s. of Eq. (\ref{eq:2-copies}) are non-zero. 

Now, we  rewrite the two-copy state as
\begin{align}
	\ket{\psi}_{D_1  A_1} \otimes \ket{\psi}_{D_2  A_2}  =  \sqrt{p_{\rm qubit}}  \,  |\Psi_{\rm qubit}\>_{(D_1A_2)(D_2A_1)}   +  \sqrt{  1-  p_{\rm qubit}}  \,  |\Psi_{\rm rest}\>_{(D_1A_2)(D_2A_1)}  \, , 
\end{align}
with  
\begin{align}
	\nonumber & p_{\rm qubit}    :  =  \alpha_{i_1}^4  + \alpha_{i_2}^4 \\
	\label{2qubit} & |\Psi_{\rm qubit}\>_{(D_1A_2)(D_2A_1)}   :=  \frac{\alpha_{i_1}^2  |\phi^{(q)}_{i_1}\>_{D_1A_2} \otimes |\phi^{(q)}_{i_1}\>_{D_2A_1} + \alpha_{i_2}^2  |\phi^{(q)}_{i_2}\>_{D_1A_2} \otimes |\phi^{(q)}_{i_2}\>_{D_2A_1}}{\sqrt{p_{\rm qubit}}}  \\
	\nonumber &|\Psi_{\rm rest}\>_{(D_1A_2)(D_2A_1)} :=   \frac{\sum_{i \in \Z_d \setminus \{i_1,i_2\}}   \alpha_i^2 \, |\phi^{(q)}_i\>_{D_1A_2}  \otimes   |\phi^{(q)}_{i }\>_{D_2A_1}      +   \sum_i \sum_{l\not =  q  } \alpha_i \, \alpha_{ i\oplus l\ominus q} \, |\phi^{(l)}_i\>_{D_1A_2}  \otimes   |\phi^{(2q\ominus l)}_{i  \oplus l  \ominus q}\>_{D_2A_1} }{\sqrt{1-p_{\rm qubit}}}\, .  \label{qubitdefinitions}
\end{align}

Now, the state $|\Psi_{\rm qubit}\>_{(D_1A_2)(D_2A_1)}$ is equivalent to a two-qubit entangled state.   Indeed, this state belongs to the tensor product space $\spc H_{\rm Alice's~qubit}   \otimes \spc H_{\rm Bob's~qubit} $, where    \begin{align}
	\spc H_{\rm Alice's~qubit}  :  = \Span\left\{  |\phi^{(q)}_{i_1}\>_{D_1A_2} \, , |\phi^{(q)}_{i_2}\>_{D_1A_2}\right\}  \qquad {\rm and} \qquad \spc H_{\rm Bob's~qubit's~qubit}  :  = \Span\left\{  |\phi^{(q)}_{i_1}\>_{D_2A_1} \, , |\phi^{(q)}_{i_2}\>_{D_2A_1}\right\}  
\end{align}
are two-dimensional subspaces of  the Hilbert spaces associated to Alice's and Bob's systems, respectively.  Moreover, every unit vector in these two subspaces is a valid pure state in our toy theory.  By self-duality of our toy theory, all orthonormal bases in these subspaces correspond to allowed measurements.  Hence, all quantum measurements in these two-dimensional subspaces can be simulated within our toy theory.

Following Refs.~\cite{gisin1991bell,POPESCU1992}, we  use the  two-qubit entanglement in the state (\ref{2qubit}) to violate the CHSH inequality. In the two-qubit subspace we use the same settings as in Refs.~\cite{POPESCU1992}: for a two-qubit state of the form
\begin{align}\label{twoqubitstate}
	|\Psi\>_{\rm Alice's~qubit, ~Bob's~qubit}  = \alpha \, \ket{ \uparrow}_{\rm Alice's~qubit}   \ket{\uparrow }_{\rm Bob's~qubit} + \beta \, \ket{ \downarrow}_{\rm Alice's~qubit} \ket{\downarrow}_{\rm Bob's~qubit's}  \,,
\end{align}
with 
\begin{align}
	\nonumber |\uparrow\>_{\rm Alice's~qubit}    &:= |\phi_{i_1}^{q}\>_{D_1A_2}   \qquad   &|\downarrow\>_{\rm Alice's~qubit}    &:=  |\phi_{i_2}^{q}\>_{D_1A_2}    \\ 
	|\uparrow\>_{\rm Bob's~qubit}    &:= |\phi^{q}_{i_1}\>_{D_2A_1}   \qquad  &|\downarrow\>_{\rm Bob's~qubit}    &:=   |\phi_{i_2}^{q}\>_{D_2A_1}  \, ,
\end{align}
Alice's  measurement  for setting $x \in \{0,1\}$ is given by  the projectors on the orthonormal basis $\{|v^{(x)}_0\>_{\rm Alice's~qubit},  |v^{(x)}_1\>_{\rm Alice's~qubit}\}$ defined as 
\begin{align}
	\nonumber 
	&|v^{(0)}_0\>_{\rm Alice's~qubit}  =  |\uparrow\>_{\rm Alice's~qubit}  \\  
	\nonumber &|v^{(0)}_1\>_{\rm Alice's~qubit}  =  |\downarrow\>_{\rm Alice's~qubit}  \\
	\nonumber  &|v^{(1)}_0\>_{\rm Alice's~qubit}  =   \frac{\ket{\uparrow}_{\rm Alice's~qubit} + \ket{\downarrow}_{\rm Alice's~qubit}}{\sqrt 2}   \\
	&|v^{(1)}_0\>_{\rm Alice's~qubit}  =   \frac{\ket{\uparrow}_{\rm Alice's~qubit} - \ket{\downarrow}_{\rm Alice's~qubit}}{\sqrt 2} \, ,
\end{align}
while Bob's  measurement for setting $y\in  \{0,1\}$ is given by the projectors on the orthonormal basis $\{|w^{(x)}_0\>_{\rm Bob's~qubit},  |w^{(x)}_1\>_{\rm Bob's~qubit}\}$ defined as 
\begin{align}
	\nonumber  &  |w_0^{(0)}\>_{\rm Bob's~qubit} := \cos\theta\ket{\uparrow}_{\rm Bob's~qubit}+\sin\theta\ket{\downarrow}_{\rm Bob's~qubit} \\ &  
	\nonumber  |w_1^{(0)}\>_{\rm Bob's~qubit} := -  \sin\theta  \ket{\uparrow}_{\rm Bob's~qubit}+  \cos\theta \ket{\downarrow}_{\rm Bob's~qubit}  \\
	\nonumber &|w_0^{(1)}\> := -\cos\theta\ket{\uparrow}_{\rm Bob's~qubit}+\sin\theta\ket{\downarrow}_{\rm Bob's~qubit} \\ 
	&|w_0^{(1)}\> :=  \sin\theta   \ket{\uparrow}_{\rm Bob's~qubit}+ \cos\theta  \ket{\downarrow}_{\rm Bob's~qubit} \, ,
\end{align}
with $2\theta:=\arctan (2\alpha\beta)$. 

With these measurement settings, the CHSH correlation  assumes the value 
\begin{align}
	\nonumber \omega_{{\rm CHSH}}^{\rm qubit}   &=\sum_{x,y}  \sum_{a,b}  (-1)^{ a+  b+ xy}  \, \left| \, \left(\< v^{(x)}_a|_{\rm Alice's~qubit} \otimes  \< w^{(y)}_b|_{\rm Bob's~qubit} \right)\,  |\Psi\>_{\rm Alice's~qubit, ~Bob's~qubit} \right|^2  \\  
	& =  2 \sqrt{1+4\alpha^2\beta^2}
\end{align}
In our case, the two-qubit  state (\ref{twoqubitstate}) is the state $|\Psi_{\rm qubit}\>_{(D_1A_2)(D_2A_1)}$ defined in Eq. (\ref{qubitdefinitions}), and therefore, we have 
\begin{align}\label{alphabeta}
	\alpha  =  \frac{\alpha_{i_1}^2}{\alpha_{i_1}^4  +  \alpha_{i_2}^4}  \qquad{\rm and}  \qquad \beta  =  \frac{\alpha_{i_2}^2}{\alpha_{i_1}^4  +  \alpha_{i_2}^4} 
\end{align}

We now extend the above measurements outside the two-qubit subspace, choosing an extension that guarantees that one of Alice's measurement settings is just a local measurement on  dit $D_1$. In the extension, Alice's  measurement for setting $x$ is given by the  two-outcome POVM  $\{P^{(x)}_0, P^{(x)}_1 \}$ defined as 
\begin{align} 
	\nonumber P_0^{(0)}  &:= |v_0^{(0)}\>\<v_0^{(0)}|_{\rm Alice's~qubit}   +  \sum_{l\not  =  q} |\phi_{i_1}^{(l)}  \>\<  \phi_{i_1}^{(l)}|_{D_1A_2}\\
	\nonumber P_1^{(0)}  & :=  |v_1^{(0)}\>\<v_1^{(0)}|_{\rm Alice's~qubit}   +  \sum_{l\not =  q}   |\phi_{i_2}^{(l)}  \>\<  \phi_{i_2}^{(l)}|_{D_1A_2}   +  \sum_{i\not \in \{i_1,i_2\}}  \sum_l   |\phi_{i}^{(l)}  \>\<  \phi_{i}^{(l)}|_{D_1A_2}  \\
	\nonumber  P_0^{(1)}  &:=  |v_0^{(1)}\>\<v_0^{(1)}|_{\rm Alice's~qubit}   +  \sum_{l\not  =  q} |\phi_{i_1}^{(l)}  \>\<  \phi_{i_1}^{(l)}|_{D_1A_2}    \\ 
	P_1^{(1)}  &:=  |v_1^{(1)}\>\<v_1^{(1)}|_{\rm Alice's~qubit}  +    \sum_{l\not =  q}   |\phi_{i_2}^{(l)}  \>\<  \phi_{i_2}^{(l)}|_{D_1A_2}   +  \sum_{i\not \in \{i_1,i_2\}}  \sum_l   |\phi_{i}^{(l)}  \>\<  \phi_{i}^{(l)}|_{D_1A_2}   \, , 
\end{align}
and Bob's measurement for setting $y$ is given by the two-outcome POVM $\{ Q_0^{(y)},  Q_1^{(y)}\}$ defined  as 
\begin{align} 
	\nonumber Q_0^{(0)}  &:= |w_0^{(0)}\>\<w_0^{(0)}|_{\rm Bob's~qubit}   +  \sum_{l\not  =  q} |\phi_{i_1 \oplus l  \ominus q}^{(2q\ominus l)}  \>\<  \phi_{i_1\oplus l  \ominus q}^{(2q\ominus l)}|_{D_2A_1}\\
	\nonumber Q_1^{(0)}  & :=  |w_1^{(0)}\>\<w_1^{(0)}|_{\rm Bob's~qubit}   +  \sum_{l\not =  q}   |\phi_{i_2\oplus l \ominus q }^{(2q\ominus l)}  \>\<  \phi_{i_2\oplus l \ominus q}^{(2q\ominus l)}|_{D_2A_1}   +  \sum_{i\not \in \{i_1,i_2\}}  \sum_l   |\phi_{i\oplus l \ominus q}^{(2q \ominus l)}  \>\<  \phi_{i\oplus l 
		\ominus q}^{(2q\ominus l)}|_{D_2A_1}  \\
	\nonumber Q_0^{(1)}  &:=  |v_0^{(1)}\>\<v_0^{(1)}|_{\rm Bob's~qubit}    +  \sum_{l\not  =  q} |\phi_{i_1 \oplus l  \ominus q}^{(2q\ominus l)}  \>\<  \phi_{i_1\oplus l  \ominus q}^{(2q\ominus l)}|_{D_2A_1}    \\ 
	Q_1^{(1)}  &:=  |v_1^{(1)}\>\<v_1^{(1)}|_{\rm Bob's~qubit}  +  \sum_{l\not =  0}   |\phi_{i_2\oplus l \ominus q }^{(2q\ominus l)}  \>\<  \phi_{i_2\oplus l \ominus q}^{(2q\ominus l)}|_{D_2A_1}   +  \sum_{i\not \in \{i_1,i_2\}}  \sum_l   |\phi_{i\oplus l \ominus q}^{(2q \ominus l)}  \>\<  \phi_{i\oplus l 
		\ominus q}^{(2q\ominus l)}|_{D_2A_1}   \, .
\end{align}
This choice of local measurements guarantees that, outside the two-qubit subspace $\spc H_{\rm Alice's~qubit} \otimes \spc H_{\rm Bob's~qubit}$, Alice's ad Bob's outcomes are perfectly correlated, and therefore achieve the optimal classical value of the  CHSH correlation.  In this way,  the total value of the CHSH correlation is  
\begin{align}
	\nonumber    \omega_{{\rm CHSH}}^{\rm total}   &=  p_{\rm qubit} \,  \omega_{{\rm CHSH}}^{\rm qubit}      +   (1-p_{\rm qubit})\,  \omega_{{\rm CHSH}}^{\rm classical}  \\
	&  =    p_{\rm qubit}  2 \sqrt{1+4\alpha^2\beta^2} + (1-p_{\rm qubit}) 2 \, ,
\end{align}
with $\alpha$ and $\beta$ as in Eq. (\ref{alphabeta}). 

To conclude, note that Alice's measurement $\{P_0^{(0)},  P_1^{(0)}\}$ represents a local measurement performed on dit $D_1$ alone.  Indeed, one has 
\begin{align}
	\nonumber P_0^{(0)}  &:= |v_0^{(0)}\>\<v_0^{(0)}|_{\rm Alice's~qubit}   +  \sum_{l\not  =  q} |\phi_{i_1}^{(l)}  \>\<  \phi_{i_1}^{(l)}|_{D_1A_2}\\
	\nonumber & =  |i_1\>\<i_1|_{D_1} \otimes |i_1\oplus q\>\<i_1\oplus q|_{A_2}  +  \sum_{l\not  =  q} |i_1 \>\<i_1|_{D_1} \otimes |i_1 \oplus l\>\<i_1\oplus l|_{A_2}\\
	&=  |i_1\>\<i_1|_{D_1}\otimes I_{A_2} \, ,
\end{align}
and $P_1^{(0)}  =  I_{D_1A_2}  -  P_{0}^{(0)}  =  \left(\sum_{i\not =  i_1}|i\>\<i|\right)_{D_1} \otimes I_{A_2}$.

\section{No-go theorems on theories that reproduce the predictions of our toy theory}

Suppose that the predictions of our toy theory are reproduced by a deeper theory that describes reality at the fundamental level.  Under mild assumptions, we now show that the states in the deeper theory cannot, in general, be decomposed into a list of individual states associated to  the classical and  anti-classical systems. In general, the state of composite systems involving classical and anti-classical subsystems must contain holistic degrees of freedom that cannot be broken down into local parts.   

We present two versions of our argument, based on slightly different frameworks and assumptions.   

\subsection{Argument 1: deeper OPT with the same compositional structure as the toy theory}  

This version of the argument adopts the OPT framework to describe the deeper theory.  The argument is based on one assumption, namely  that the deeper theory  is an OPT that respects the compositional structure of the original theory. This type of assumption has been recently discussed and formalized in  the study of  ontological models and Bell inequalities, see e.g. \cite{gheorghiu2020ontological,schmid2020structure}.  

The compositional assumptions used in our argument are the following: 
\begin{itemize}
	\item every system $S$ in the original theory is associated to a system $\map D (S)$ in the deeper theory, which is meant to describe the underlying degree of freedom giving rise to system $S$.   For example, for a classical bit $B$, the underlying degree of freedom $\map  D(B)$ could be the electric charge  inside a capacitor, with the bit value being $1$ if the amount of charge is above a threshold value. As a special case, one could  have $\map D(S)  =  S$, for every system $S$, meaning that the systems in the deeper theory are the same systems as in the original theory. In general, the original system $S$ and the underlying system $\map D(S)$ may have a  state spaces of different dimensions, as in the example of the classical bit and the electric charge in a capacitor.  
	
	\item    composite systems of the form $S_1 S_2$ are associated to composite systems of the form $\map D(S_1)  \map D(S_2)$.  
	\item every process  $\map M$ in the toy theory corresponds to  an underlying process  $\map D(\map M)$ in the deeper theory.
	\item  local processes of the form $\map M \otimes \map N$ are mapped into local processes of the form $\map D(\map M)\otimes \map D(\map N)$. 
\end{itemize}

We now show that, under the above assumptions,  the deeper theory cannot always assign individual states to the degrees of freedom associated to classical systems. Consider a composite system $B_1B_2A_1A_2$, consisting of two bits and two anti-bits. In the deeper theory, this system will be described by a composite system consisting of subsystems $\map D(B_1)$, $\map D(B_2)$,  $\map D(A_1)$, and $\map D(A_2)$. Let us denote by $\lambda_{\map D(B_1)  \map D(B_2) \map D(A_1)  \map D(A_2)}$ a possible pure state of this composite system. 

Consider now the Bell scenario used in the previous section of this Supplemental Material: systems $B_1A_1$ and $B_2 A_2$ are in the entangled state $|\Psi\>$,  Alice measures systems $B_1A_2$, and Bob measures systems $B_2A_1$.  Here, the product state $\Psi_{B_1A_1} \otimes \Psi_{B_2A_2}$  in the toy theory  corresponds to a product state $ \map D(\Psi_{B_1A_1}) \otimes \map D(\Psi_{B_2A_2})$ in the deeper theory, where  $\map D(\Psi_{BA})$ is a (generally mixed) state of the system $\map D(A) \map D(B)$, with $A  \in  \{A_1,A_2\}$ and $B\in  \{B_1,B_2\}$. According to the deeper theory, the probability that Alice and Bob get outcomes $a$ and $b$, respectively, when their settings are $x$ and $y$, respectively, is   
\begin{align}
	p_{\rm deeper}  (a,b|x,y)  := \Big( \map D  (P_a^x)\Big|_{\map D(B_1)  \map D (A_2) } \otimes  \Big( \map D  (Q_b^y) \Big|_{\map D(B_2)  \map D(A_1)} \,      \Big|\map D(\Psi)\Big)_{\map D(B_1)  \map D(A_1)} \otimes \Big|\map D(\Psi)\Big)_{\map D(B_1) \map  D(A_1)}  \, ,        \end{align}
where  $(  P^x_a)_a$ and $(  Q^y_b)_b$ are the POVMs representing Alice's and Bob's measurements in the toy theory, respectively.   Here we used the  notation $|\rho)$ and $(E|$ for an arbitrary state $\rho$ and an arbitrary effect $E$ in the deeper theory, and the notation $(E|\rho)$ to represent the pairing between states and effects in the deeper OPT (for more background about the OPT framework and notation, we refer the reader to \cite{purificationT-CDP,inf-derivation-CDP, capitolo-opt, qt-from-principles-CDP}.

The requirement that the deeper theory reproduces the predictions of the toy theory amounts to the equality  
\begin{align}
	p_{\rm deeper}  (a,b|x,y)  =  \Tr  \Big(    [P_a^x]_{B_1A_2 } \otimes  [Q_b^y]_{B_2 A_1} \,      [\Psi]_{B_1A_1} \otimes [\Psi]_{B_2A_2} \Big) \, ,        
\end{align}
for every $a,b,x,y$.  

Now, suppose that, at the pure state level, the deeper theory can  assign  individual ontic states to the degrees of freedom associated to classical systems. For mixed states, this condition implies that the state $\map D( \Psi )_{BA}$ of a generic bit $B$ and a generic anti-bit $A$ is separable, namely
\begin{align}\label{fundamentallysep}
	\map D( \Psi )_{BA}    =  \int \, \d \lambda_B \, \d \lambda_A \,  p(\lambda_B,\lambda_A)   \,  \lambda_{\map D(B)} \otimes \lambda_{\map D(A)} \, .
\end{align}
where  $\lambda_B$ and $\lambda_A$ are pure states of systems $\map D(B)$ and system $\map D(A)$, respectively, and $p(\lambda_B, \lambda_A)$ is a joint probability distribution. This decomposition implies that the overall state $ \map D(\Psi_{B_1A_1}) \otimes \map D(\Psi_{B_2A_2})$ is separable with respect to all possible bipartitions.  Hence, it cannot lead to any Bell inequality violation.    This  conclusion  is in contradiction with the fact that the state  $ \Psi_{B_1A_1} \otimes \Psi_{B_2A_2}$ violates a Bell inequality in the toy theory.  To avoid the contradiction, the decomposition (\ref{fundamentallysep}) should not be possible.  

The above argument proves that it is not  possible to  assign individual pure states to the  physical systems underlying classical systems  in any deeper OPT that reproduces  the predictions of our toy theory while maintaining the same compositional structure.  A way out, of course, is to give up the assumption that the deeper theory respects the  compositional structure of the original toy theory. To some extent, however, this conclusion would be even more radical than the impossibility to assign definite individual pure states to the classical systems: even the classical systems themselves and the notion of local operations performed on them would not correspond in any direct way to  systems and local processes taking place in  the underlying reality.   Even in this case, a realistic interpretation of classical systems  would be difficult to maintain in a world described by our toy theory.  

\subsection{Ontological model}

The second version of our argument is based the framework of ontological models \cite{hardy2004quantum, spekkens2005contextuality, leifer2014quantum}.   For the composite system $B_1B_2A_1A_2$, the underlying reality is described by an ontic state $\lambda_{B_1B_2A_1A_2}$, and the predictions of the toy theory are reproduced by averaging the ontic state over a suitable  probability distribution $p (\lambda_{B_1B_2A_1A_2})$.   In the ontological model, the local measurements performed by Alice and Bob correspond to an underlying physical process  described by a response function, {\em i.e.}  a conditional probability distribution  $q(a,b| \lambda_{B_1B_2A_1A_2}  ,x,y)$ that  specifies the probability of the outcomes $(a,b)$ for every given settings $(x,y)$ and for every given ontic state $\lambda_{B_1B_2A_1A_2}$.  

We say that a conditional probability distribution is a {\em valid} response  function, if it corresponds to a physical process allowed in by ontological model. The difference between the ontological model considered here and the OPT considered in the previous subsection is that here we do not assume any particular form for the valid response functions: for example, we do not require   $q(a,b| \lambda_{B_1B_2A_1A_2}  ,x,y)$  to be the product of  two local response functions associated to Alice's and Bob's measurements.  For experiments involving both bits and anti-bits, we do not assume that the valid response functions satisfy the constraints of local realism. 

As usual in the study of Bell nonolocality, the correlations between Alice's and Bob's outcomes are quantified by expressions of the form  
\begin{align}\label{omega}
	\omega  =  \sum_{x,y,a,b}   r(x,y)   \,  \int  \,d  \lambda_{B_1B_2A_1A_2}  \, p (\lambda_{B_1B_2A_1A_2})   \,  q(a,b| \lambda_{B_1B_2A_1A_2}  ,x,y)  \,  \omega  (a,b,x,y) \, ,  
\end{align}
where $r(x,y)$ is the probability distribution for the settings, and $\omega  (a,b,x,y)$ is a random variable.

We now show that, for an ontological model reproducing the correlations of our toy theory,  the ontic state cannot be  decomposed as  
\begin{align}\label{decomposed}
	\lambda_{B_1B_2A_1A_2}  =  (\lambda_{B_1 B_2} ,  \lambda_{A_1A_2}) \, ,
\end{align}
where $\lambda_{B_1 B_2}$ is an ontic state associated to the  composite   $B_1B_2$ and $\lambda_{A_1A_2}$ is an ontic state associated to the  composite $A_1A_2$. In other words, it is not possible to provide a description of reality that separates the bits from the anti-bits.  

The argument follows from two mild assumptions: 

\begin{enumerate}
	\item  If $q(a,b| \lambda_{B_1B_2A_1A_2}  ,x,y)$  is a  valid response function  for system $B_1B_2A_1A_2$ and $\lambda_{B_1B_2A_1A_2}  =  (\lambda_{B_1B_2} ,  \lambda_{A_1A_2})$, then  $q_{\lambda_{B_1B_2}} (a,b| \lambda_{A_1A_2}, x,y)  :   =  q (a,b| \lambda_{B_1B_2}, \lambda_{A_1A_2}  ,x,y)$ is a valid response function for system $A_1A_2$.      Moreover, if  $q(a,b| \lambda_{B_1B_2A_1A_2}  ,x,y)$ and $q_{\lambda_{B_1B_2}} (a,b| \lambda_{A_1A_2}, x,y)$  have the same interpretation in terms of local experiments:  if $(a,b| \lambda_{B_1B_2A_1A_2}  ,x,y)$  is the probability  outcomes $(a,b)$ of two local experiments with settings $(x,y)$, then so is    $q_{\lambda_{B_1B_2}} (a,b| \lambda_{A_1A_2}, x,y)$.  
	\item composite systems consisting only of classical systems, or only of anti-classical systems, satisfy the constraints of local realism.  
\end{enumerate}

The first assumption is natural,  because one can think of the state  $\lambda_{B_1B_2}$  as part of the environment of the degrees of freedom associated to system $A_1A_2$: if $q(a,b| \lambda_{B_1B_2A_1A_2}  ,x,y)$ describes a valid process acting on all the degrees of freedom associated to system $B_1B_2A_1A_2$, then  $q_{\lambda_{B_1B_2}} (a,b| \lambda_{A_1A_2}, x,y) $ should describe an effective process acting only on the degrees of freedom associated to  $A_1A_2$.  Moreover, if the initial process described local measurements performed by Alice and Bob at spacelike separated locations, so should do the effective process $q_{\lambda_{B_1B_2}} (a,b| \lambda_{A_1A_2}, x,y)$: if the inputs $(x,y)$ (the outputs $(a,b)$)  of the original process $q(a,b| \lambda_{B_1B_2A_1A_2}  ,x,y)$   are received (produced)  at two spacelike separated locations, then the inputs  (outputs) of the effective process $q_{\lambda_{B_1B_2}} (a,b| \lambda_{A_1A_2}, x,y)$  are (received) produced at the same spacelike separated locations.

The second assumption is also natural,  since classical and anti-classical composites, individually considered, are described by classical  theory, which satisfies the constraint of local realism. Our assumption amounts to the fact that, {\em even if classical theory is not fundamental}, one would expect the  reality underlying classical theory to still satisfy the constraints of local realism.   If this were not the case, the  local realism of classical theory would be an artifact of the average over the ontic states, and would not reflect the structure of the underlying reality.

Let us now proceed to the argument. We know that our toy theory violates Bell inequalities for a system of two bits and two anti-bits. Then,  there must exist at least one ontic state such that the correlation  
\begin{align}\label{omegalambda}
	\omega_{\lambda_{B_1B_2A_1A_2}}  :=  \sum_{x,y,a,b}   r(x,y)   \,  q(a,b| \lambda_{B_1B_2A_1A_2}  ,x,y)  \,  \omega  (a,b,x,y) \, ,  
\end{align}
exceeds the maximum value compatible with local realism. 

Suppose that the ontic state can be broken down as $\lambda_{B_1B_2A_1A_2}  =  (\lambda_{B_1B_2},   \lambda_{A_1A_2} )$.  Then,  the correlation  (\ref{omegalambda}) can be rewritten as 
\begin{align}\label{antibitsonly}
	\omega_{\lambda_{B_1B_2A_1A_2}}  =  \sum_{x,y}   r(x,y)   \,      q_{\lambda_{B_1B_2}}(a,b| \lambda_{A_1A_2}  ,x,y)  \,  \omega  (a,b,x,y) \, ,  
\end{align}
where $q_{\lambda_{B_1B_2}}(a,b| \lambda_{A_1A_2}  ,x,y)$ is, by our first assumption, a valid response function describing a physical process  acting on the degrees of freedom associated to system  $A_1 A_2$.

Our first assumption also guarantees that the response function $q_{\lambda_{B_1B_2}}(a,b| \lambda_{A_1A_2}  ,x,y)$ describes two local experiments performed by Alice and Bob at two spatially separated locations. Since the process $q_{\lambda_{B_1B_2}}(a,b| \lambda_{A_1A_2}  ,x,y)$ refers to the degrees of freedom associated to two anti-bits, it must satisfy the constraints of local realism:  hence, it should factorize as 
\begin{align}
	q_{\lambda_{B_1B_2}}(a,b| \lambda_{A_1A_2}  ,x,y)  =q^{(1)}_{\lambda_{B_1 B_2}}  (a| \lambda_{A_1A_2}  ,x)  \,  q^{(2)}_{\lambda_{B_1B_2}}  (b| \lambda_{A_1A_2}  ,y)  \, .
\end{align}
As a consequence, the correlation  $\omega_{\lambda_{B_1B_2A_1A_2}}$ in Eq. (\ref{antibitsonly}) cannot violate any Bell inequality, in contradiction with the hypothesis.

To avoid the contradiction,  the decomposition $\lambda_{B_1B_2A_1A_2}  =  (\lambda_{B_1 B_2},  \lambda_{A_1A_2})$ should not be possible, that is, it should not be possible to break the ontic state of the composite system $B_1B_2A_1A_2$  into a part corresponding to the bits and a part corresponding to the anti-bits.  We refer to this fact as {\em ontic inseparability}, in agreement with the terminology used in  Ref. \cite{howard1985einstein}.   Our argument proves, in particular, that the ontic state associated to the composite system  cannot be broken down into  individual ontic states associated to the two bits $B_1$ and $B_2$, and into the two-anti-bit composite $A_1A_2$, that is, it cannot be decomposed as $\lambda_{B_1B_2A_1A_2}  =  (\lambda_{B_1},  \lambda_{B_2} ,  \lambda_{A_1A_2})$.   Even more so, the ontic state cannot be broken down into  a list of ontic states of the individual components, {\em i.e.}  it cannot be decomposed as $\lambda_{B_1B_2A_1A_2}  =  (\lambda_{B_1},  \lambda_{B_2} ,  \lambda_{A_1},  \lambda_{A_2})$.  

\section{Proof of consistency of conditional states\\ (Theorem \ref{theo:consistency} of this Supplemental Material)}\label{sec:consistency}

The proof of Theorem \ref{theo:consistency}
is based on a series of lemmas. For $q\in\Z_d$,  we denote by   
\begin{align}
	\Pi_{D A}^{q}  :=  \sum_{i\in \Z_d}  \,  |i\>\<i|_{D} \otimes |i\oplus q \>\<i\oplus q|_A
\end{align}
the  projector on the subspace $\spc H_{DA}^{q}  \subset \spc H_D \otimes \spc H_A$. 
For $\st q \in \Z_d^m$,  we denote by
\begin{align}\label{PIhpiq}
	\Pi_{\st D \st A}^{(\pi, \st q )}  :  =  \Pi_{D_1A_{\pi (1)}}^{(q_1)} \otimes \cdots \otimes \Pi_{D_mA_{\pi (m)}}^{(q_m)}   \, ,
\end{align} 
the projector on $\spc H^{\pi,  \st q}_{\st D \st A} $.

\begin{lemma}\label{lemma:purecompress}
	Let $S$ be a composite of type $(m,m)$, and let $DA$ be a subsystem of $S$  consisting of a single dit $D$ and a single anti-dit $A$. 
	Then, for every pure state $|\psi\>_S  \in \spc H_s$,
	and  every  $q \in  \Z_d$, the vector 
	\begin{align}
		(\Pi^q_{DA}  \otimes I_{ S \setminus DA})\,  |\psi\>_S    
	\end{align}
	is proportional to a valid pure state.  
\end{lemma}

\Proof 
Proposition \ref{prop:alternative} ensures that every pure state $|\psi\>_S \in \spc H_S$ can written as   
\begin{equation}\label{simplelabelling}
	\begin{aligned}
		\ket{\psi}_S =  \sum_{i_1,\dots, i_m}  \lambda_{i_1 \dots i_m}   ~|i_1\>_{D_1} \otimes \cdots \otimes |i_m\>_{D_m} \otimes   |i_1 \oplus q_1\>_{A_{1}} \otimes \cdots \otimes |i_m\oplus q_m\>_{A_{m}} \, ,  
	\end{aligned}
\end{equation}
for a  suitable vector $\st q  =  (q_1, \dots,  q_m) \in \Z_d^m$ and a  suitable labelling of the dits and anti-dits.   Using this  labelling, the dit $D$ and the anti-dit $A$ can be written as  $D=  D_u$ and $A= A_v$ for suitable integers $u$ and $v$.  
If $u=v$, then we have 
\begin{align}
	(\Pi^q_{D_uA_v}  \otimes I_{ S \setminus D_uA_v})  \ket{\psi}_S  =  \delta_{q, q_u} \,  |\psi\>_S  \, ,
\end{align}
meaning that $(\Pi^q_{D_uA_v}  \otimes I_{ S \setminus D_uA_v})  \ket{\psi}$ is either zero or equal to $|\psi\>_S$.  In either case, it is  proportional to a valid pure state.

Let us now consider the case $v\not  =  u$. 
In this case, we have 
\begin{align}
	\nonumber &  (\Pi^q_{D_uA_v}  \otimes I_{ S \setminus D_uA_v})  \ket{\psi_S}\\
	\nonumber &  
	\quad   =(\Pi^q_{D_uA_v}  \otimes I_{ S \setminus D_uA_v})  \Bigg( \sum_{i_1,\dots,i_u ,\dots, i_{v} , \dots,  i_m}  \lambda_{i_1 \dots i_m}   ~|i_1\>_{D_1} 
	\otimes \cdots \otimes 
	|i_u\>_{D_u}
	\otimes \cdots \otimes 
	|i_{v}\>_{D_{v}}
	\otimes \cdots \otimes 
	|i_m\>_{D_m} \\
	&  \qquad  
	\qquad \otimes   |i_1 \oplus q_1\>_{A_{1}} \otimes \cdots \otimes 
	|i_u \oplus q_u\>_{A_{u}}
	\otimes \cdots \otimes 
	|i_{v} \oplus q_{v}\>_{A_{v}}
	\otimes \cdots \otimes 
	|i_m\oplus q_m\>_{A_{m}}\Bigg) \, .
\end{align}
Now, notice that the projector $\Pi^q_{D_uA_v}$   annihilates all the terms in the sum except the ones that satisfy the relation $i_{v} \oplus q_{v}  =  i_u  \oplus q$,  which is equivalent to
\begin{align}
	i_{v}    =  i_u  \oplus q \ominus  q_{v}  \,.
\end{align}
Hence, we have 
\begin{align}
	\nonumber 
	&(\Pi^q_{D_uA_v}  \otimes I_{ S \setminus D_uA_v})  \ket{\psi} \\
	\nonumber 
	&\quad   = \sum_{i_1,\dots,i_u ,\dots, i_{v-1} ,i_{v+1},   \dots,  i_m}  \lambda_{i_1 \dots i_m}   ~|i_1\>_{D_1} 
	\otimes \cdots \otimes 
	|i_u\>_{D_u}
	\otimes \cdots \otimes 
	|i_u  \oplus q \ominus  q_v  \>_{D_v}
	\otimes \cdots \otimes 
	|i_m\>_{D_m} \\
	\label{ultima} &  \qquad  
	\qquad \otimes   |i_1 \oplus q_1\>_{A_1} \otimes \cdots \otimes 
	|i_u \oplus q_u\>_{A_u}
	\otimes \cdots \otimes 
	|i_u \oplus q\>_{A_v}
	\otimes \cdots \otimes 
	|i_m\oplus q_m\>_{A_m}\, .
\end{align}
One can see that, up to normalization, the r.h.s. is a valid pure state of system $S$:  every dit is paired with an anti-dit, and paired dits/anti-dits belong to subspaces of a given type.  Specifically, dit $D_u$ is paired with anti-dit $A_v$ and belongs to the subspace $\spc H^q_{D_uA_v}$, dit $D_{v}$ is paired with anti-dit $A_{u}$ and belongs to the subspace $\spc H^{q_u  \oplus q_{v} \ominus q}_{D_vA_u}$, while dit $D_j$ with $j\in \{1,\dots, m\} \setminus  \{u,  v\}$ is paired with anti-dit $A_{j}$ and belongs to the subspace  $\spc H^{q_j}_{D_jA_j}$. 
Hence,  one has   
\begin{align}
	(\Pi^q_{D_uA_v}  \otimes I_{ S \setminus D_uA_v})  \ket{\psi} \in \spc H_{\st D \st A}^{\pi , \st q'} = \spc H_{D_1 A_{\pi (1)}}^{  q_1'} \otimes   \cdots \otimes  \spc H_{D_m A_{\pi (m)}}^{  q_m'} \, ,
\end{align}
where $\pi \in  {\sf S}_m$ is the permutation that exchanges $u$ and $v$, and $\st q' \in \Z_d^m$ is the vector defined by $q_u'  =  q_u  \oplus q$ and $q_v'  :  =  q_v  \oplus q_u  \ominus q$, and $q_j'  =  q_j\, ,\forall  j   \in  \{1,\dots,  m\} \setminus\{u,v\}$.    
\qed

\medskip

\begin{lemma}\label{lem:labelling1}
	Let $S$ be a composite of type $(m,m)$, let $|\psi\>_S  \in \spc H_S$ be a pure state of system $S$, and let  $S =  D_1\cdots D_m  A_1\cdots A_m$ be a labelling of the dits/anti-dits in $S$ such that, for every $j\in  \{1,\dots,  m\}$, dit $ D_j$ is paired with anti-dit $A_j$ in the state $|\psi\>_S$.  
	Let $u$ and $v$ be two integers in $\{1,\dots, m\}$ and let $D_uA_v$ be the corresponding subsystem of $S$. Then, there exists a new labelling $S =  D'_1\cdots D'_m  A'_1\cdots A'_m$ such that 
	\begin{enumerate}
		\item for every $j\in  \{1,\dots,  m\}$, dit $D'_j$ is paired with dit $A'_j$ in any  vector of the form $(\Pi^q_{D_uA_v}  \otimes I_{ S \setminus DA})    \,  |\psi\>_S$ with $q\in\Z_d$,
		\item $D_j'=  D_j$ for every $j\in \{1,\dots,  m\}$,
		\item $A_{u}'  =A_v$ and $A_{v}'  = A_u$
		\item  $A_j'=A_j$ for every $j\in \{1,\dots,  m\}\setminus \{u,v\}$. 
	\end{enumerate}
\end{lemma}

\Proof  With the initial labelling $S  = D_1\cdots D_m  A_1\cdots A_m$, the state $|\psi\>_S$     can be written as in Eq. (\ref{simplelabelling}) and the dit/anti-dit pair $DA$ can be written as $D= D_u$ and $A=A_v$ for suitable integers $u$ and $v$.     If $u=v$,  then there is nothing to prove:  the action of the projector $\Pi^q_{D_uA_v}$ does not alter the pairing of dits with anti-dits, and the original labelling already has the desired properties.  In other words, the lemma is proved by simply setting  $D_j'  =  D_j$ and $A_j'  = A_j$ for every $j\in  \{1,\dots,  m\}$.  

If $u\not=  v$,   the state $(\Pi^{q}_{D_uA_v}  \otimes I_{ S \setminus D_u A_v})    \,  |\psi\>_S$  takes the form (\ref{ultima}).  From  the discussion in the lines below Eq. (\ref{ultima}) it is evident that  the dit  $D_j$ and the anti-dit $A_j$ are paired for every $j\in \{1,\dots, m\}\setminus \{u,v\}$. Then, a new labelling with the desired properties can be built by setting  $D'_j:= D_j \, , \forall j\in  \{1,\dots,  m\}$,  $A_u': =  A_v$,   $A_v'  :=  A_u$,    and $A_j'  = A_j$ for every $j\in \{1,\dots, m\}\setminus \{u,v\}$.  \qed  

\medskip 

\begin{lemma}\label{lemma:labellingm}
	Let $S$ be a composite of type $(m,m)$, let $|\psi\>_S  \in \spc H_S$ be a pure state of system $S$, and let  $S =  D_1\cdots D_m  A_1\cdots A_m$ be a labelling of the dits/anti-dits in $S$ such that, for every $j\in  \{1,\dots,  m\}$, dit $ D_j$ is paired with anti-dit $A_j$ in the state $|\psi\>_S$.   For an integer $n\le m$, let $\{u_1,\dots,  u_{n}\}\subseteq \{1,\dots,  m\}$ and  $\{v_1,\dots,  v_{n}\}\subseteq \{1,\dots,  m\}$ be two subsets  of cardinality $n$, and let  $T_n:=  D_{u_1} \cdots D_{u_n} A_{v_1}  \cdots A_{v_n}$ be the corresponding subsystem of $S$.   Then, there exists a new labelling $S =  D'_1\cdots D'_m  A'_1\cdots A'_m$ such that 
	\begin{enumerate}
		\item for every $j\in  \{1,\dots,  m\}$, dit $D'_j$ is paired with anti-dit $A'_j$ in any  vector of the form 
		\begin{align}\left(\Pi^{q_1}_{D_{u_1} A_{v_1}} \otimes \cdots \otimes \Pi^{q_{n}}_{ D_{u_{n}}  A_{v_{n}}}  \otimes I_{ S \setminus T_n}\right)    \,  |\psi\>_S
		\end{align}  with $ (q_1,\dots,  q_{n}) \in\Z_d^{ n}$,
		\item   $D_j'  =  D_j$ for every $j\in  \{1,\dots ,  m\}$, 
		\item $  A'_{u_x}=  A_{v_x}  $  and $  A'_{v_x}=  A_{u_x}$  for every $x\in\{1, \dots, n\}$.
		\item   $A_j'  =  A_j$ for every $j\in  
		\{1,\dots ,  m\}\setminus\{  u_1,\dots,  u_n,v_1,\dots, v_n\}$. 
	\end{enumerate}
\end{lemma}

\Proof The proof is by induction on $n$.  For $n=1$, the desired statement has already been  proven in Lemma \ref{lem:labelling1}. 

Now, we assume that the statement holds for a given $n \in  \{1, \dots,  m-1\}$ and show that it must hold also for $n+1$.  Let $\{u_1,  \dots,  u_{n+1}\} \subseteq  \{1,\dots,  m\}$ and $\{v_1,  \dots,  v_{n+1}\} \subseteq  \{1,\dots,  m\}$ be two subsets of cardinality $n+1\le m$.  
For every class $(q_1, \dots, q_{n+1}) \in  \Z_d^{ (n+1)}$, one has the relation  
\begin{align}
	\nonumber &  \left(\Pi^{q_1}_{D_{u_1} A_{v_1}} \otimes \cdots \otimes  \Pi^{q_{n+1}}_{ D_{u_{n+1}}  A_{v_{n+1}}}  \otimes I_{ S \setminus T_{n+1}}\right)    \,  |\psi\>_S\\
	\nonumber &=\left( \Pi^{q_{n+1}}_{ D_{u_{n+1}}  A_{v_{n+1}}}  \otimes I_{ S \setminus D_{u_{n+1}}  A_{v_{n+1}}}\right) \,   \left(\Pi^{q_1}_{D_{u_1} A_{v_1}} \otimes \cdots \Pi^{q_{n}}_{ D_{u_{n}}  A_{v_{n}}}  \otimes I_{ S \setminus T_n}\right)    \,  |\psi\>_S \\
	& = \left( \Pi^{q_{n+1}}_{ D_{u_{n+1}}  A_{v_{n+1}}}  \otimes I_{ S \setminus D_{u_{n+1}}  A_{v_{n+1}}}\right) \,   |\psi_{n}\>_S  \qquad  |\psi_n\>_S:  = \left(\Pi^{q_1}_{D_{u_1} A_{v_1}} \otimes \cdots \Pi^{q_{n}}_{ D_{u_{n}}  A_{v_{n}}}  \otimes I_{ S \setminus T_n}\right)    \,  |\psi\>_S   \, .
\end{align}  
The induction hypothesis, applied to the state $|\psi\>_S$, guarantees that there exists a new labelling $S=  D_1'  \cdots  D_m'  A_1'  \cdots  A_m'$ such that 
\begin{enumerate}
	\item for every $j\in  \{1,\dots,  m\}$, dit $D'_j$ is paired with anti-dit $A'_j$ in the  vector $|\psi_n\>_S$, 
	\item   $D_j'  =  D_j$ for every $j\in  \{1,\dots ,  m\}$,
	\item   $    A'_{u_x}= A_{v_x}$ and $    A'_{v_x}= A_{u_x}$ for every $x\in\{1, \dots, n\}$.
	\item   $    A'_{j}= A_{j}$ for every $j\in\{1, \dots, m\} \setminus \{  u_1, \dots,  u_n, v_1,\dots,  v_n\}$.
\end{enumerate}
By further applying the induction hypothesis to the state $|\psi_n\>_S$, we obtain another labelling  $S=  D_1^{\prime \prime}  \cdots  D_m^{\prime \prime}  A_1^{\prime \prime}  \cdots  A_m^{\prime \prime}$ such that   
\begin{enumerate}
	\item for every $j\in  \{1,\dots,  m\}$, dit $D_j^{\prime \prime}$ is paired with anti-dit $A_j^{\prime \prime}$ in the vector 
	$\left(\Pi^{q_{n}}_{ D_{u_{n+1}}  A_{v_{n+1}}}    \otimes I_{ S \setminus D_{u_{n+1}}  A_{v_{n+1}}}\right) \,   |\psi_{n}\>_S$,
	\item   $D_j^{\prime \prime}  =  D_j'$ for every $j\in  \{1,\dots ,  m\}$,  
	\item $A_{u_{n+1}}^{\prime \prime}  =  A'_{v_{n+1}}$ and $A_{v_{n+1}}^{\prime \prime}  =  A'_{u_{n+1}}$,
	\item $A_j^{\prime \prime}  =  A_j'$  for every $j\in  \{1,\dots, m\} \setminus\{ u_{n+1},v_{n+1}\}$.
\end{enumerate}  
Note that the decomposition $S=  D_1^{\prime \prime}  \cdots  D_m^{\prime \prime}  A_1^{\prime \prime}  \cdots  A_m^{\prime \prime}$ has all the desired properties:  
\begin{enumerate}
	\item for every $j\in  \{1,\dots,  m\}$, dit $D^{\prime \prime}_j$ is paired with anti-dit $A^{\prime \prime}_j$ in the vector
	\begin{align}\left(\Pi^{q_1}_{D_{u_1} A_{v_1}} \otimes \cdots \otimes \Pi^{q_{n+1}}_{ D_{u_{n+1}}  A_{v_{n+1}}}  \otimes I_{ S \setminus T_{n+1}}\right)    \,  |\psi\>_S\,,
	\end{align}  
	\item   $D_j'  =  D_j$ for every $j\in  \{1,\dots ,  m\}$, 
	\item $  A'_{u_x}=  A_{v_x}  $  and $  A'_{v_x}=  A_{u_x}$  for every $x\in\{1, \dots, n+1\}$.
	\item   $A_j'  =  A_j$ for every $j\in  
	\{1,\dots ,  m\}\setminus\{  u_1,\dots,  u_{n+1},v_1,\dots, v_{n+1}\}$.
\end{enumerate} 
Hence, the validity of the desired statement for an integer $n\le m-1$ implies its validity for the successive integer $n+1$. This completes the proof by induction. 
\qed

\medskip  
\begin{lemma}\label{lem:twoprojectors}
	Let $S =  D_1\cdots D_m  A_1\cdots A_m$ be a composite of type $(m,m)$, and let $|\psi\>_S  \in \spc H_S$ be a pure state of system $S$.  For an integer $n\le m$, let $\{u_1,\dots,  u_{n}\}\subseteq \{1,\dots,  m\}$ and  $\{v_1,\dots,  v_{n}\}\subseteq \{1,\dots,  m\}$ be two subsets  of cardinality $n$, and let  $T_n:=  D_{u_1} \cdots D_{u_n} A_{v_1}  \cdots A_{v_n}$ be the corresponding subsystem of $S$.   Then, there exists a bijective function $s:  \{1,\dots, m\} \setminus \{u_1,\dots,  u_n\} \to  \{1,\dots, m\} \setminus \{v_1,\dots,  v_n\}$ such that, for every vector $ \st q=  (q_1,\dots,  q_n)\in  \Z_d^{ n}$ there exists another vector $\st l = (l_j)_{j\in \{1,\dots, m\} \setminus \{u_1,\dots,  u_n\}   } \in  \Z_d^{ (m-n)}$ satisfying the condition 
	\begin{align}\label{twoprojectors}
		\left(\Pi_{T_n}^{\st q}  \otimes I_{S\setminus T_n}\right)  \,|\psi\>_S =    \left(\Pi_{T_n}^{\st q}  \otimes \Pi_{S\setminus T_n}^{\st l} \right)  \,|\psi\>_S 
	\end{align}
	with 
	\begin{align}
		\Pi^{\st q}_{T_n}  :  =\bigotimes_{x\in  \{1,\dots,  n\}}    \Pi^{q_x}_{D_{u_x}A_{v_x}}    \qquad  {\rm and}  \qquad     \Pi^{\st l}_{S\setminus T_n}  :  =  \bigotimes_{  j\in  \{ 1,\dots, m \}\setminus\{  u_1,\dots   u_n\}}   \Pi^{l_j}_{D_jA_{s(j)}}  \,.  
	\end{align}  
\end{lemma}

\Proof Without loss of generality, let us assume that, for every $j\in  \{1,\dots, m\}$,  dit $D_j$ is paired with anti-dit $A_j$ in  the state $|\psi\>_S$ (if this is not the case, one can just relabel the dits and anti-dits of $S$, as shown in Proposition \ref{prop:alternative}).    Then,  we apply Lemma \ref{lemma:labellingm}, thus obtaining a new labelling  $S=  D_1^{ \prime}  \cdots  D_m^{ \prime}  A_1^{ \prime}  \cdots  A_m^{ \prime}$  such that 
\begin{enumerate}
	\item for every $j\in  \{1,\dots,  m\}$, dit $D^{\prime}_j$ is paired with anti-dit $A^{\prime }_j$ in the vector
	\begin{align}\label{projectedpsi}\left(\Pi^{q_1}_{D_{u_1} A_{v_1}} \otimes \cdots \otimes \Pi^{q_{n}}_{ D_{u_{n}}  A_{v_{n}}}  \otimes I_{ S \setminus T_n}\right)    \,  |\psi\>_S =  \left(\Pi_{T_n}^{\st q}  \otimes I_{S\setminus T_n}\right)  \,|\psi\>_S =:   |\psi'\>_S\,,
	\end{align}  
	\item   $D_j'  =  D_j$ for every $j\in  \{1,\dots ,  m\}$, 
	\item $  A'_{u_x}=  A_{v_x}  $  and $  A'_{v_x}=  A_{u_x}$  for every $x\in\{1, \dots, n\}$.
	\item   $A_j'  =  A_j$ for every $j\in  
	\{1,\dots ,  m\}\setminus\{  u_1,\dots,  u_n,v_1,\dots, v_n\}$.
\end{enumerate}
Now, the subsystem $S\setminus T_n$ can be decomposed as $S\setminus T_n  =   \bigotimes_{j\in  \{1,\dots, m\} \setminus\{  u_1,\dots,  u_n\}}  D'_{j} A_j'$.  Since every dit $D'_j$ is paired to the corresponding anti-dit $A'_j$ in the vector $|\psi'\>_S$,  there exists a vector $(l_{j})_{j\in {j\in  \{1,\dots, m\} \setminus\{  u_1,\dots,  u_n\}} }$ such that 
\begin{align}\label{almost}
	|\psi'\>_S  =  \left( I_{T_n}  \otimes \Pi_{S\setminus T_n}^{\st l} \right)  \, |\psi'\>_S \, .
\end{align}
with 
\begin{align}
	\Pi_{S\setminus T_n}^{\st l}   : =   \bigotimes_{{j\in  \{1,\dots, m\} \setminus\{  u_1,\dots,  u_n\}}}  \Pi_{D_j'A_j'}^{ l_j } \,.
\end{align}
Recalling that $  A'_{v_x}=  A_{u_x}$  for every $x\in\{1, \dots, n\}$ and $A_j'  =  A_j$ for every $j\in  
\{1,\dots ,  m\}\setminus\{  u_1,\dots,  u_n,v_1,\dots, v_n\}$, we can define the bijective function $s:  \{ 1,\dots,  m\} \setminus \{u_1,\dots,  u_n\}  \to  \{ 1,\dots,  m\} \setminus \{v_1,\dots,  v_n\}$ as 
\begin{align}
	\nonumber s(v_x)  &=  u_x  \qquad   &\forall x\in\{1,\dots,  n\}\\
	s(j)  & =  j  \qquad& \forall j\in \{ 1,\dots,  m\}\setminus \{u_1,\dots,  u_n, v_1,\dots,  v_n\} \, .
\end{align}
and rewrite the projector $\Pi_{S\setminus T_n}^{\st l} $ as 
\begin{align}
	\Pi_{S\setminus T_n}^{\st l}    =   \bigotimes_{{j\in  \{1,\dots, m\} \setminus\{  u_1,\dots,  u_n\}}}  \Pi_{D_j A_{s(j)}}^{ l_j } \,.
\end{align}
This observation, together with Eq. (\ref{almost}),  concludes the proof. \qed  

\medskip  

\begin{lemma}\label{lem:consistencymmpure}
	Let $S$  be a composite of type $(m,m)$ and let $T$ be a subsystem of $S$, of type $(n,n)$ with $n\le m$.  For every pure state  $|\psi\>_S  \in \spc H_S$  and every pure state $|\phi\>\in  \spc H_T$, the vector  $(\<\phi|_T \otimes I_{S\setminus T}) \,  |\psi\>_S$ is proportional to a pure state of system $S\setminus T$.  
\end{lemma}

\Proof  Let $S=  D_1\cdots D_m  A_1\cdots A_m$ be an arbitrary  decomposition of $S$.    Since $|\phi\>_T$ is a pure state of system $T$, every dit in $T$, say $D_{u_x}$, must be paired to a suitable anti-dit in $T$, say $A_{v_x}$, for $x\in \{1,\dots, n\}$.  Hence, there exists a vector $\st q  =  (q_1,\dots, q_n )\in  \Z_d^{n}$ such that 
\begin{align}
	|\phi\>_T   =  \Pi_{\st q}^T  \, |\phi\>_T  \qquad       \Pi^{\st q}_{T}  :  =\bigotimes_{x\in  \{1,\dots,  n\}}    \Pi^{q_x}_{D_{u_x}A_{v_x}}    
\end{align}
Then, Lemma \ref{lem:twoprojectors} implies that there exists a bijective function $s:  \{1,\dots,  m\} \setminus \{u_1,  \dots,  u_n\} \to \{1,\dots,  m\} \setminus \{v_1,  \dots,  v_n\} $ such that  one has 
\begin{align}
	\left(  \Pi^{\st q}_{T} \otimes I_{S\setminus T} \right)   |\psi\>_S   =   \left(  \Pi^{\st q}_{T} \otimes  \Pi^{\st l}_{S\setminus T} \right)\,|\psi\>_S   \qquad {\rm with}  \qquad     \Pi^{\st l}_{S\setminus T}  :  =  \bigotimes_{  j\in  \{ 1,\dots, m \}\setminus\{  u_1,\dots   u_n\}}   \Pi^{l_j}_{D_jA_{s(j)}}  \,.  
\end{align}
Hence, we have  
\begin{align}
	\nonumber (\<\phi|_T \otimes I_{S\setminus T}) \,  |\psi\>_S  &  =(\<\phi|_T \otimes I_{S\setminus T})   \left(  \Pi^{\st q}_{T} \otimes I_{S\setminus T} \right)   |\psi\>_S    \\
	\nonumber &=  (\<\phi|_T \otimes I_{S\setminus T})  \left(  \Pi^{\st q}_{T} \otimes  \Pi^{\st l}_{S\setminus T} \right)\,|\psi\>_S   \\
	\nonumber 
	&  =   (\<\phi|_T \otimes I_{S\setminus T})  \left(  I_T \otimes  \Pi^{\st l}_{S\setminus T} \right)\,|\psi\>_S\\
	&=   \Pi^{\st l}_{S\setminus T} \,    (\<\phi|_T \otimes I_{S\setminus T}) \,|\psi\>_S\, .
\end{align}
Hence, the vector $ (\<\phi|_T \otimes I_{S\setminus T}) \,|\psi\>_S$ is invariant under the action of the projector $\Pi^{\st l}_{S\setminus T}$. By definition of the set of pure states (Definition \ref{def:purestatesmm}), this condition guarantees that $ (\<\phi|_T \otimes I_{S\setminus T}) \,|\psi\>_S$   is proportional to a valid pure state of system $S\setminus  T$. \qed

\begin{lemma}\label{lem:consistencymnpure}
	Let $S$  be a composite of arbitrary type $(m,n)$ and let $T$ be a subsystem of $S$, of type $(m',n')$ with $m'\le m$ and $n'\le n$.  For every pure state  $|\psi\>_S  \in \spc H_S$  and every pure state $|\phi\>\in  \spc H_T$, the operator  
	\begin{align}
		(\<\phi|_T \otimes I_{S\setminus T}) ~ |\psi\>\<\psi|_S  ~ (|\phi\>_T \otimes I_{S\setminus T}) 
	\end{align}
	is proportional to a pure state of system $S\setminus T$.  
\end{lemma}

\Proof   Suppose that $m'\ge n'$.  In that case, we can add $m'-n'$ anti-dits to both $S$ and $T$, thus obtaining two systems $S'$ and $T'$ of type $(m+m'-n',  n+m'-n')$  and $(m',m')$ respectively.  If $m>n$  ($m<n$), add $m-n$  anti-dits ($n-m$ dits) to $S'$, obtaining a new system $S^{\prime \prime}$ of type $(k,k)$, with $k=\max\{m,n\}  +  |m'-n'|$.  Then, define the pure states
\begin{align}
	\nonumber |\phi'\>_{T'}  &=  |\phi\>_T  \otimes |\st r\>_{T'\setminus T}\\
	|\psi^{\prime \prime}\>_{S^{\prime \prime}}  &=  |\psi\>_S  \otimes   |\st r\>_{T'\setminus T}  \otimes |\st s\>_{S^{\prime \prime}  \setminus [S  \,   (T'\setminus T)]}
	\, ,
\end{align}
where $|\st r\>_{T'\setminus T}$ is a computational basis state of the $m'-n'$ anti-bits added to $T$, and $|\st s\>_{S^{\prime \prime}\setminus [S (T'\setminus T)]}$ is a computational basis state of remaining the dits and anti-dits added to $S$.  

With the above definition, we have 
\begin{align*}
	\nonumber   (\<\phi|_T \otimes I_{S\setminus T}) ~ |\psi\>\<\psi|_S  ~ (|\phi\>_T \otimes I_{S\setminus T})  &  =  \Big(\<\phi|_T \otimes  \<\st r|_{T'\setminus T}  \otimes  I_{S\setminus T}\Big)~   \Big(|\psi\>\<\psi|_S  \otimes |\st r\>\<\st r|_{T'\setminus T}  \Big)  ~      \Big(|\phi\>_T \otimes |\st r\>_{T'\setminus T} \otimes I_{S\setminus T}\Big)\\
\end{align*}
and therefore
\begin{align}
	\nonumber 
	&\Bigg(  (\<\phi|_T \otimes I_{S\setminus T}) ~ |\psi\>\<\psi|_S  ~ (|\phi\>_T \otimes I_{S\setminus T})\Bigg)  ~    \otimes~  \Bigg( |\st s\>\<\st s|_{S^{\prime \prime}  \setminus [S  \,   (T'\setminus T)]}\Bigg)  
	\\
	\nonumber & \qquad  =  (\<\phi|_T\otimes  \<\st r|_{T'\setminus T}  \otimes  I_{S^{\prime \prime}\setminus T'})~   (|\psi\>\<\psi|_S  \otimes |\st r\>\<\st r|)_{T'\setminus T}   \otimes  |\st s\>\<\st s|_{S^{\prime \prime}  \setminus [S  \,   (T'\setminus T)]} )  ~      (|\phi\>_T \otimes |\st r\>_{T'\setminus T} \otimes I_{S^{\prime \prime}\setminus T'})\\
	&  \qquad =  (\<\phi'|_{T'} \otimes I_{S^{\prime \prime}\setminus T'}) ~ |\psi'\>\<\psi'|_{S^{\prime \prime}}  ~ (|\phi'\>_{T'} \otimes I_{S^{\prime \prime}\setminus T'}) \, .   \label{ultimostato}
\end{align}
Now, Lemma \ref{lem:consistencymmpure} guarantees that the r.h.s. of Eq. (\ref{ultimostato}) is proportional to a valid pure state of system $S^{\prime \prime} \setminus T'$.  

To conclude, note that 
\begin{align}
	(\<\phi|_T \otimes I_{S\setminus T}) ~ |\psi\>\<\psi|_S  ~ (|\phi\>_T \otimes I_{S\setminus T})   =  \Tr_{S^{\prime \prime}  \setminus [S  \,   (T'\setminus T)]}  \left[  (\<\phi'|_{T'} \otimes I_{S^{\prime \prime}\setminus T'}) ~ |\psi'\>\<\psi'|_{S^{\prime \prime}}  ~ (|\phi'\>_{T'} \otimes I_{S^{\prime \prime}\setminus T'})  \right] \, .
\end{align}
Since the state spaces are closed under partial trace (Theorem \ref{theo:marginal}), we conclude that the r.h.s. of the above equation is proportional to a valid state of system $S\setminus T$.  The state is pure, because the l.h.s. of the above equation is rank-one. \qed  

\medskip  

We are finally ready to prove Theorem \ref{theo:consistency}.  

\medskip  

{\bf Proof of Theorem \ref{theo:consistency}.}  
For the proof, we adopt a notation that is consistent with the notation of the previous lemmas: instead of denoting the composite system by $ST$, we will denote it by $S$, and we will denote its subsystems by  $T$ and $S\setminus T$.    

Let $\rho_{S}  \in \St (S)$ be an arbitrary state of system $S$ and let $P_T  \in  \Eff (T)$ be an arbitrary effect on subsystem $T$. The operators $\rho_S$ and $P_T$ can be decomposed as   $\rho_S  =  \sum_{ i}  \, \lambda_i  \,  |\psi_i\>\<\psi_i|_S$ and $P_T  =  \sum_j  \, \mu_j  \,  |\phi_j\>\<\phi_j|_T$, where $(\lambda_i)_i$ and $(\mu_j)_j$ are positive coefficients, $|\psi_i\>_S$ are pure states of $S$, and  $|\phi_j\>$ are pure states of $T$.  Using this decomposition, we obtain  
\begin{align}
	\nonumber \Tr_{T}   [   (P_T  \otimes I_{S\setminus T}) \,  \rho_S ]   &  =  \sum_{i,j}  \lambda_i  \,  \mu_j  \,     \Tr_{T}   [   (|\phi_j\>\<\phi_j|_T  \otimes I_{S\setminus T}) \,  |\psi_i\>\<\psi_i|_S ]  \\
	&  = \sum_{i,j}  \lambda_i  \,  \mu_j  \,     (\<\phi_j|_T  \otimes I_{S\setminus T})~  |\psi_i\>\<\psi_i|_S ~ (|\phi_j\>_T  \otimes I_{S\setminus T})\,.
\end{align}
By Lemma \ref{lem:consistencymnpure}, each summand in the r.h.s. is proportional to a valid pure state. Then, convexity of the set of pure states and the fact that the trace of the r.h.s. is less than 1, implies that the r.h.s is a valid (generally subnormalized) pure state. \qed

\end{document}